\documentclass[traditabstract]{aa} %
\usepackage{amsmath}
\usepackage{graphicx}
\usepackage{epstopdf}
\usepackage{deluxetable}

\makeatletter
\def\@to{to}
\makeatother

\usepackage{txfonts}
\usepackage{mathrsfs}
\usepackage{natbib}
\bibpunct{(}{)}{;}{a}{}{,}
\usepackage{hyperref}
\newcommand{\Msun}{~M_\odot}

\newcommand{\kms}{\rm ~km~s^{-1}}
\newcommand{\ergs}{\rm ~erg~s^{-1}}

\newcommand{\ml}{~\Msun ~\rm yr^{-1}}
\newcommand{\mll}{\Msun ~\rm yr^{-1}}

\begin{document}

\title{The Carnegie Supernova Project II. The shock wave revealed through the fog: The strongly interacting Type~IIn SN 2013L\thanks{This paper includes data gathered with the 6.5 meter Magellan Telescopes located at Las Campanas Observatory, Chile.}$^{,}$\thanks{Photometry and spectra presented in this paper are available on \href{https://wiserep.weizmann.ac.il/}{WISeREP}.}}

\author{
F. Taddia\inst{1,2} 
\and
M. D. Stritzinger\inst{1}
\and
C. Fransson\inst{2}
\and 
P. J. Brown\inst{3}
\and 
C. Contreras\inst{4}
\and 
S. Holmbo\inst{1}
\and
T. J. Moriya\inst{5,6}
\and
M. M. Phillips\inst{4}
\and
J. Sollerman\inst{2}
\and
N. B. Suntzeff\inst{3}
\and
C. Ashall\inst{7}
\and
C. R. Burns\inst{8}
\and 
L. Busta\inst{4}
\and
A. Campillay\inst{4}
\and
S. Castell\'{o}n\inst{4}
\and
C. Corco\inst{4}
\and
F. Di Mille\inst{4}
\and
C. Gall\inst{9,1}
\and
C.  Gonz\'{a}lez\inst{4}
\and 
E. Y. Hsiao\inst{7}
\and
N. Morrell\inst{4}
\and
A. Nyholm\inst{2}
\and
J. D. Simon\inst{8}
\and
J. Ser\'{o}n\inst{10,4}}

\institute{
Department of Physics and Astronomy, Aarhus University, Ny Munkegade 120, DK-8000 Aarhus C, Denmark
\and
 The Oskar Klein Centre, Department of Astronomy, Stockholm University, AlbaNova, 10691 Stockholm, Sweden
 \and 
 George P. and Cynthia Woods Mitchell Institute for Fundamental Physics and Astronomy, Texas A\&M University, Department of Physics and Astronomy, College Station, TX,
 77843, USA
 \and
 Carnegie Observatories, Las Campanas Observatory, Casilla 601, La Serena, Chile
 \and
 National Astronomical Observatory of Japan, National Institutes of Natural Sciences, 2-21-1 Osawa, Mitaka, Tokyo 181-8588, Japan
 \and
 School of Physics and Astronomy, Faculty of Science, Monash University, Clayton, VIC 3800, Australia
 \and
 Department of Physics, Florida State University, Tallahassee, FL 32306, USA
 \and
 Observatories of the Carnegie Institution for Science, 813 Santa Barbara St., Pasadena, CA 91101, USA
 \and 
DARK, Niels Bohr Institute, University of Copenhagen, Lyngbyvej 2, 2100 Copenhagen, Denmark
 \and 
Cerro Tololo Inter-American Observatory, National Optical Astronomy Observatory, Casilla 603, La Serena, Chile
}
 
\date{Received; accepted}

\abstract{We present ultra-violet (UV) to mid-infrared (MIR) observations of the long-lasting Type~IIn supernova (SN) 2013L obtained by the \textit{Carnegie Supernova Project II} (CSP-II)  beginning two days after discovery and extending until +887 days (d). 
The SN reached a peak $r$-band absolute magnitude of $\approx-$19~mag and an even brighter UV peak, and its light curve evolution resembles that of SN~1988Z. 
The spectra of SN~2013L are dominated by hydrogen emission features, characterized by three components attributed to different emission regions. A unique feature of this Type~IIn SN is that, apart from the first epochs, the blue shifted line profile is dominated by the macroscopic velocity of the expanding shock wave of the SN.  We are therefore able to trace the evolution of the shock velocity in the dense and partially opaque circumstellar medium (CSM), from $\sim 4800 \kms$ at +48~d, decreasing as $t^{-0.23}$ to $\sim 2700  \kms$ after a year. 
We performed spectral modeling of both the broad- and intermediate-velocity components of the H$\alpha$ line profile. The high-velocity component is consistent with emission from a radially thin, spherical shell located behind the expanding shock with emission wings broadened by electron scattering. We propose that the intermediate component originates from preionized gas from the unshocked dense CSM with the same velocity as the narrow component, $\sim 100 \kms$, but also that it is broadened by electron scattering. 
These features provide direct information about the shock structure, which is consistent with model calculations.
The spectra exhibit broad \ion{O}{i} and [\ion{O}{i}] lines that emerge at $\gtrsim$+144~d and broad \ion{Ca}{ii} features. The spectral continua and the spectral energy distributions (SEDs) of SN~2013L after +132~d are well reproduced by a two-component black-body (BB) model; one component represents emitting material with a temperature between 5$\times$10$^3$ to 1.5$\times$10$^4$~K (hot component) and the second component is characterized by a temperature around 1--1.5$\times$10$^3$~K (warm component). The warm component dominates the  emission at very late epochs ($\gtrsim$~+400~d), as is evident from both the last near infrared (NIR) spectrum and MIR observations obtained with the \textit{Spitzer Space Telescope}. Using the BB fit to the SEDs, we constructed a  bolometric light curve that was  modeled together with the unshocked CSM velocity and the shock velocity derived from the H$\alpha$ line modeling. The circumstellar-interaction model of the bolometric light curve reveals a mass-loss rate history with large values ( $1.7\times 10^{-2} - 0.15 \ml$) over the $\sim $25 -- 40 years before explosion, depending on the radiative efficiency and anisotropies in the CSM. The drop in the light curve at $\sim 350$ days and the presence of electron scattering wings at late epochs indicate an anisotropic CSM. The mass-loss rate values and the unshocked-CSM velocity are consistent with the characteristics of a massive star, such as a luminous blue variable (LBV) undergoing strong eruptions, similar to $\eta$ Carina. 
Our analysis also suggests a scenario where pre-existing dust grains have a distribution that is characterized by a small covering factor.}

\keywords{supernovae: general -- supernovae: individual: SN~2013L}

\authorrunning{Taddia et al.}
\titlerunning{A CSP-II study of the Type~IIn SN~2013L}

\maketitle

\section{Introduction}

Type IIn supernovae (SNe~IIn) were introduced  as a class by \citet{schlegel90}, who noted the presence of narrow hydrogen emission lines superposed on blue spectra. 
The emission lines that are characteristic of SNe~IIn are generated from the interaction between rapidly expanding SN ejecta and a hydrogen-rich circumstellar medium (CSM), which drives the formation of a forward shock into the CSM and a reverse shock into the ejecta \citep[see, e.g.,][]{chevalier94}. Multiple components can form the emission lines. These include a narrow ($v_{FWHM} \sim 100$~km~s$^{-1}$) component originating  from unshocked CSM, ionized by the high-energy photons emitted from the shock region. In addition, an intermediate component ($v_{FWHM} \sim 1000$~km~s$^{-1}$) has been observed to arise from the region between the forward shock and the reverse shock. A broad component ($v_{FWHM} \sim 10\,000$~km~s$^{-1}$) can arise from the ionized expanding SN ejecta, or it is more likely due to electron-scattering broadening of the narrower component by an optically thick CSM \citep[e.g.,][]{chugai01,huang17}.

The picture is further complicated by the distribution of the CSM and the presense of dust. 
For example, the CSM  could be aspherical and/or asymmetric \citep[e.g.,][]{stritzinger12}, while the presence of dust \citep[e.g.,][]{fox11} in the ejecta and/or in the CSM could modify the emission line profiles. Dust also produces an excess of red flux as it redistributes bluer flux to near-infrared (NIR) and mid-infrared (MIR) wavelengths \citep[see, e.g.,][]{gall14}.

A variety of light-curve shapes and luminosity levels characterize the various SN~IIn populations \citep[see review in e.g., ][]{smith16}. In \citet{taddia13IIn} and \citet{taddia15met}, three main subgroups are schematized as namely long-lasting SNe~IIn or 1988Z-like objects, fast-declining SNe~IIn-L or 1998S-like objects, and plateau SNe~IIn-P or 1994W-like objects. \citet{nyholm19} identify two groups of slow and fast rising SNe IIn in their Palomar Transient Factory sample. The large range of luminosities goes from faint ($M_r\sim-16$ mag) SNe IIn to
 superluminous type II SNe (SLSNe~II) that appear to be extremely bright SNe~IIn with $M_r\sim-22$ mag \citep{nyholm19}. 
SN~IIn light curves can be used, together with information from the spectral lines, such as CSM velocity, shock velocity, and ejecta velocity, to estimate the mass-loss rate, the ejecta mass, and the explosion energy of the SN progenitor that produces the CSM via its stellar winds and/or episodic mass-loss events \citep[e.g.,][]{moriya13,moriya14}.

\citet{taddia15met} find that different SN~IIn subgroups seem to occur in different environments, with long-lasting SNe~IIn occurring at lower metallicity 
than fast-declining SNe~IIn. Long-lasting SNe~IIn have similar environmental properties as those of SN impostors  that are believed to be produced by luminous-blue-variable (LBV) eruptions. Transients likely arising from significant stellar eruptions  sometimes occur prior to the birth of SNe~IIn. Examples of such objects include the outbursts observed prior to and at the position of SN~2006jc \citep{pastorello07} and  several SNe~IIn studied by PTF/iPTF \citep{ofek14b,nyholm17}.
SN~2009ip has recently been identified as the prototype of a new SN~IIn class, which is characterized by a history of variability prior to two significant outbursts occurring a few weeks apart \citep{pastorello18}. The light curves of SN~2009ip \citep{margutti14} and iPTF13z \citep{nyholm17} also exhibited variability during their main outburst. This  might be related to the final explosion of the star, although it is a matter open to debate  in the case of SN~2009ip \citep[e.g.,][]{fraser13}.

The variety of SN~IIn properties suggests the presence of multiple progenitor channels. LBVs have been invoked as possible progenitors, as they expel significant amounts of mass during their eruptive phases at velocities similar to those observed in SN~IIn spectra \citep{kiewe12,taddia13IIn}. In the traditional stellar evolution scenario, however, LBVs are not supposed to be the terminal phase in the lives of massive stars, but they only represent  a transitional phase as they  evolve and eventually become Wolf Rayet (WR) stars. 

\citet{smith15asocial} argue that LBVs actually arise from relatively isolated environments, suggesting  that they might originate from lower mass stars in binary systems. This result has recently been disputed by \citet{aadland18}, whose work favors the traditional stellar-evolution scenario for LBVs.
Red supergiants with superwinds \citep{smith09_RSG} and yellow hypergiants (YHGs, \citealp{smith14}) have also been suggested to produce transients, such as SN~1995N \citep{fransson02} and PTF11iqb \citep{smith15iqb}. Electron-capture SNe from super-AGB stars have been suggested to produce SNe-like SN~1994W \citep{mauerhan13} and SN impostors that are similar to intermediate luminosity red transient (ILRT) SN~2008S \citep{botticella09}. Wolf-Rayet stars are known to produce CSM interacting SNe, such as the so-called SNe~Ibn, which exhibit helium features and lack hydrogen features \citep{pastorello08}. 

In this paper we focus on the long-lasting Type~IIn SN~2013L, whose bright emission was observed for more than 1500 days. This object is also characterized by the presence of multicomponent, asymmetric emission lines, late-time NIR and MIR excess, and very blue early spectral energy distributions (SEDs) dominated by UV emission. 
SN~2013L was also studied by \citet[][hereafter A17]{andrews17}, who provide $BgVri$ photometry, with most of the epochs obtained between +30~d and +192~d after discovery. They also obtained six extra $R$, $V$, and $i$ band points at later epochs; the last point was obtained at $+$1509~d and they analyze early optical and NIR spectra. A17 also provide late-time (up to +1509~d) spectra covering the H$\alpha$ region.

Here, we present additional data and an analysis of SN~2013L based on observations obtained by the \textit{Carnegie Supernova Project} (CSP-II; \citealt{phillips19}). This includes optical  $uBgVri$-band photometry extending from +2~d after discovery out to +696~d. Furthermore, we present new, multiple epoch NIR $YJH$-band photometry, extending from +4~d to +887~d and one epoch of $K_s$ band at +739~d. We also present UV photometry obtained from the \textit{Neil Gehrels Swift} Observatory, extending from +7~d to +43~d (see also \citealp{delarosa16}). New optical spectra are also presented as are a number of new NIR spectra that were obtained as part of the CSP-II NIR spectroscopy program \citep{hsiao19}. Our aim is to understand the properties of the CSM and the underlying progenitor system of SN~2013L. This comprehensive and high-quality dataset enables us to study SN~2013L in detail, which only a handful of SNe~IIn in the literature  match \citep{stritzinger12,fransson14}.  

This paper is structured as follows: In Sect.~\ref{sec:basic}
we provide basic information on SN~2013L and its host galaxy. In Sect.~\ref{sec:dataacq} we describe data acquisition and reduction. In Sect.~\ref{sec:lc}, we provide an introduction of the SN light curves and colors.
This is followed by the presentation and analysis of the spectra in Sect.~\ref{sec:spec}. 
SEDs and the bolometric light curve are constructed in Sect.~\ref{sec:bolo}. The bolometric properties along with the H$\alpha$ profiles  are modeled in Sect.~\ref{sec:model}, and the results on SN mass loss and dust emission are provided in Sect.~\ref{sec:discussion}. Finally, our conclusions are presented in Sect.~\ref{sec:conclusion}.

\section{SN~2013L and its host galaxy}
\label{sec:basic}
SN~2013L was discovered by \citet{monard13} in ESO 216-39 with an unfiltered apparent magnitude of 15.6. First designated as the possible supernova PSN J11452955-5035531, it was  classified by the Public ESO Spectroscopic Survey of Transient Objects (PESSTO; \citealp{smartt15}) as a young SN~IIn \citep{atel}. The previous nondetection of SN~2013L (with a limiting unfiltered apparent magnitude of 19) dates 18.957 days prior to discovery, which occurred on 2013 Jan 22.025 UT (JD 2456314.525; \citealp{monard13}). We use days (d) relative to discovery to define the phase of the SN throughout this paper. 

\begin{figure*}
\centering
\includegraphics[width=18cm]{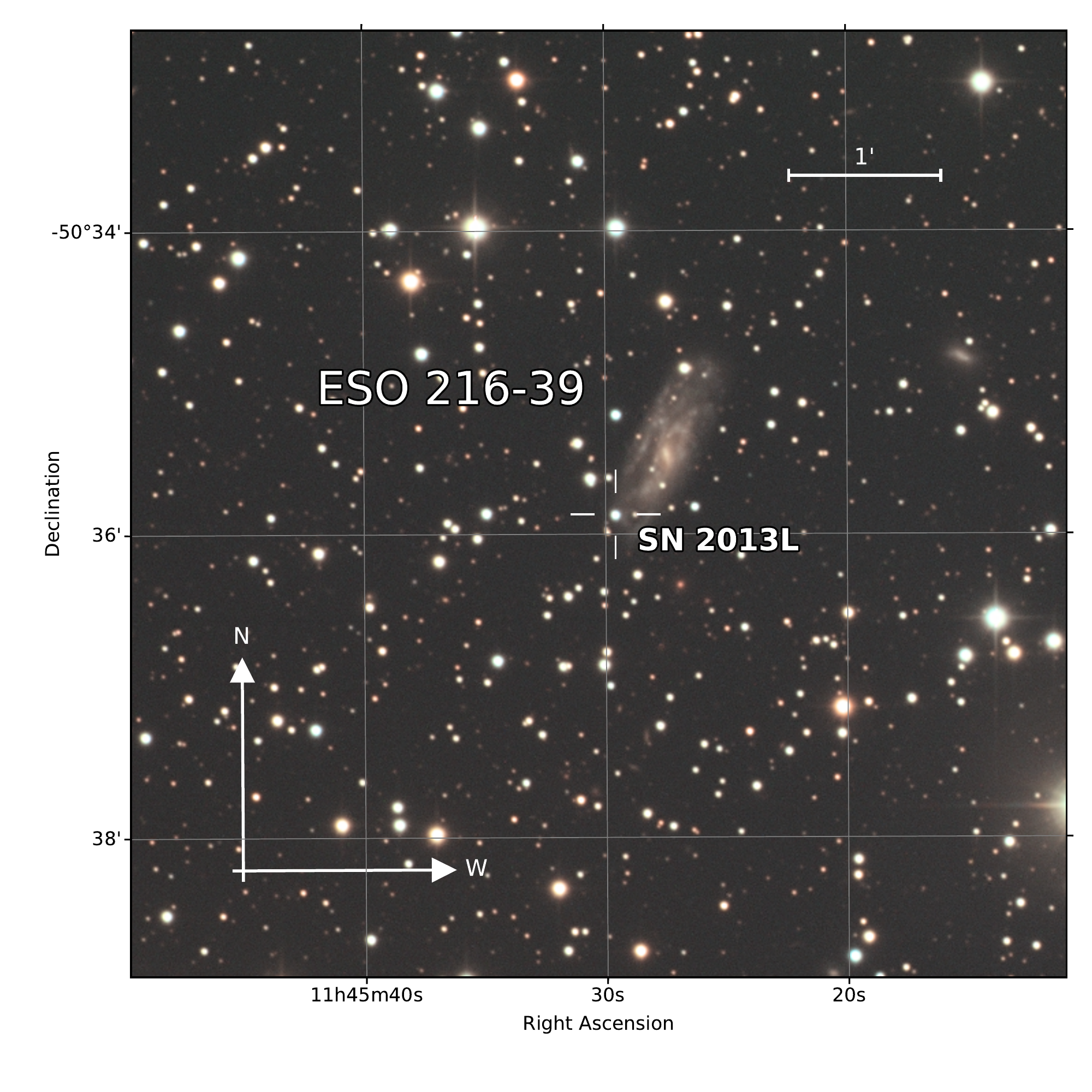}
\caption{\label{FC} Finding chart for SN~2013L, from a combination of all of the $BVgri$ images obtained with the Swope telescope between JD 2456316.7 and JD 2456448.6.}
\end{figure*}

The coordinates of SN~2013L are R.A.(J2000.0)$=$
11h45m29s.55, Decl.(J2000.0) $= -50^\circ$35$'$53\farcs1. The SN is located 20$\arcsec$ east and 24$\arcsec$ south of the galaxy core. A finding chart of SN~2013L in ESO 216-39 is provided in Fig.~\ref{FC}. 
In \citet{taddia15met}, we computed the deprojected distance of SN~2013L from its nucleus, which turned out to be 1.09 times the galaxy radius (r$_{25}$), that is, its location lies in the outskirts of the host. The host-galaxy center is characterized by a N2 \citep{pettini04} metallicity of 12+log(O/H)$=$8.68$\pm$0.18~dex, which when assuming a typical $-$0.47~dex~r$_{25}^{-1}$ metallicity gradient \citep{pilyugin04} implies that SN~2013L was located in a subsolar metallicity environment, that is, 12+log(O/H)$=$8.17$\pm$0.23~dex. This is typical of long-lasting Type~IIn~SNe, such as SN~1988Z and SN impostors \citep{taddia15met}.

ESO 216-39 is located at z$=$0.01738 (A17; NED\footnote{\href{http://ned.ipac.caltech.edu}{http://ned.ipac.caltech.edu}} lists a slightly lower redshift (0.016992$\pm$0.000150), which corresponds to a distance of 72$^{+8.5}_{-5}$ Mpc (A17). The distance modulus is therefore 34.287$^{+0.242}_{-0.156}$~mag. 
According to \citet{schlafly11}, the Milky Way extinction is $A^{MW}_V = 0.367$~mag when assuming a \citet{fitzpatrick99} reddening law that is characterized by $R_V = 3.1$.
The host-galaxy extinction is neglected in A17, and we assume the same in this work. Moreover, a high-resolution spectrum obtained with the Clay ($+$ MIKE) telescope on 
 $+$10 d (see Sect.~\ref{sec:dataacq} and  Sect.~\ref{sec:spec}) shows no indication of prevalent \ion{Na}{i}~D absorption or DIB features. Specifically, the DIB 5780~\AA\ feature is not detectable and the \ion{Na}{i}~D absorption line has a total equivalent width of 0.14$\pm$0.07~\AA. This suggests minimal host-galaxy dust reddening. Indeed, according to Fig.~9 of \citet{phillips13}, this EW value implies $A_V^{host} \lesssim 0.2$ mag.
 In the following, by extinction correction, we mean only Milky-Way extinction correction.

\section{Data acquisition and reduction}
\label{sec:dataacq}

The CSP-II obtained 49 nights of optical $uBgVri$  photometry and 21 nights of NIR ($YJH$) photometry. Optical imaging was obtained with the Henrietta Swope 1.0-m telescope ($+$SITe3 direct CCD camera),  whereas NIR imaging was mainly obtained with the du Pont 2.5-m telescope ($+$RetroCam), while also two nights worth of observations were obtained with the FourStar camera \citep{persson13} on the Magellan Baade 6.5-m telescope. In addition, a single, late $K_s$-band observation was obtained with FourStar.
We also obtained photometric observations with the UltraViolet Optical Telescope (UVOT; \citealp{roming05}) on the Neil Gehrels \textit{Swift} Observatory \citep{gehrels04} in the UV ($uw1$, $um2$, $uw2$) and in the optical ($UBV$) over the course of 19 different nights. 

Reduction of the optical and NIR data was performed as described in \citet{krisciunas17}. 
Given that the SN was still present in our most recent attempt to obtain deep host-galaxy images, we did not perform template subtraction on the science images. However, the background at the location of SN~2013L is quite smooth and low, enabling us to robustly measure its magnitude evolution.
Point-spread function (PSF) photometry of the SN was computed and calibrated relative to a local sequence of stars in the field of SN~2013L. The local sequence itself was  calibrated relative to the standard-star fields  observed over multiple photometric
nights from \citet{landolt92} ($BV$) and \citet{smith02} ($ugri$). 
 The NIR $J$- and $H$-band local sequences were calibrated relative to the standard stars from \citet{persson98}, while the $Y$-band local sequence was calibrated relative to $Y$-band  magnitudes of \citeauthor{persson98} standards presented in \citet{krisciunas17}. Finally, $K_{s}$-band photometry of SN~2013L was calibrated relative to 2MASS stars.
 Photometry of the local sequences  in the standard system are listed in Table~\ref{SN13L_opt_locseq} (optical) and Table~\ref{SN13L_nir_locseq} (NIR), while  photometry of SN~2013L in the CSP-II natural system is listed in Table~\ref{tab:phot} (optical) and Table~\ref{tab:photnir} (NIR).

Swift images were reduced using aperture photometry and subtraction of the underlying host-galaxy count rates following the method described by \citet{brown14} and by using the zero-points of \citet{breeveld11}. The Swift UVOT images and photometry are available in the Swift Optical Ultraviolet Supernova Archive (SOUSA; \citealp{brown14}), and they are listed in Table~\ref{tab:photUVOT}. 

\begin{figure}
\includegraphics[width=18cm]{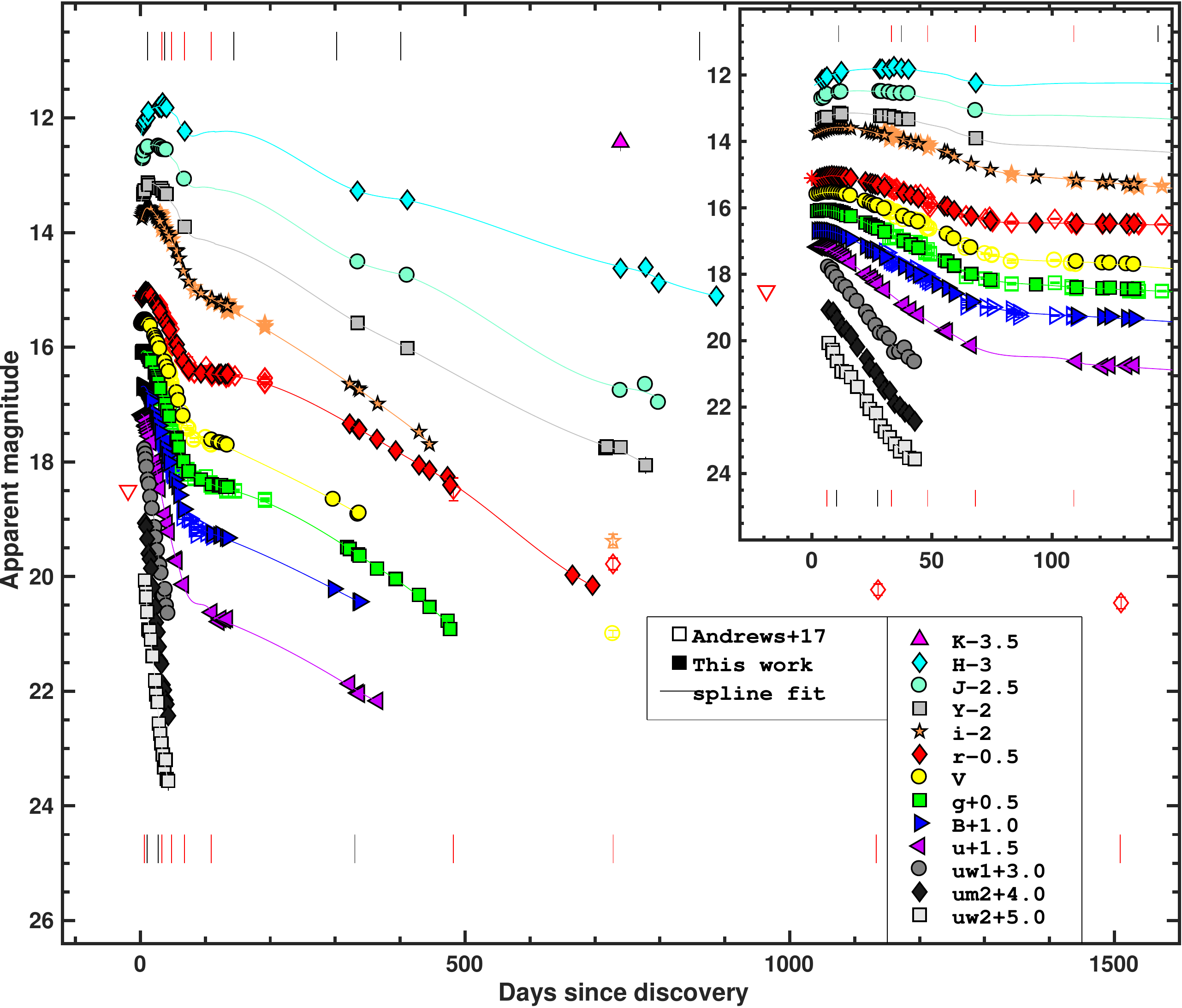}
\caption{\label{LC} CSP-II $uBgVriYJHK_s$-band and  \textit{Swift} $uvw2, uvm2, uvw1$-band light curves of SN~2013L  (filled symbols).  Light curves were shifted for clarity by the amounts shown in the legend. The photometry is also included from A17 (open symbols). Solid lines correspond to  tension spline fits to the CSP-II photometry. Epochs of spectral observations are indicated with vertical segments (NIR spectra at the top and optical spectra at the bottom; in black our new spectra, in red those already published by A17). The nondetection limit in an unfiltered/$r$ band is indicated by a red triangle, while unfiltered/$r$ band discovery and confirmation magnitudes are shown by red stars. The top-right inset shows a zoomed in view of the light curves over the first 150 days of evolution. }
\end{figure}

The resulting light curves of SN~2013L are shown in Fig.~\ref{LC}. The $BgVri$ photometry and two epochs of  MIR photometry
that are presented in A17 are also included in Fig.~\ref{LC}. 
SN 2013L is one of the best observed SNe~IIn to date, with data extending from UV to MIR wavelengths.

We used all of the available optical spectra of SN~2013L. This includes four spectra from the ESO Very Large Telescope (VLT) equipped with the X-Shooter spectrograph and a single NTT ($+$EFOSC) spectrum.  To these, we added a single spectrum obtained with the du Pont ($+$B\&C spectrograph) telescope and a single medium-resolution spectrum taken with the MagE spectrograph \citep{marshall08} on the Magellan Clay telescope. 
A single high-resolution spectrum was also taken with the Magellan Clay telescope that is equipped with the Magellan Inamori Kyocera Echelle (MIKE) spectrograph \citep{bernstein03}. This spectrum was used to check the host extinction (see Sect.~\ref{sec:basic}) and the early CSM velocity (see Sect.~\ref{sec:spec}). We also used the four late-time spectra presented by A17.
In addition to the four X-Shooter spectra covering the NIR, we added six previously unpublished NIR spectra obtained with the Magellan Baade telescope equipped with the Folded Port Infrared Echellette (FIRE; \citealt{simcoe13}). Our full NIR spectral time-series  extends from $+$11~d to $+$861~d.

\begin{figure}
\includegraphics[width=18cm]{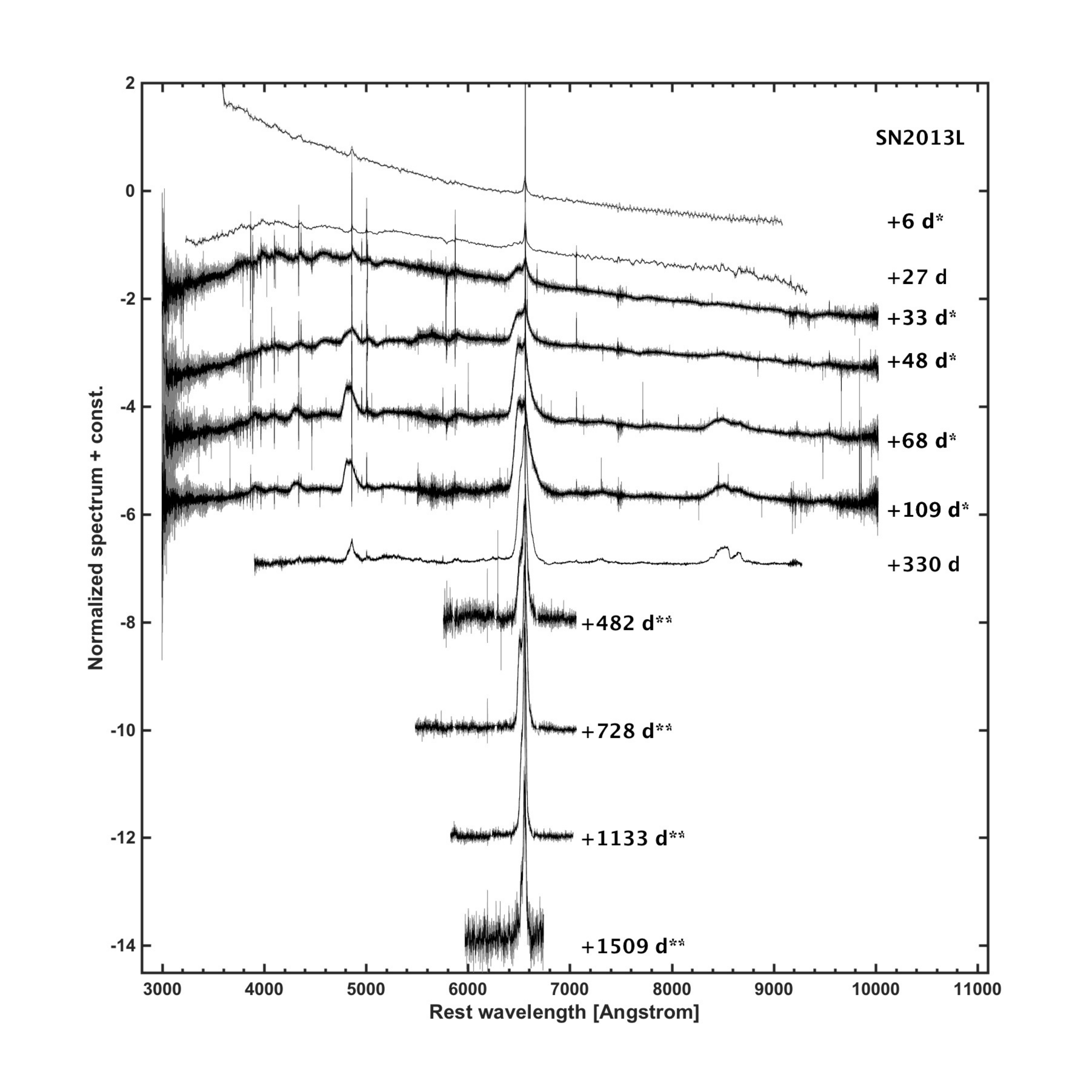}
\caption{\label{specopt}Visual-wavelength spectroscopic time-series of SN~2013L. Indicated next to each spectrum is the phase, i.e., days from discovery. The spectra that were already published in \citet{andrews17} and publicly available are marked by a star next to their phases, the four late-time spectra from A17 are marked by two stars. Each spectrum was normalized by its median and shifted by a constant for better visualization.}
\end{figure}

\begin{figure}
\includegraphics[width=18cm]{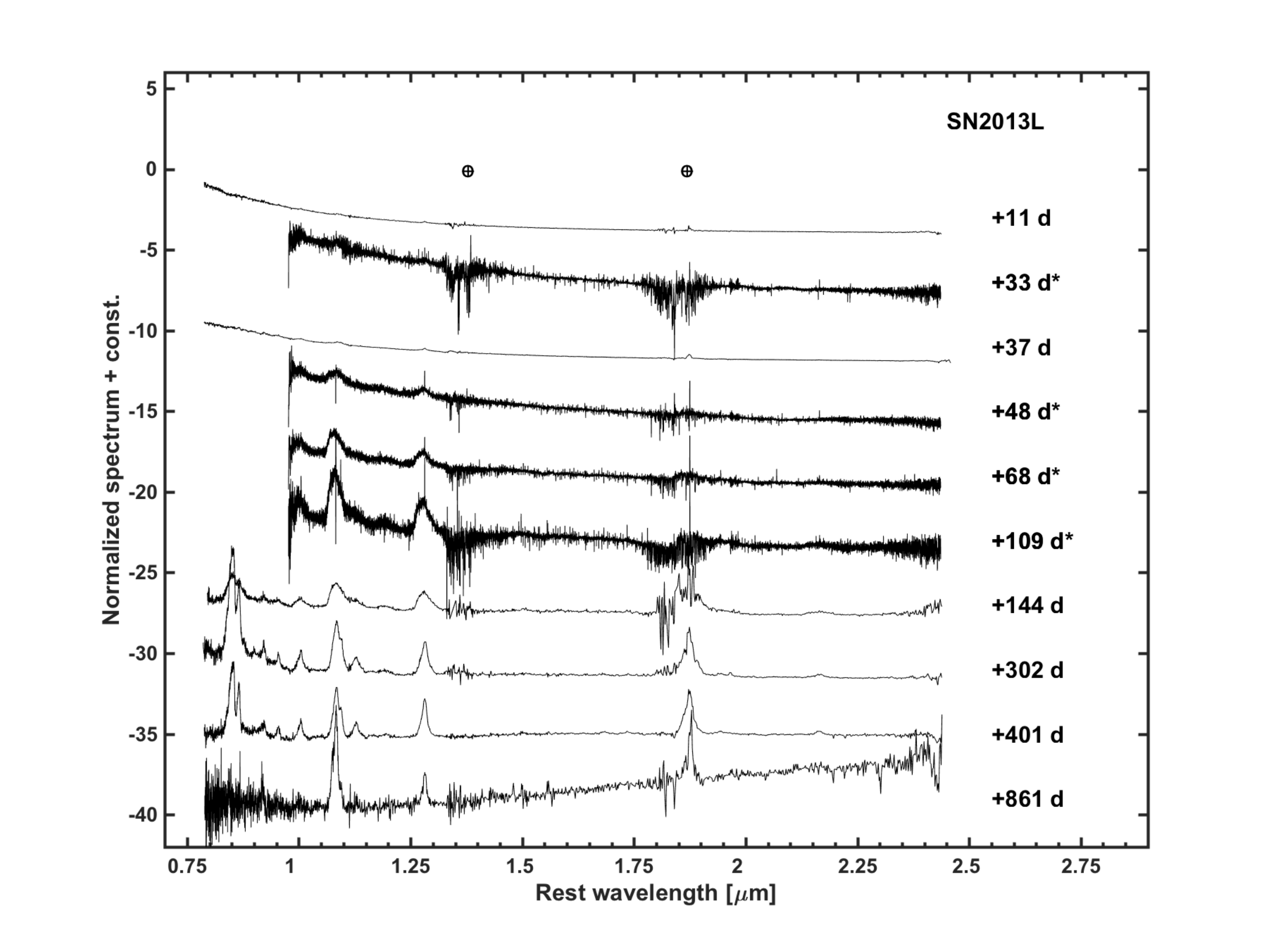}
\caption{\label{specnir}NIR spectroscopic time-series of SN~2013L obtained with FIRE and X-shooter. The phase of each spectrum relative to discovery is indicated next to each spectrum. The spectra that were already published in \citet{andrews17} are marked by a star next to their phases. Each spectrum was normalized by its median and shifted by a constant for a better visualization. We marked the telluric bands at the top.}
\end{figure}

Optical spectra were reduced in the standard manner, including bias and flat corrections, wavelength calibration to an arc lamp, and flux calibration to a spectroscopic standard star \citep[see][]{hamuy06}. We then absolute-flux calibrated the spectra to the CSP-II photometry by performing synthetic ($r$ in the optical and $J$ in the NIR, except the last spectrum where we used the $H$ band) photometry to compute the appropriate scaling of each spectrum in order to match the observed corresponding broad-band magnitude. The X-shooter data were first processed using the {\tt esorex} pipeline and then extracted and telluric calibrated using our own python scripts. 
The FIRE spectra were reduced using the {\tt firehose} software package developed by \citet{simcoe13}. The reduction steps are described by \citet{hsiao19}. 
A log of spectroscopic observations is given in Table~\ref{tab:spectra} and the full optical and NIR time-series are plotted in  Fig.~\ref{specopt} and Fig.~\ref{specnir}, respectively. 

\section{Photometry}
\label{sec:lc}
The definitive photometry of SN~2013L is plotted in Fig.~\ref{LC}. The UV to NIR light curves spanning a total of 13 filters ($uw2$, $um2$, $uw1$, $u$, $B$, $g$, $V$, $r$, $i$, $Y$, $J$, $H$, $K_s$) are  displayed from bottom to top. 
To appreciate the density of early coverage, the first 200 days of evolution are shown within an inset and also include photometry published by A17.
 To facilitate a good match between our photometry and that of A17, we introduced small shifts to the latter; specifically, we added 0.1, 0.15, 0.2, 0.12, and 0.2 mag to the $BgVri$-band light curves, respectively. The shifted light curves agree well with ours at each epoch and in each filter. 

The CSP-II light curves commenced within two days after discovery and cover the flux evolution with a high cadence as the SN reached maximum light.  
The time of maximum occurs first in the bluer bands followed in time with the redder bands. The peak was not observed in the UV bands from $uw2$ to $u$ due to the rapid drop in UV-emission. 
The light curves are fit with tension splines to better characterize their shape as well as to infer the epoch and magnitude at maximum light. 
The epochs of peak and their apparent maximum magnitudes are listed in Table~\ref{tab:peak}.

After the peak, the optical light curves show a rapid decline until about 80 days, when they begin to decline more slowly. The same appears to be true for the NIR, where we have less data covering a 75-day period after the peak. We report the magnitude difference between 80 days and the peak  ($\Delta m_{80}$) in the optical and NIR in Table~\ref{tab:peak}, as well as the magnitude difference between +80~d and +300~d ($\Delta m_{80-300}$). The early decline is faster in the bluer bands. We note that the UV light curves from Swift, which extend to +43 days, also decline very rapidly after the peak and even more rapidly than the $u$-band light curve.
The later decline is different in the different bands; the $r$ band declines more slowly due to the prevalent H$\alpha$ emission. The $g$-, $r-,$ and $i$-band decline after 300 days is faster than the rate measured between +80~d and +300~d. However, after about 700 days when the SN was followed in the $R$ band by A17, the emission appears to flatten out. 
SN~2013L has a long-lasting light curve, similar to a few other SN~IIn events in the literature, and in the following we compare SN~2013L to the data sets of these similar objects. 

\begin{figure}
\includegraphics[width=18cm]{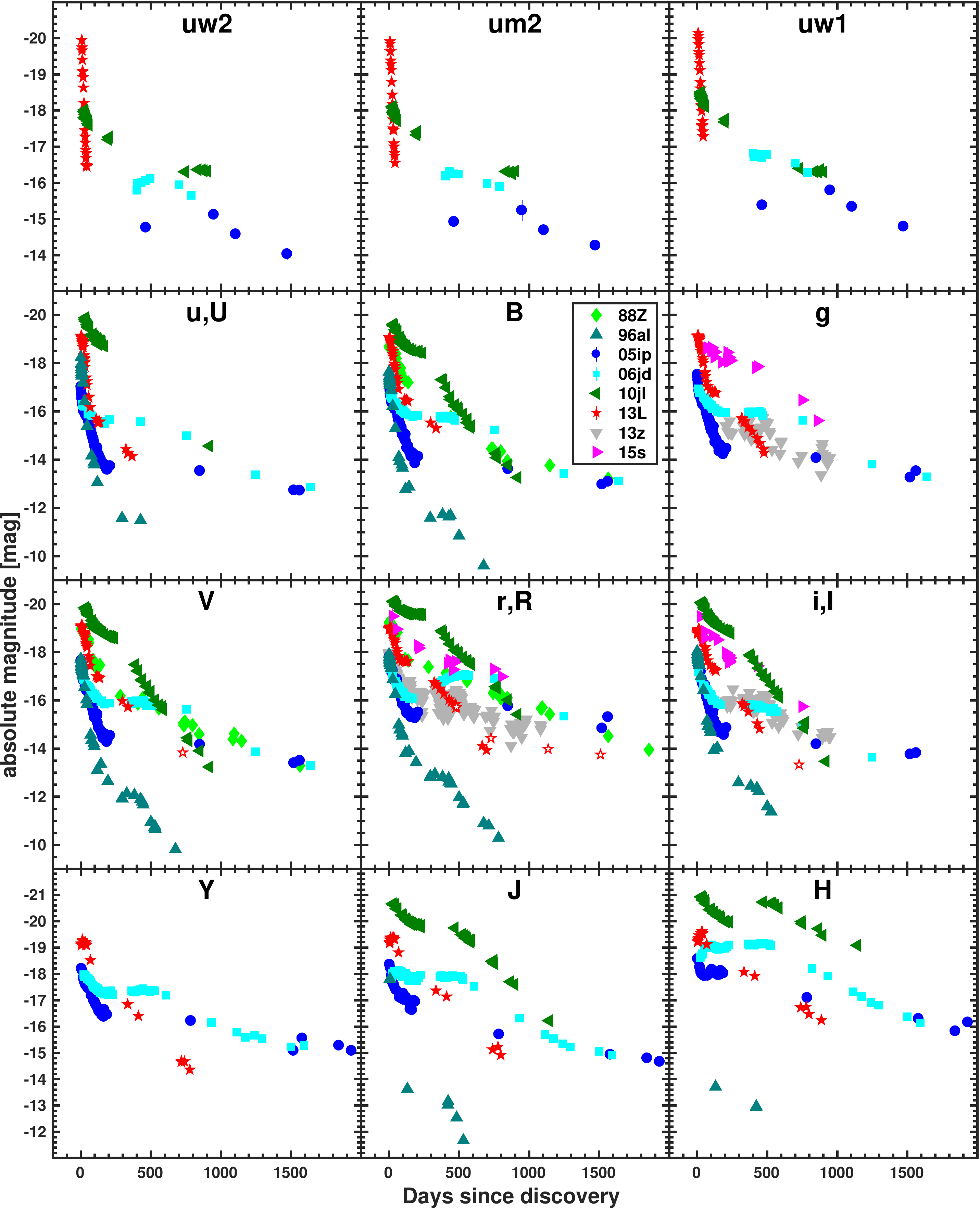}
\caption{\label{absmag}Absolute magnitude light curves of SN~2013L covering UV to NIR wavelengths as well as those of the long-lasting Type~IIn SN~1988Z  \citep{turatto93}, SN~1996al  \citep{benetti16},  SNe~2005ip and 2006jd  \citep{stritzinger12}, SN~2010jl  \citep{fransson14}, iPTF13z  \citep{nyholm17}, and KISS15s \citep{kokubo19}.}
\end{figure}

Using the light curves and their peak apparent magnitudes from Table~\ref{tab:peak}, the extinction, and the distance from Sect.~\ref{sec:basic}, we computed the absolute magnitude light curves and their peaks, which we plotted
in Fig.~\ref{absmag} and report in Table~\ref{tab:peak}, respectively.
With a peak $r$-band magnitude of $-$19.06$^{+0.24}_{-0.16}$~mag, SN~2013L is among the most luminous long-lasting SNe~IIn in the optical, together with SN~1988Z \citep[$-$19.26$\pm$0.3~mag;][]{turatto93} and SN~2010jl \citep[$-$20.12$^{+0.17}_{-0.19}$~mag;][]{fransson14,gall14}.
The early (first $\sim$130 days) decline rate is rather similar to that of SN~1988Z, SN~2005ip, SN~2006jd \citep{stritzinger12}, and iPTF13z \citep{nyholm17}, whereas SN~2010jl and KISS15s \citep{kokubo19} decline slower and SN~1996al declines much faster. Later on, in particular after 300 days, the $r$-band light curve of SN~2013L declines linearly at a similar rate compared  to SN~2010jl (which is 2 mag brighter at that phase) and SN~1996al (which is 4 mag fainter at that phase). SN~2006jd and iPTF13z develop a bumpy light curve (only in the $r$ band for SN~2006jd), while the light curves of SN~2005ip indicate nearly constant emission over a period of hundreds of days. 

The NIR ($YJH$) emission peaks around $-$19.3 to $-$19.5 mag, which is rather luminous for a SN~IIn; it is 1~mag brighter than SNe~2005ip and 2006jd in  the $Y$ and $J$ bands, but fainter than SN~2010jl by about 1~mag. In the $H$ band, SN~2013L declines rather quickly when compared to SNe~2005ip and 2006jd over the first +600~d. In fact, these other long-lasting SNe~IIn are characterized by a strong NIR excess due to dust emission from just a few weeks after discovery, which is not immediately evident in SN~2013L over the first +130~d. We return to this  point  in Sect.~\ref{sec:bolo}. 
In SN~2010jl, the fast optical decline after +400~d corresponds to an increase in the $H$ band, but this is not observed in SN~2013L.  

\begin{figure}
\includegraphics[width=18cm]{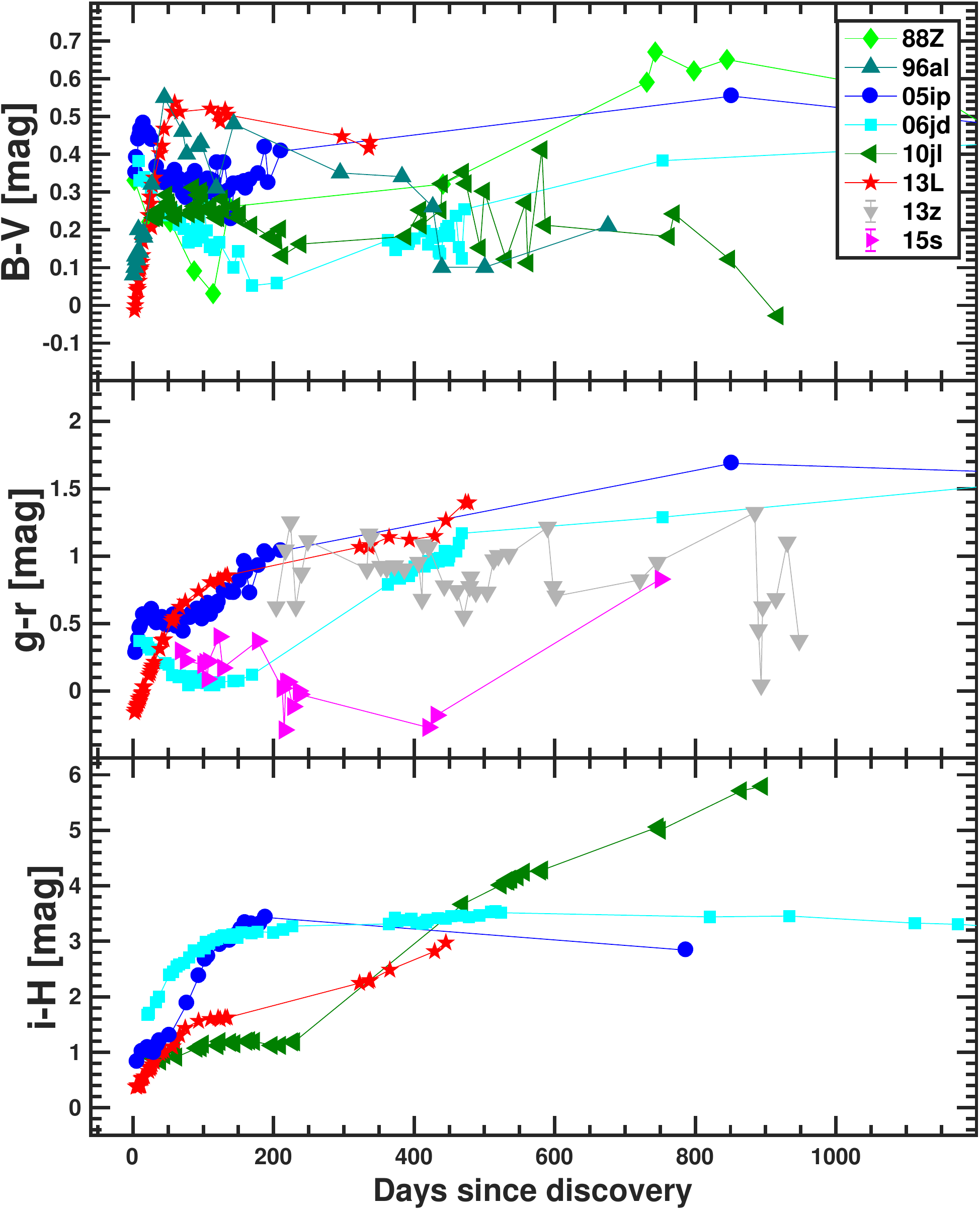}
\caption{\label{color}$B-V$, $g-r$, and $i-H$ color evolution of SN~2013L as compared to that of other long lasting SNe~IIn (SN~1988Z from \citealp{turatto93}, SN~1996al from \citealp{benetti16}, SNe~2005ip and 2006jd from \citealp{stritzinger12}, SN~2010jl from \citealp{fransson14}, iPTF13z from \citealp{nyholm17}, KISS15s from \citealp{kokubo19}).}
\end{figure}

The $g-r$ color of SN~2013L plotted in Fig.~\ref{color} evolves from  blue to red colors beginning from the onset of our first day of observations and continues to do so to almost +500~d. 
This  evolution is rather fast until +150~d, after which the color evolution slows down. The same is seen in the $B-V$ color curve up to +150~d; however, after +150~d, the $B-V$ color becomes slightly bluer. The $g-r$ color evolution of SN~2013L is similar to that of SN~2005ip, whereas the $B-V$ evolution is similar to that of SN~1996al. On the other hand, SN~2006jd, SN~1988Z, and SN~KISS15s show a different evolution,  evolving toward the blue over the first $+$200~d ($+$400~d for KISS15s).
The $i-H$ color evolution of SN~2013L, which is shown in the bottom panel of Fig.~\ref{color}, confirms that SN~2013L exhibits a smaller NIR excess compared to SN~2005ip and SN~2006jd as well as that SN\ 2013L is more comparable to the evolution of SN~2010jl at epochs earlier than $\sim$ $+$450~d. 
The similarities of its optical-NIR color evolution to objects, such as  SN~2010jl, is consistent with  dust emission.

\section{Spectroscopy}
\label{sec:spec}

The visual-wavelength spectral time series of SN~2013L extending from +6~d to +1509~d is plotted in Fig.~\ref{specopt}. The four later spectra, between +482~d and +1509~d, are from A17 and they only cover the wavelength region centered around H$\alpha$. 
The spectra obtained out to $+$48~d exhibit a blue continuum superposed with narrow Balmer emission lines, which are the only features departing from a smooth black-body (BB) shape. While the spectra become redder, the H$\alpha$ emission feature grows significantly in strength, emerging by +27~d and more clearly at +33~d. Similar evolution is exhibited by both the H$\beta$ and $H\gamma$ emission features. 
The \ion{Ca}{ii} NIR triplet emerges in emission from the continuum by +48~d and grows in strength over time as compared to the continuum.
In summary, SN~2013L shows typical SN~IIn spectral properties at optical wavelengths.

The NIR spectra (see Fig.~\ref{specnir}) also cover a large time span, extending from +11~d to +861~d. Until +37~d, they exhibit a BB continuum with narrow Paschen emission features. By +48~d, broad \ion{He}{i}~$\lambda$10830 and P$\beta$~$\lambda$12822 features emerge.
By +144~d, the spectral continuum shows a clear flattening toward the red, which becomes quite prevalent in the last spectrum. These late-time NIR spectra present a number of emission lines in addition to P$\beta$ and \ion{He}{i}, which we discuss below.

In Fig.~\ref{lineID} we show a closer inspection of the various spectral features. To do so, we selected spectra after +109~d, where the lines begin to dominate over the continuum and become easier to identify. At optical wavelengths,
all the Balmer lines down to H$\delta$ are clearly identified and characterized by broad and similarly shaped emission profiles; see also Fig.~\ref{linecomp} and Fig.~\ref{fitHa}. At the blue end of the spectrum, relatively broad features that are compatible with  \ion{Fe}{ii}~$\lambda\lambda$5018, 5169 are identified. \ion{He}{i}~$\lambda\lambda$5876, 7065 features are also  identified and confirmed by the presence of a conspicuous \ion{He}{i}~$\lambda$10830 feature in the NIR. Weak [\ion{O}{i}]~$\lambda$6300 is observed (see also Fig.~\ref{OI}) as well as [\ion{O}{i}]~$\lambda$5577. \ion{O}{i}~$\lambda$7774 and \ion{O}{i}~$\lambda$8446 are only marginally detected, and they are blended with a strong \ion{Fe}{ii} line and with the \ion{Ca}{ii} NIR triplet (see also Fig.~\ref{CaII}), respectively. The forbidden [\ion{Ca}{ii}]~$\lambda$7293,7324  emission lines are  also present. 
The NIR spectra are dominated by Paschen lines, from P$\alpha$ down to P$\eta$. We note that P$\gamma$ is blended with \ion{He}{i}~$\lambda$10830. Bracket lines from Br$\gamma$ to Br$\eta$ are also observed to be much weaker than the Paschen series. A bright, broad \ion{O}{i}~$\lambda$11287 line is observed during late epochs, as is evident in  Fig.~\ref{OI}.

\subsection{Balmer lines}
\label{sec_Balmer}
Close inspection of the different hydrogen lines reveals remarkably similar features and they evolve in step with time. 
This is demonstrated in Fig.~\ref{linecomp} (left panel) where the H$\alpha$, H$\beta,$ and P$\beta$  lines in the $+$6~d to $+$330~d spectra are plotted in velocity space and scaled to match their peaks. In doing so, the spectra were first corrected for extinction and then the continuum was removed through the subtraction of a first order polynomial function matched to the continuum. 
Clearly, the line profiles and their time evolution are nearly identical. 
A transition from only narrow to a narrow plus a broad component is observed in the +33~d spectrum and significant asymmetric profiles characterize each of the features. 
In the top-right panel of Fig.~\ref{linecomp}, we plotted the total luminosity of H$\alpha$ versus that of the continuum in the H$\alpha$ region. The line luminosity increases over time, starting from $\sim$10$^{40}$ erg~s$^{-1}$, and peaks ($\sim$2$\times$10$^{41}$ erg~s$^{-1}$) at +109~d. At +330~d, it has almost the same value as the peak. Following that, it decreases until it reaches  $\sim$10$^{40}$ erg~s$^{-1}$ at +1509~d. The continuum decreases with time, and the flux from the line already dominates after +68~d.

In the bottom-right subpanel, we show how the values of H$\alpha$/H$\beta$ and H$\beta$/H$\gamma$ increase with time after the first +48~d. The behavior of SN~2013L in terms of H$\alpha$ and continuum luminosity is rather similar to that of SN~2010jl \citep{fransson14}, and this also applies when we look at the Balmer line ratios, which we can see in the lower-right panel of Fig.~\ref{linecomp}.

We further characterize the line profiles to derive information on geometry and velocities in Fig.~\ref{fitHa}. Following A17, and using H$\alpha$ after extinction correction and continuum subtraction, we fit the profiles with the sum of two functions to reproduce the broad component. A Lorentzian is fit to the part of the profile symmetric with respect to the zero velocity, and a Gaussian is fit to the bluer part of the profile. The sum of these two functions gives a perfect fit to the profiles, as already shown by A17. We confirm that this also works on our +330~d spectrum, which is not covered by A17.  We show the best fit in the left panel of Fig.~\ref{fitHa}, and the  evolution of the velocities that were derived by these fits are in the right panel. The Lorentzian, which is centered around zero velocity at all times, shows a FWHM that increases from $\sim$2000 km~s$^{-1}$ to $\sim$5000 km~s$^{-1}$ in the time range between $+$27~d and $+$109~d.
Then the FWHM velocity of the  Lorentzian component drops to $\sim$3000 km~s$^{-1}$ at $+$330~d  and down to $\sim$1000 km~s$^{-1}$ at +1509~d. The Gaussian, reproducing the blue part of the H$\alpha$ profile, is blue-shifted by $\sim$4000 km~s$^{-1}$ at early epochs. The blue-shift slowly decreases down to $\sim$1000 km~s$^{-1}$ at +1509~d. The FWHM of the Gaussian drops with time from $\sim$6000 km~s$^{-1}$ at early epochs to $\sim$2500 km~s$^{-1}$ by +330~d and later epochs. 

The H$\alpha$ total luminosity, which we describe above, is dominated by the broad Lorentzian component, and both the Lorentzian and the Gaussian luminosity behave similarly.
The narrow component of H$\alpha$ (see Fig.~\ref{fitHa}, bottom-right panel) has a decreasing minimum absorption velocity ($V_{min}^{P-Cyg}$, measured by A17 and by us, assuming a velocity of zero at the emission peak of H$\alpha$) between $-$68~km~s$^{-1}$ and $-$130~km~s$^{-1}$. 

For an estimate of the CSM velocity from the narrow H$\alpha$ component ($\rm V_w$), one should use the blue-velocity-at-zero-intensity (BVZI) of this narrow component, assuming zero velocity at a H$\alpha$-rest wavelength, as in Fig.~\ref{narrowHa}. This velocity is constant around 120~km~s$^{-1}$ during the first +109~d. It becomes higher (up to 240~km~s$^{-1}$) at later epochs. The resolution of the spectra (see A17, their Table~3) is sufficient to resolve these narrow lines to obtain their velocities (FWHM resolutions of XSHOOTER and IMACS are 40 and 57 km~s$^{-1}$, respectively). 
Values of wind velocity in the range of 120--240~km~s$^{-1}$ are also observed in other SNe~IIn \citep[see, e.g.,][]{stritzinger12,taddia13IIn,fransson14,gall14}. 

The first spectrum where we were able to measure the narrow P~Cygni minimum is the high-resolution MIKE spectrum taken at phase $+$10~d. The NTT (+6~d) and du Pont (+27~d) spectra do not have the resolution to measure the P~Cygni absorption profile.
Figure~\ref{narrowHa} also shows a transition in the shape of the narrow H$\alpha$, with the emission part dominating at early phases and decreasing with time until the absorption component dominates after +330~d. 

The decomposition of the broad profile of H$\alpha$ into one Lorentzian and one Gaussian component reproduces the line shape well, which was already shown by A17 and is confirmed here. However, we also applied a different, more physically-motivated decomposition to the line. 
This is based on the observation that the H$\alpha$ profile can also be the sum of a broad, flat, skewed component, which decreases its FWHM with time, and of a narrower component centered at zero velocity whose relative strength increases over time. This is well illustrated in Fig.~\ref{lineHa_scale}, where we normalized the continuum-subtracted H$\alpha$ profiles to the flux of the ``shoulder'' of the broad, skewed component. This shoulder moves toward lower velocities with time. 
This decomposition is suggested above all by the spectral modeling we present in Sect.~\ref{sec:linemodel}, which reproduces the broad,  flat component with the emission from a narrow emitting shell. While this would give a boxy profile if the medium is optically thin, electron scattering smooths this profile and occultation results in a skewed blue-shifted profile (Sect.~\ref{sec:linemodel}). 

In Fig.~\ref{fitHa3} (left panel) we fit the H$\alpha$ profiles as a combination of a broad skewed Gaussian with two different HWHM (half-width-at-half-maximum) (blue) and a central, narrower Gaussian (green) function.
The central Gaussian in the first spectrum is too weak to be seen. Here, we ignored the narrow P~Cygni region, as in Fig.~\ref{fitHa}. The centroid of the two components and their FWHM and HWHM  are shown in the top panels. The narrower Gaussian is basically centered at zero velocity, while the skewed Gaussian moves to lower velocities from +100~d to +1500~d. 
The
FWHM of the broad component decreases with time, except for the first two epochs where it is more difficult to measure. 
The FWHM of the central Gaussian is rather constant, but it decreases in the last two spectra. 
The luminosities of the two components behave similarly, rising to a peak and then declining.
In the bottom-right panel, we report the BVZI of the narrow P~Cygni, which is discussed above and shown in Fig.~\ref{narrowHa}.

Finally, we measured the fluxes of three bright hydrogen lines from the NIR spectra and we show their similar evolution in Fig.~\ref{fluxNIR}.  These lines rise rapidly in flux to reach their peak between $+$100~d and $+$200~d.

\subsection{Lines from heavier elements and continuum}
\label{sec_Other}

Other interesting features present in the spectra of SN~2013L are the oxygen lines. In Fig.~\ref{OI}, we see broad [\ion{O}{i}]~$\lambda$6300 beginning from +109~d and lasting onward. The same is true for \ion{O}{i}~$\lambda$11287, which becomes very strong by +302~d. The FWHM velocity of \ion{O}{i}~$\lambda$11287 at this phase is around 4300~km~s$^{-1}$, which is obtained by fitting a Gaussian to the continuum-subtracted line after extinction correction and absolute calibration. The \ion{O}{i}~$\lambda$11287 persists until the last spectrum at $+$861~d. The FWHM of the \ion{O}{i}~$\lambda$11287 line almost matches the FWHM of the H$\alpha$ profile (see Fig.~\ref{OIvsHa}).

\begin{figure}
\includegraphics[width=18cm]{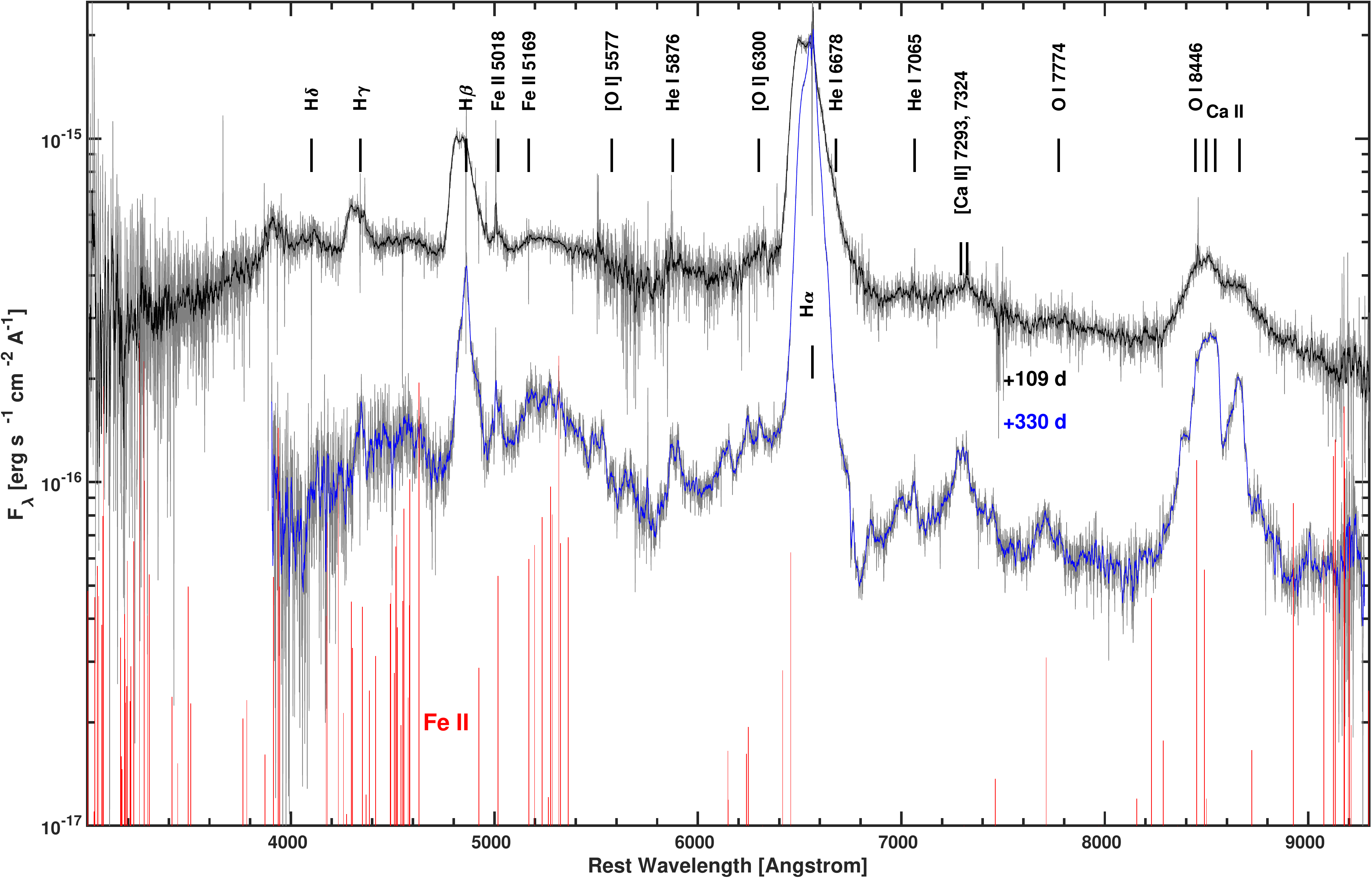}

\includegraphics[width=18cm]{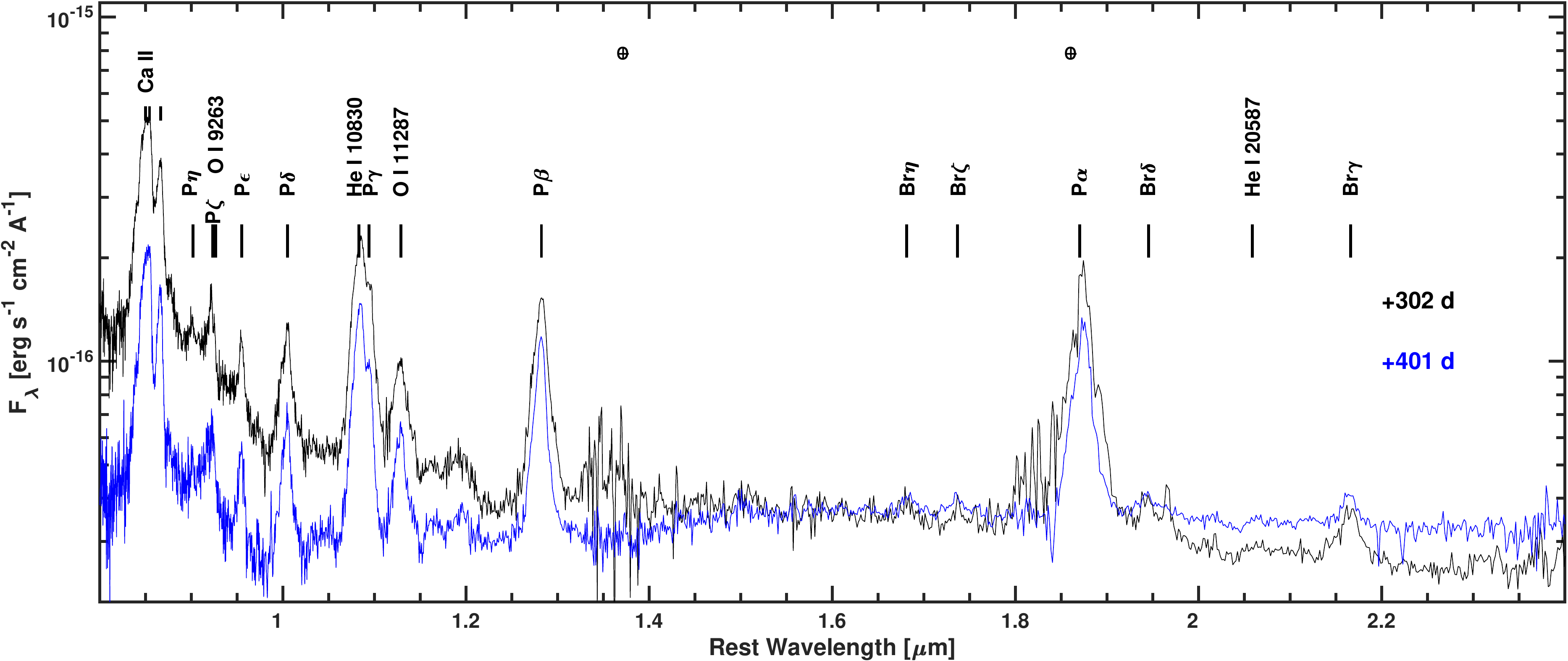}
\caption{\label{lineID}Visual (top) and NIR (bottom) spectral line identifications for SN~2013L. It is important to note that the spectra in the top panel were smoothed (dark lines) for presentation purposes. In the optical, we plotted the expected positions for the \ion{Fe}{ii} lines (red) for a model with Fe II  Ly${\alpha}$ and Ly${\beta}$ pumping, from \citet{sigut03}. These lines can explain the blue excess in the spectrum at late epochs.}
\end{figure}

\begin{figure}
$\begin{array}{cc}
\includegraphics[width=9cm]{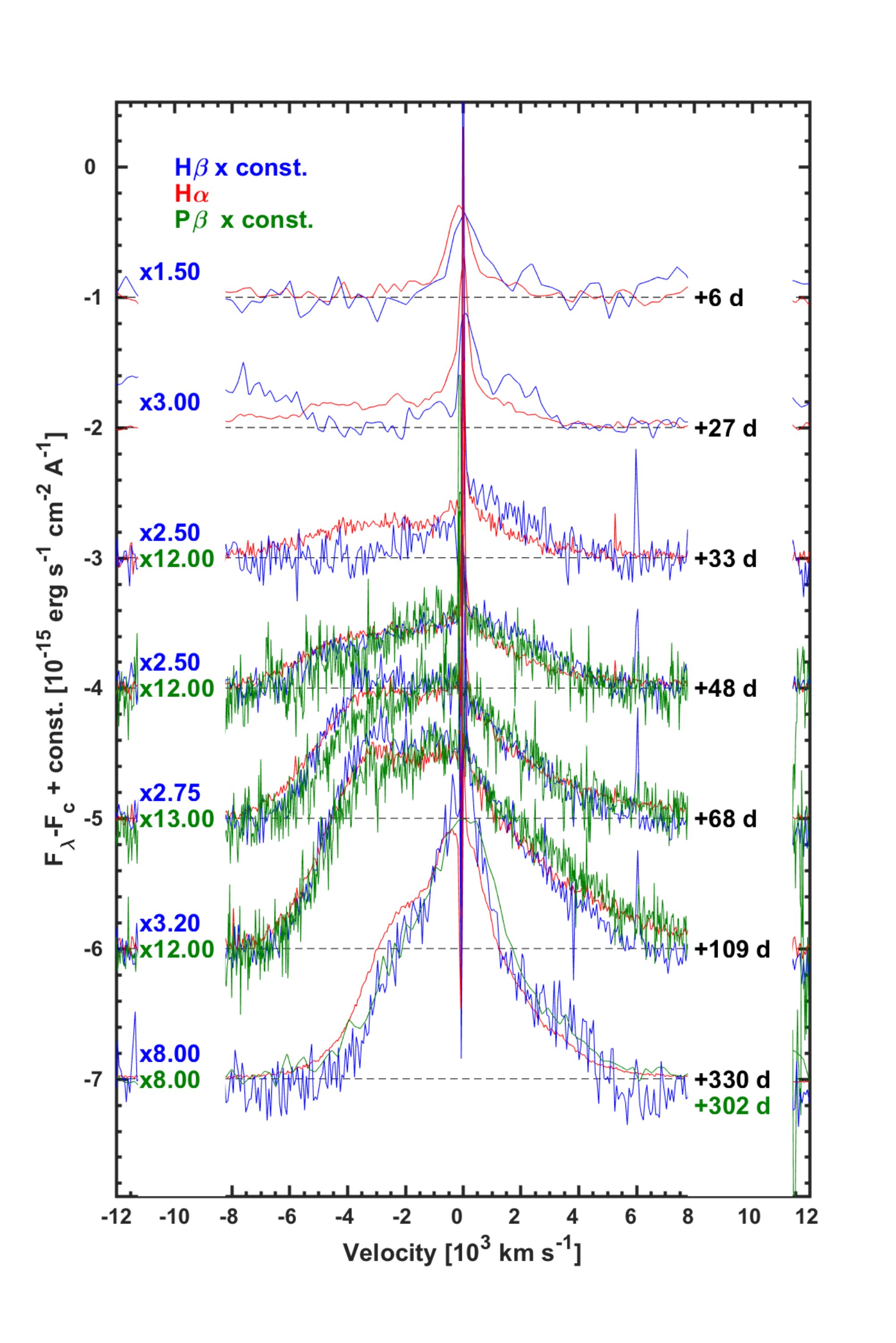}&
\includegraphics[width=8cm]{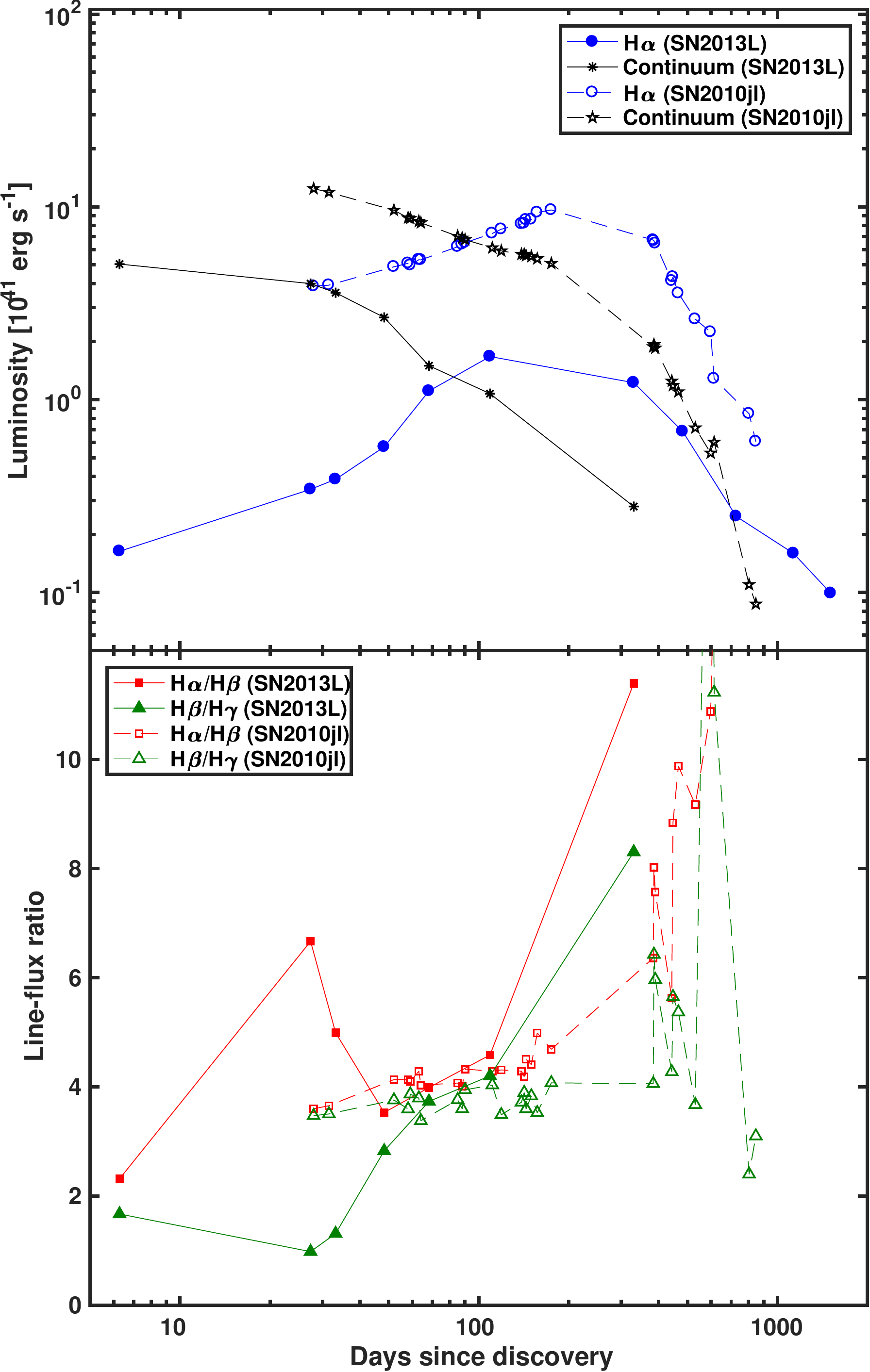}
\end{array}$
\caption{\label{linecomp}(Left panel) H$\alpha$, H$\beta$, and P$\beta$ line profiles compared to one another and plotted in velocity space. Each line has been corrected for extinction and  continuum subtracted. H$\beta$ and P$\beta$ lines were scaled by the constants indicated on the left to match the intensity of H$\alpha$. The spectral phases in days since discovery are reported on the right. The shape of the line profiles is remarkably similar at all epochs. (Top-right panel) H$\alpha$ luminosity and continuum luminosity in the H$\alpha$ region as a function of time for SN~2013L and SN~2010jl \citep{fransson14}. (Bottom-right panel) H$\alpha$/H$\beta$ and H$\beta$/H$\gamma$ flux ratio as a function of time for SN~2013L and SN~2010jl.}
\end{figure}

\begin{figure}
$\begin{array}{cc}
\includegraphics[width=9cm]{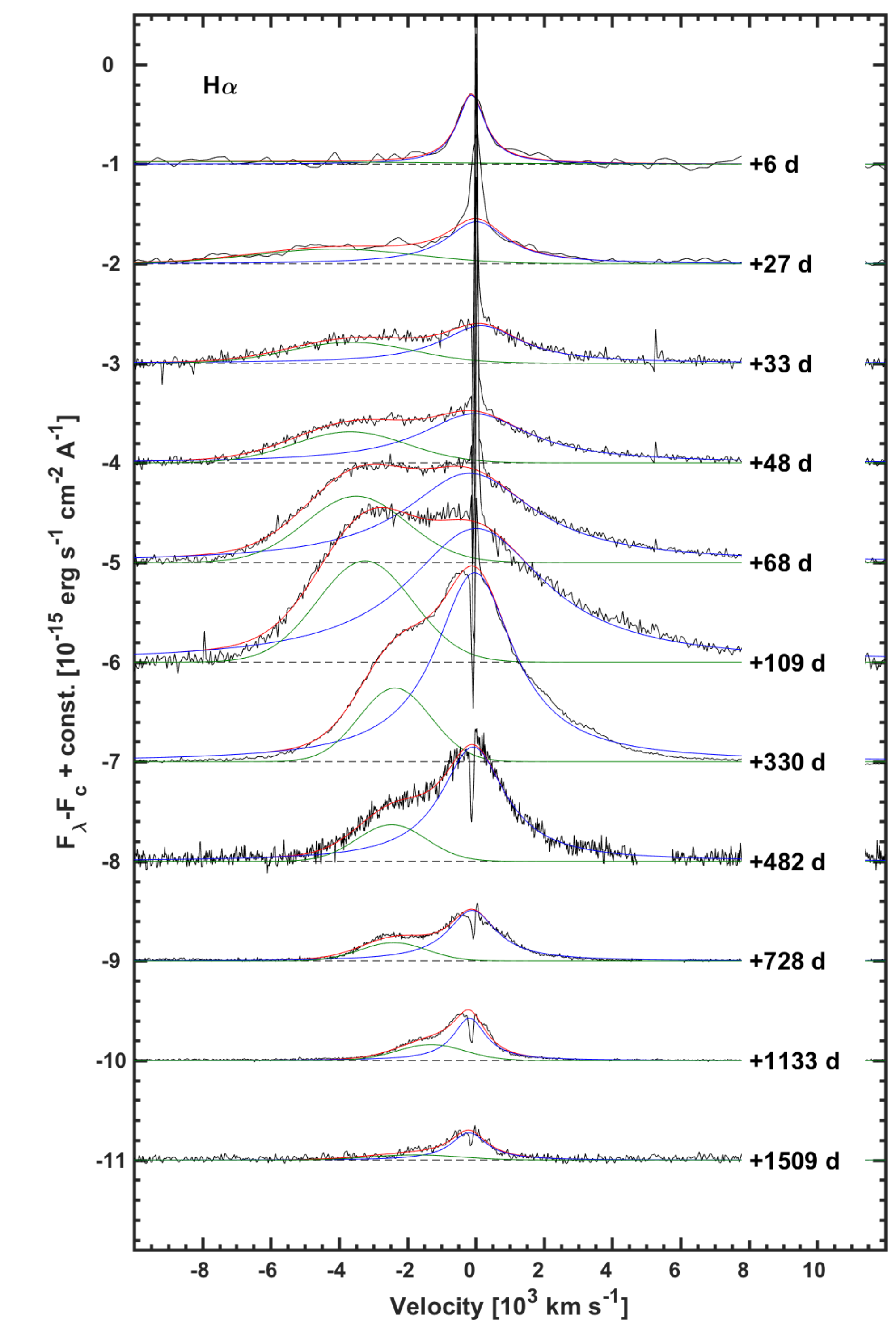}&
\includegraphics[width=9cm]{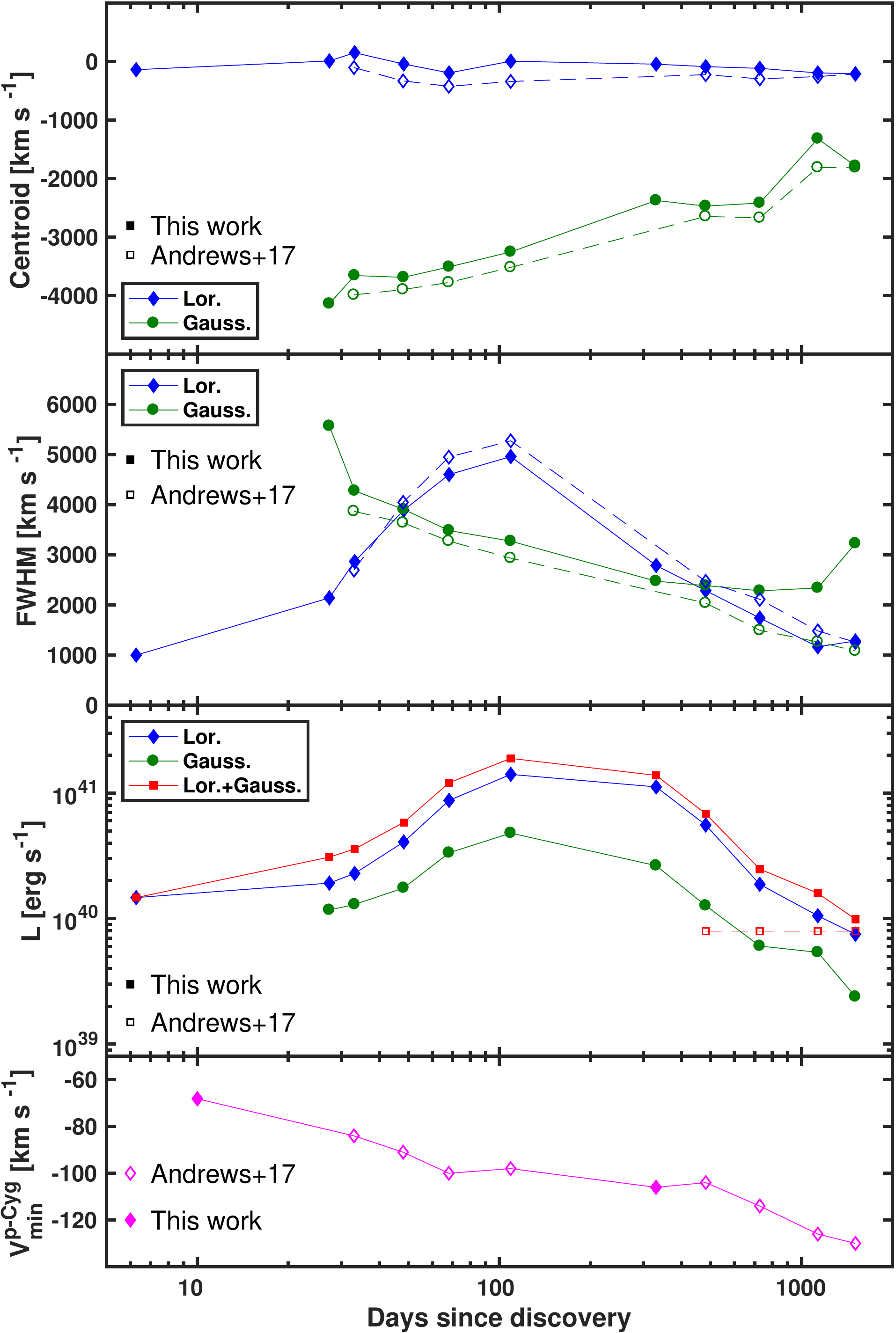}\\
\end{array}$
\caption{\label{fitHa}(Left panel) Continuum-subtracted and extinction corrected H$\alpha$ profiles (black) fitted by a sum of a Lorentzian (blue) and a Gaussian (green) function to reproduce the broad components. The total fit is shown in red. The spectral phases are shown on the right next to each spectrum. The wavelength range around the narrow component was excluded from the fit. (Right panel) Velocities and luminosities of the H$\alpha$ components. The centroid and the FWHM of the Gaussian and of the Lorentzian are shown in the top panels, their luminosities and the total luminosity, as well as the velocities of the narrow component are shown in the bottom panels. Here for the narrow component, we considered the velocity of the absorption minimum assuming zero velocity corresponding to the emission peak, such as in A17. We report our results as well as those of A17 to show the close agreement.} 
\end{figure}

\begin{figure}
\includegraphics[width=10cm]{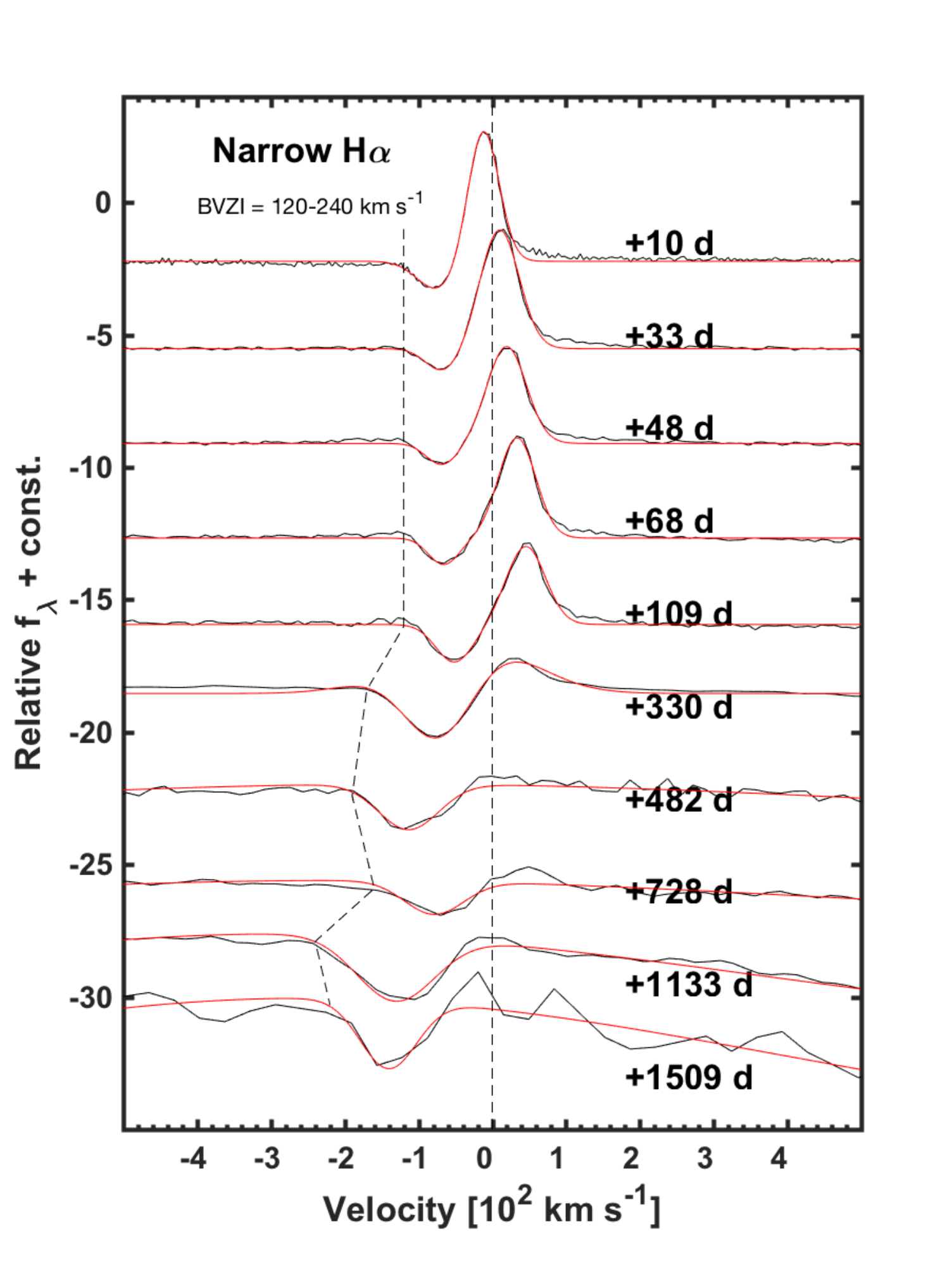}
\caption{\label{narrowHa}The narrow component of  H$\alpha$ plotted in velocity space. The spectral phases in days since discovery are reported on the right. The zero velocity was chosen to correspond to 6562.8~\AA\ in the rest frame. We plotted (in red) the best fit to the line profile, which was achieved by fitting two Gaussian components, one for the absorption and one for the emission. The emission dominates at early epochs, while absorption becomes stronger at later epochs. We marked the BVZI with a dashed segmented line that ranges from $-$120 to $-$240~km~s$^{-1}$.}
\end{figure}

Following A17, the \ion{Ca}{ii} NIR triplet from +68~d to +330~d (shown in Fig.~\ref{CaII}) is well-represented by a combination of three H$\alpha$ profiles (one for each of the three \ion{Ca}{ii} lines), which were scaled by three different constants to fit the line profile. 
In the spectra at +302~d and +330~d, there is a small excess in the bluer part of the line that might be compatible with \ion{O}{i}~$\lambda$8446, which emerges at the same time as \ion{O}{i}~$\lambda$11287 and also \ion{O}{i}~$\lambda$7774 (see Fig.~\ref{lineID}). It might also be compatible with a \ion{Fe}{ii} line, as shown in Fig.~\ref{lineID}. Also the blue excess between 4000~\AA\ and 5400~\AA\ can be explained by a forest of \ion{Fe}{ii} lines (see Fig.~\ref{lineID}). Neither of the \ion{O}{i} or  \ion{Ca}{ii} lines show any velocity shift and have a FWHM that is compatible with the central component of H$\alpha$. They are therefore likely to arise in the CSM.

In addition to the line analysis, it is of interest to study the evolution of the spectral continuum, especially in the NIR. Plotted in the inset of Fig.~\ref{specfit} are all the optical spectra after applying an extinction correction and an absolute flux calibration relative to $r$-band photometry; they were then fit with a BB function. Here we exclude the region of the spectra dominated by strong emission lines. 
A single BB function provides a good representation of the continuum.
In the NIR, where we observed spectra to later epochs, a single BB function provides a reasonable fit until $+$109~d. However, the $+$144~d spectrum shows a better fit for a two-component BB function, with a hot component fitting the bluer portion of the spectrum and a warm component dominating  the flux above 20\,000~\AA. The same is true for the later spectra, where the warm BB component becomes relatively stronger.
In the last NIR spectrum, the second BB component fitting the red portion of the spectrum is so strong that the continuum is more prevalent above 20\,000~\AA\ than in the blue. 
From the BB fit, estimates of the temperature, radius, and luminosity were derived (see Sect.~\ref{sec:bolo}) together with the same quantities derived from fitting SEDs constructed from broad-band photometry.

\section{SEDs and constructing the bolometric light curve}

\label{sec:bolo}
The spectral continuum fit discussed in Sect.~\ref{sec_Other} and shown in Fig.~\ref{specfit} reveals an interesting double BB component emerging at late epochs. We now turn to photometry and build SEDs extending from UV to NIR wavelengths. From these, we further examine the different underlying emission regions and also construct a bolometric light curve. 

Using the tension splines fitted to the photometry presented in Fig.~\ref{LC}, we obtained interpolated UV to NIR photometry from +7~d to +335~d, when photometry was obtained in all of the CSP-II bands extending from $u$ to $H$. Between +7~d and +43~d, we also obtained interpolated UVOT photometry.
Next, appropriate AB offsets that transform the CSP-II natural system to the standard system were added to the photometry \citep[see][their Table~16]{krisciunas17}. The photometry was then corrected for
 reddening and  then converted into monochromatic fluxes at the effective wavelength of each filter. 
Six examples of our SEDs are plotted from +10~d to +335~d in Fig.~\ref{SEDfit}. Here, we selected epochs where the NIR fluxes are directly available from the observations, and not just interpolated. Up through +68~d, the SEDs clearly exhibit a shape that is well-matched by a BB function, with the exception of the $r$ band (red diamonds), which begins to deviate from the best BB fit due to prevalent H$\alpha$ emission. The $r$-band flux was therefore not included in the BB fits, which are shown with a black line. The UVOT fluxes in the UV (white, black, and gray symbols) nicely follow the BB shape, indicating that at least at early epochs, the UV is not strongly suppressed by line blanketing. The NIR fluxes (light gray, green-cyan, and cyan points) fall on the BB Rayleigh-Jeans tail until +68~d; however, at later epochs such as +335~d, they show an excess that is reproduced well by a second BB function (red line). The BB temperature drops from an initial 10\,700 K at +10~d to 5800--6100 K between +68~d and +335~d. The second component in the NIR has a lower temperature of about 1400~K at +335~d.
 At +335~d, the $r$-band flux point is well above the SED continuum as a result of the prevalent H$\alpha$ feature. We performed BB fits to the SEDs from +7~d to +335~d, and we notice that the second BB component gives a better fit to the SED only starting from +132~d, which is in agreement with the spectral fits shown in Fig.~\ref{specfit}, which do not show a second component at +109~d but do so at +144~d. 

At very late epochs (+630~d to +850~d), when only $K_s$ band and Spitzer photometry at 3.6 and 4.5 $\mu$m (see A17) are available, the NIR (and MIR) excess is even stronger, as already shown in the last NIR spectrum in Fig.~\ref{specfit}. This excess is highlighted in Fig.~\ref{spitz}, where the full SED is shown, including all of the flux points and  our  last NIR spectrum. A sum of two BBs is confirmed to match the emission including the MIR flux points. 

The BB fits to the SEDs are useful to build a bolometric light curve. 
In order to obtain the luminosity,  the BB fluxes were multiplied by 4$\pi$D$^{2}$ where D is the SN distance. The luminosity from the two BB components is shown in the top panel of Fig.~\ref{LTR}, as well as their sum. The warm (NIR) component emerges only after +132~d. We also show the luminosity when adding the remaining H$\alpha$ flux as the $r$-band monochromatic flux was excluded from the BB fit. 
For comparison purposes, we show the luminosity derived from the BB fit to the spectra. The optical spectral fits give very good agreement in luminosity, the NIR spectral fits are not accurate in reproducing the hot BB luminosity in the optical given the different wavelength coverage, but they provide an estimate for the warm BB luminosity in the NIR.

Even though it was discovered before optical peak, due to the huge UV flux, the bolometric luminosity at early epochs indicates that SN~2013L was discovered after the bolometric peak. The observed bolometric peak is at $\gtrsim$3$\times$10$^{43}$~erg~s$^{-1}$.
The luminosity drops to 2$\times$10$^{42}$~erg~s$^{-1}$, one order of magnitude, in $\sim$100~days. Subsequently its luminosity declines more slowly. The warm BB luminosity contributing to the total luminosity is rather constant with time, at about  1$\times$10$^{42}$~erg~s$^{-1}$, and it begins to dominate at roughly +300~d to +400~d after discovery. Then it largely dominates by +800~d when the hot, optical BB luminosity is 3--4$\times$10$^{40}$~erg~s$^{-1}$ and the warm, NIR BB luminosity is at 7$\times$10$^{41}$~erg~s$^{-1}$.

Plotted in the central and bottom subpanels of Fig.~\ref{LTR} are the temperatures of the hot and warm BB components obtained from our BB function fits to our spectrophotometry. The hot BB temperature drops from 15\,000~K right after discovery to about 6000~K at +1000~d, and then it stays constant. The NIR warm BB temperature is at $\sim$1000--1700~K from +132~d until +861~d.

The BB radius for the hot component expands until +70 days (from 1$\times$10$^{15}$ cm to 2.5$\times$10$^{15}$ cm), then it begins to slowly decrease, reaching 3$\times$10$^{14}$~cm at +800~d. The warm NIR BB radius stays rather constant at 1--3$\times$10$^{16}$ cm. 

Although the BB fits give a good agreement with the observed SED, this may, especially at later epochs, underestimate the luminosity and also the radius of the hot component. This is illustrated by the simulation of SN 2010jl in \citet[][their Fig. 13]{dessart15}, where the model has a considerable UV contribution compared to a BB fit of only the optical range. This in turn leads to a large underestimate of the bolometric luminosity. We return to this in Sect.~\ref{sec:csmmodel}. In addition, when scattering dominates, the radius of the photosphere is significantly different from that derived from a BB-fit.

\section{Modeling}
\label{sec:model}

In the following, we detail a model consisting of various emission features driven from the interaction between the rapidly expanding supernova ejecta with circumstellar material. One of the goals of the ascribed model is to estimate the mass-loss rate and hence to clarify the progenitor scenario. To begin, the H$\alpha$ profile was modeled; in the following, a model of the bolometric light curve is provided, making use of information derived from the spectra. 

\subsection{Explosion epoch}
\label{sec:explo}

In order to accurately apply a CSM-interacting model to describe SN~2013L, it is first required to estimate its explosion epoch. We know that the last nondetection occurred 19 days before discovery (see Sect.~\ref{sec:basic}) and given the magnitude limit (19 mag), the SN certainly exploded within this 19 days interval.
Considering the radius provided in Fig.~\ref{LTR}, 
we can extrapolate to $r_{BB}=0$ by fitting  the first +10~d after discovery with a straight line  since its evolution appears linear. This provides an estimate of the explosion epoch of $-$15.0~d prior to the discovery epoch, which is within the interval determined by the last nondetection epoch. If instead of fitting only up to $+$10~d, we rather fit to $+$30~d with a second order polynomial (the evolution resembles a parabola), we obtain an explosion epoch on $-$15.6~d prior to discovery, that is, very similar to the first estimate. In the following, the explosion epoch is inferred to have occurred on $-$15~d, that is, JD$_{\rm explo}=$2456299.5$^{+15}_{-4}$. The radius fit is shown in Fig.~\ref{fitradius}. A caveat to our explosion time estimate is the possibility of a large ($10^{14}-10^{15}$~cm) progenitor radius, which would imply a later explosion date. 

\subsection{Modeling of the spectral lines}
\label{sec:linemodel}

\begin{figure}
\includegraphics[width=18cm]{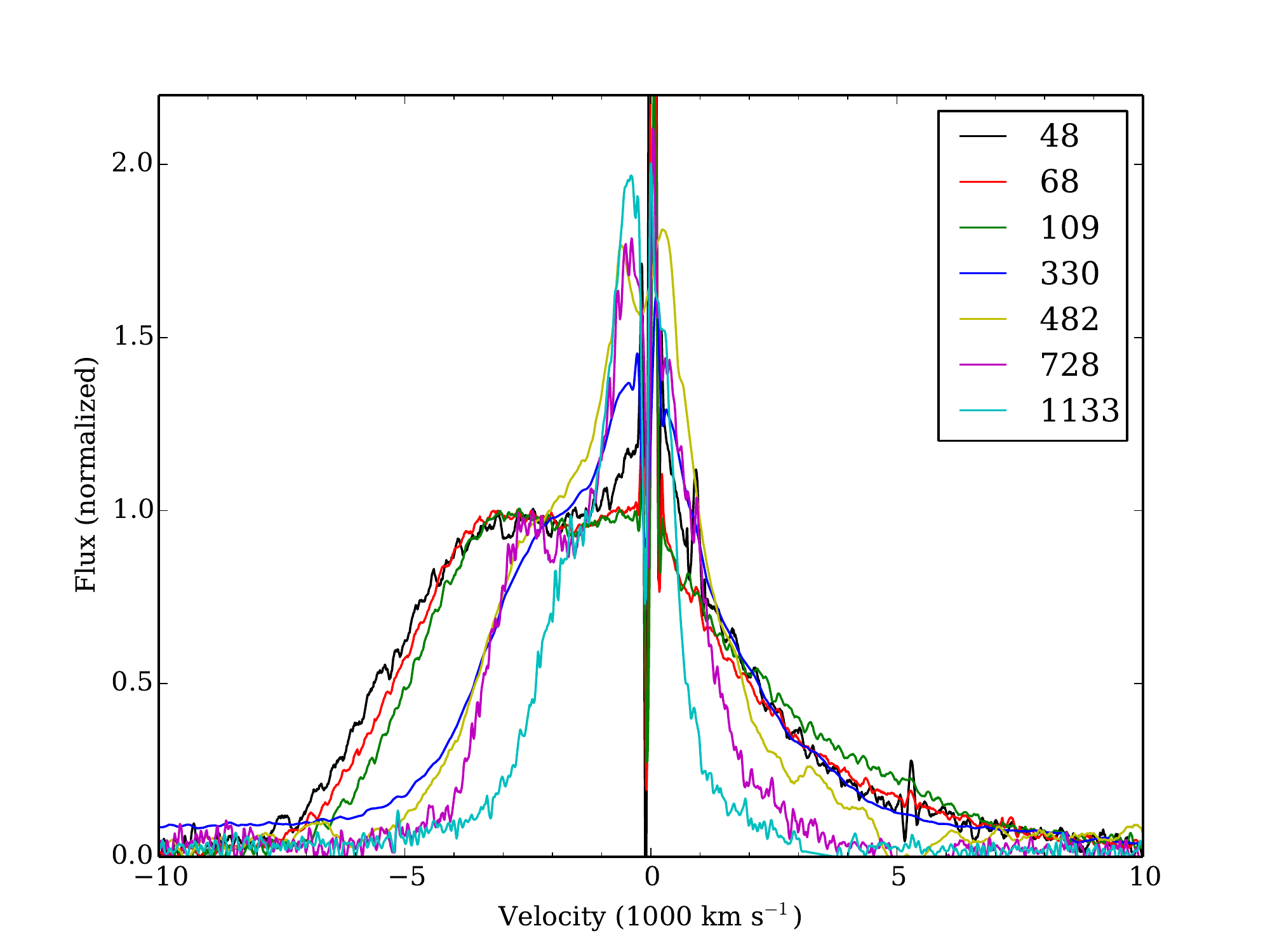}
\caption{\label{lineHa_scale}
Evolution of the continuum subtracted H$\alpha$ line profile in velocity space. The flux was normalized to the flux of the flat part of the profile between the shoulder and the central component. The spectra from +482~d to +1133~d are from A17.}
\end{figure}

\begin{figure}
$\begin{array}{cc}
\includegraphics[width=9cm]{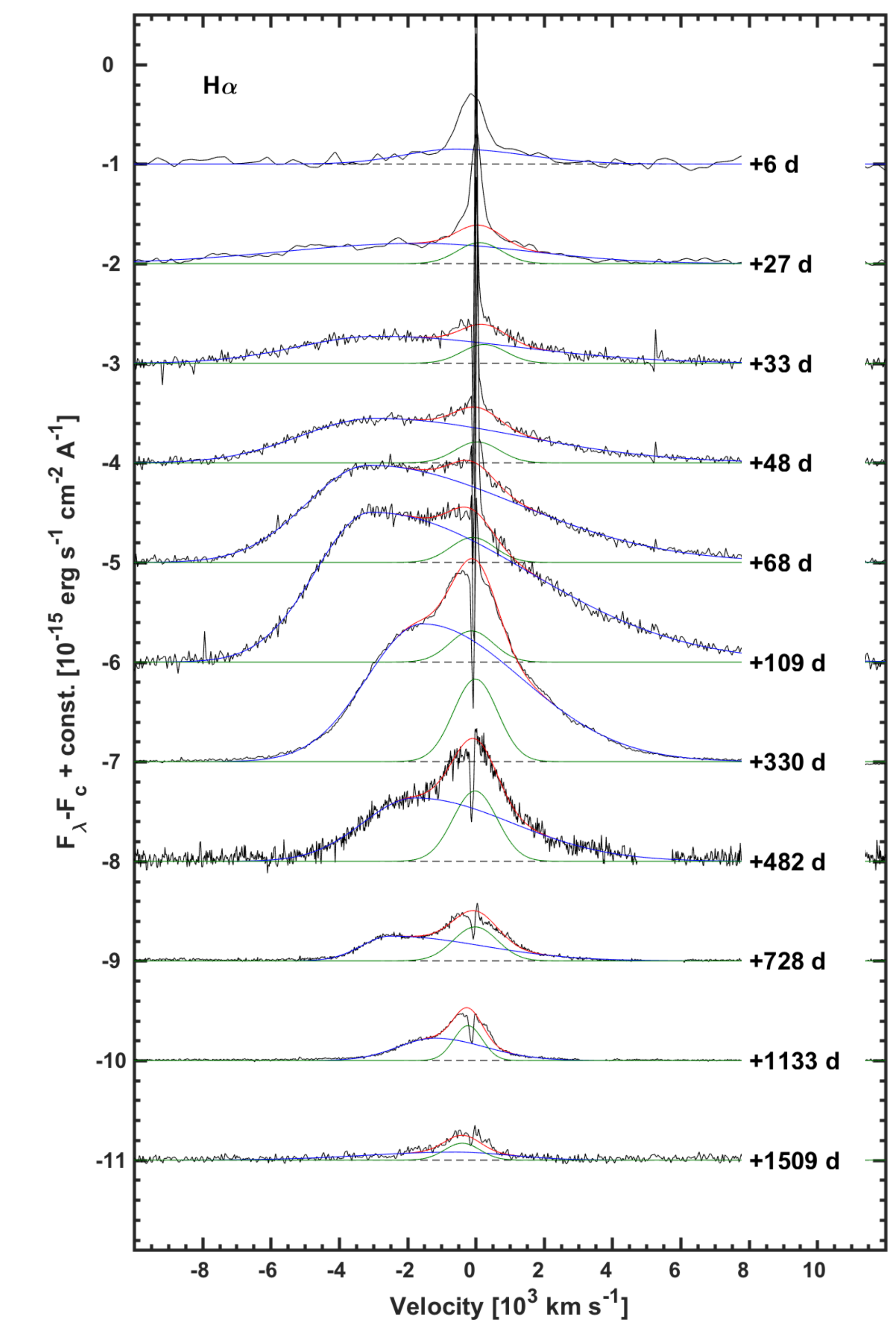}&
\includegraphics[width=9cm]{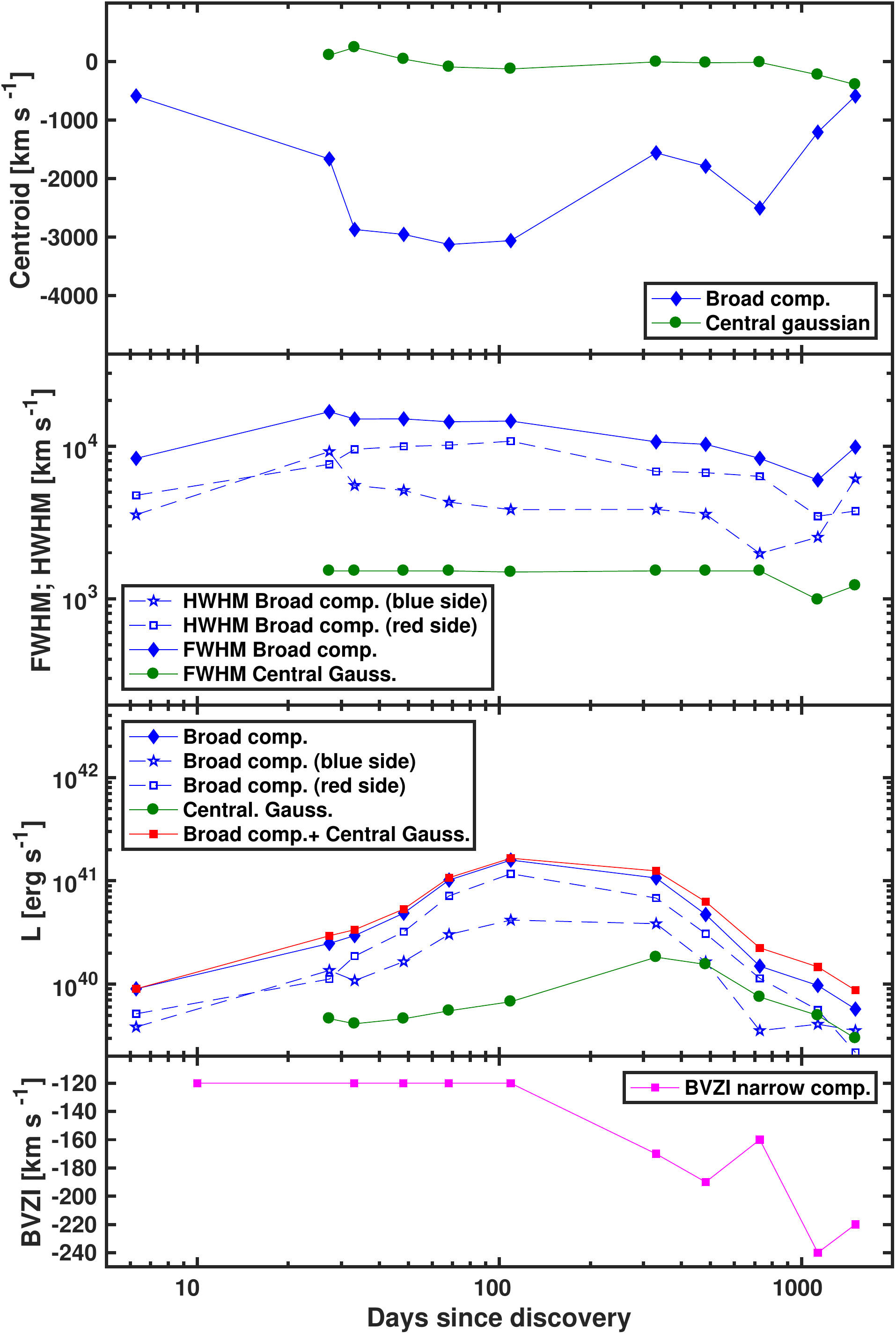}\\
\end{array}$
\caption{\label{fitHa3} Continuum-subtracted and extinction corrected H$\alpha$ profiles (black), this time fitted by a sum of a skewed, broad Gaussian (blue) and a narrower Gaussian (green) function to reproduce the broad components. The total fit is shown in red. The spectral phases are shown on the right next to each spectrum. The wavelength range around the narrow component was excluded from the fit. (Right panel) Velocities and luminosities of the H$\alpha$ components. The centroid and the FWHM and HWHM of the skewed broad Gaussian (blue) and of the central Gaussian (green) are shown in the top panels, their luminosities and the total luminosity,  as well as the velocity of the narrow component are shown in the bottom panels. Here, for the narrow component, we consider the blue-velocity-at-zero-intensity (BVZI) assuming zero velocity at the H$\alpha$ rest wavelength to obtain a better estimate of the CSM velocity.} 
\end{figure}

\begin{figure}
$\begin{array}{cc}
\includegraphics[width=8.3cm]{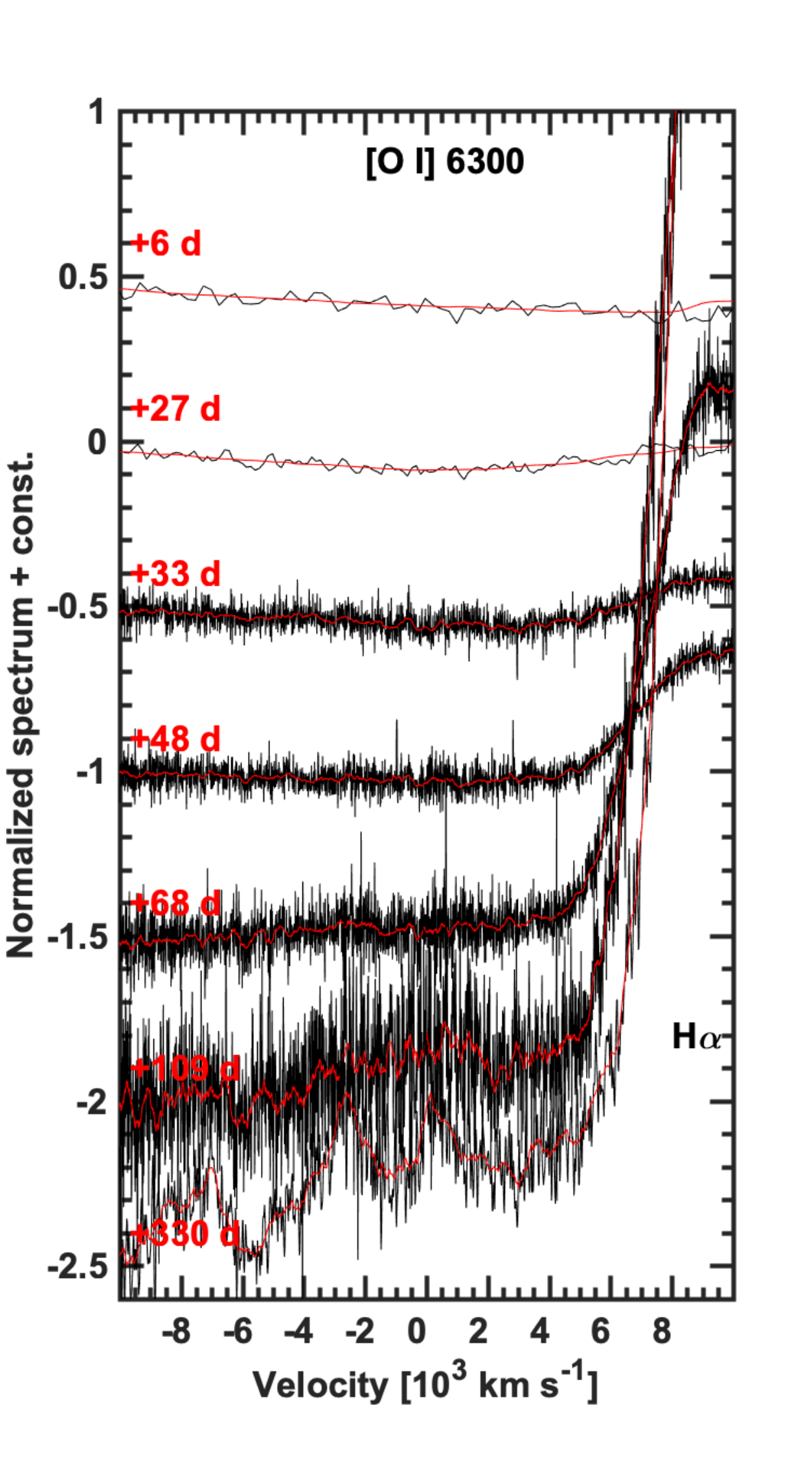}&
\includegraphics[width=9cm]{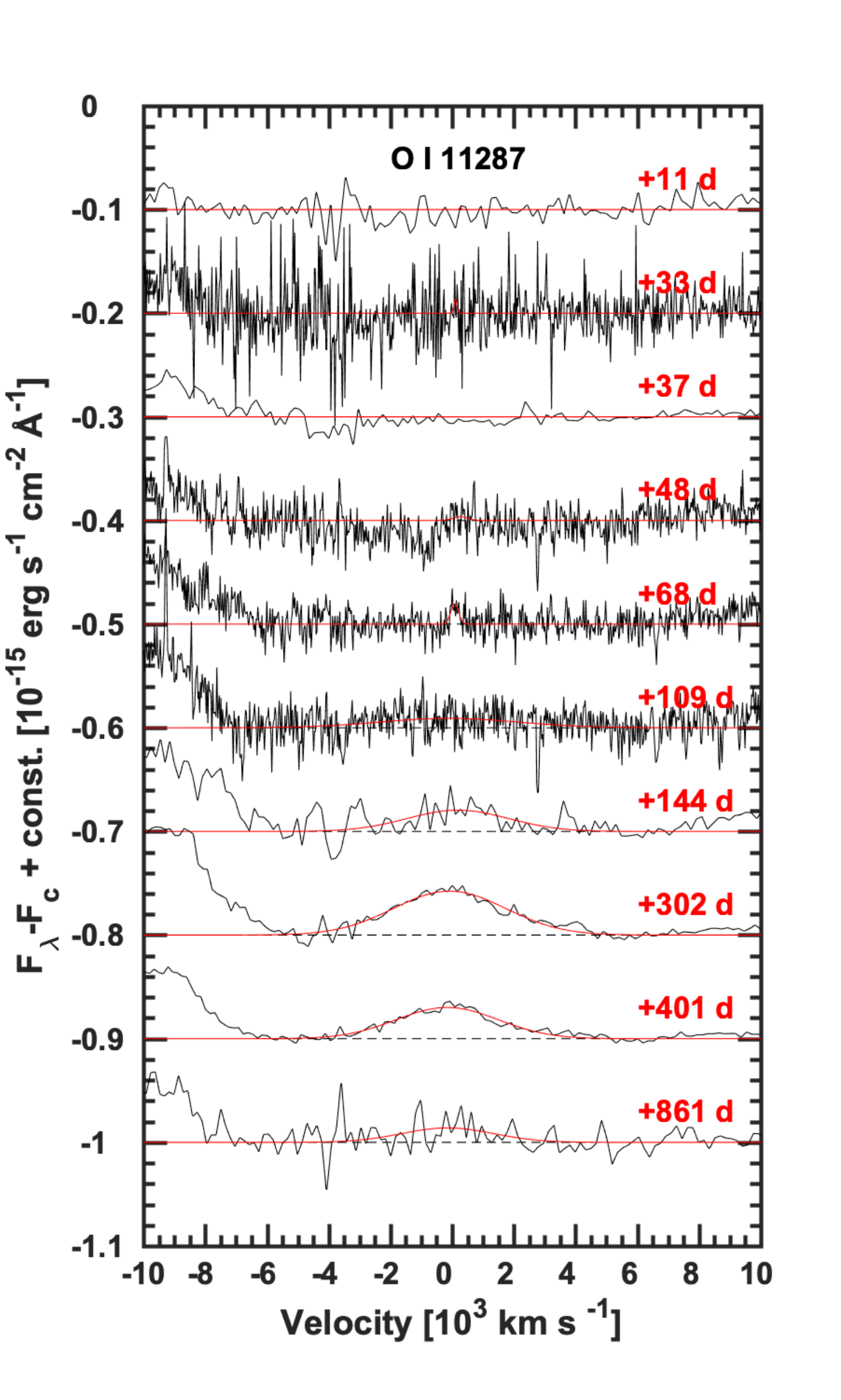}\\
\end{array}$
\caption{\label{OI} (Left panel) Normalized spectra in velocity space around the [\ion{O}{i}]~$\lambda$6300 wavelength regions. The original spectrum in black and the smoothed spectrum in red are shown to better guide the eye to follow the line profile evolution. The spectral phases are reported in red next to each spectrum. The emergence of the oxygen line at 6300~\AA\ at late epochs is evident. We note that the line at $-$3000~km~s$^{-1}$ is \ion{Fe}{ii}. (Right panel) Continuum subtracted and extinction corrected spectra in velocity space around the wavelength region of \ion{O}{i}~$\lambda$11827. This line appears at late epochs and it is well fit by a single Gaussian profile of $\sim$4000 km~s$^{-1}$ FWHM (shown in red). The phases of each spectrum are reported next to them.}
\end{figure}

\begin{figure}
\includegraphics[width=9cm]{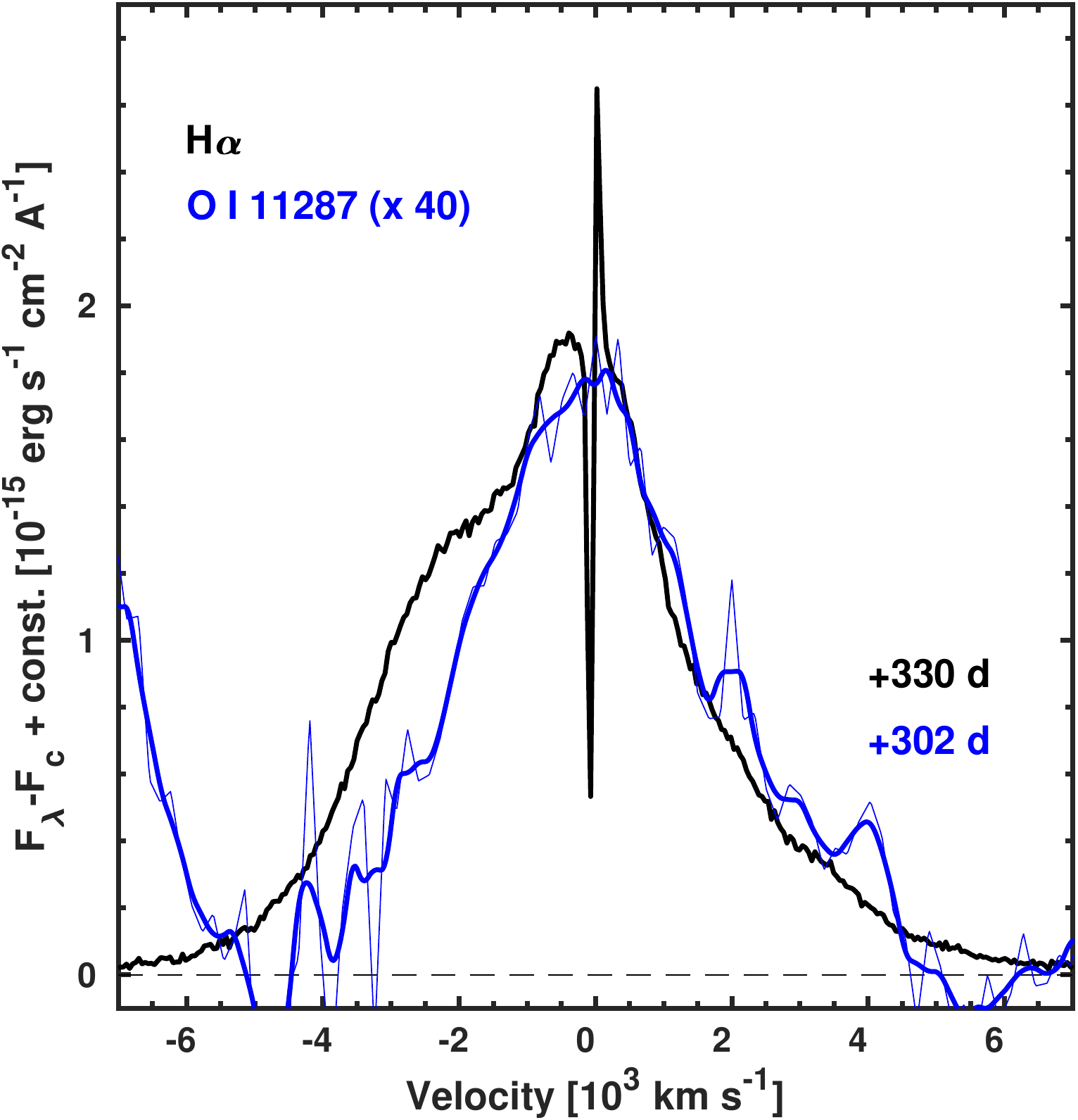}
\caption{\label{OIvsHa}Continuum-subtracted H$\alpha$  and \ion{O}{i}~$\lambda$11287 at +300 to +330~d. The NIR \ion{O}{i} line, which was scaled to match the peak of H$\alpha$, shows a broad profile that is almost as broad as H$\alpha$.}
\end{figure}

\begin{figure}
\includegraphics[width=9cm]{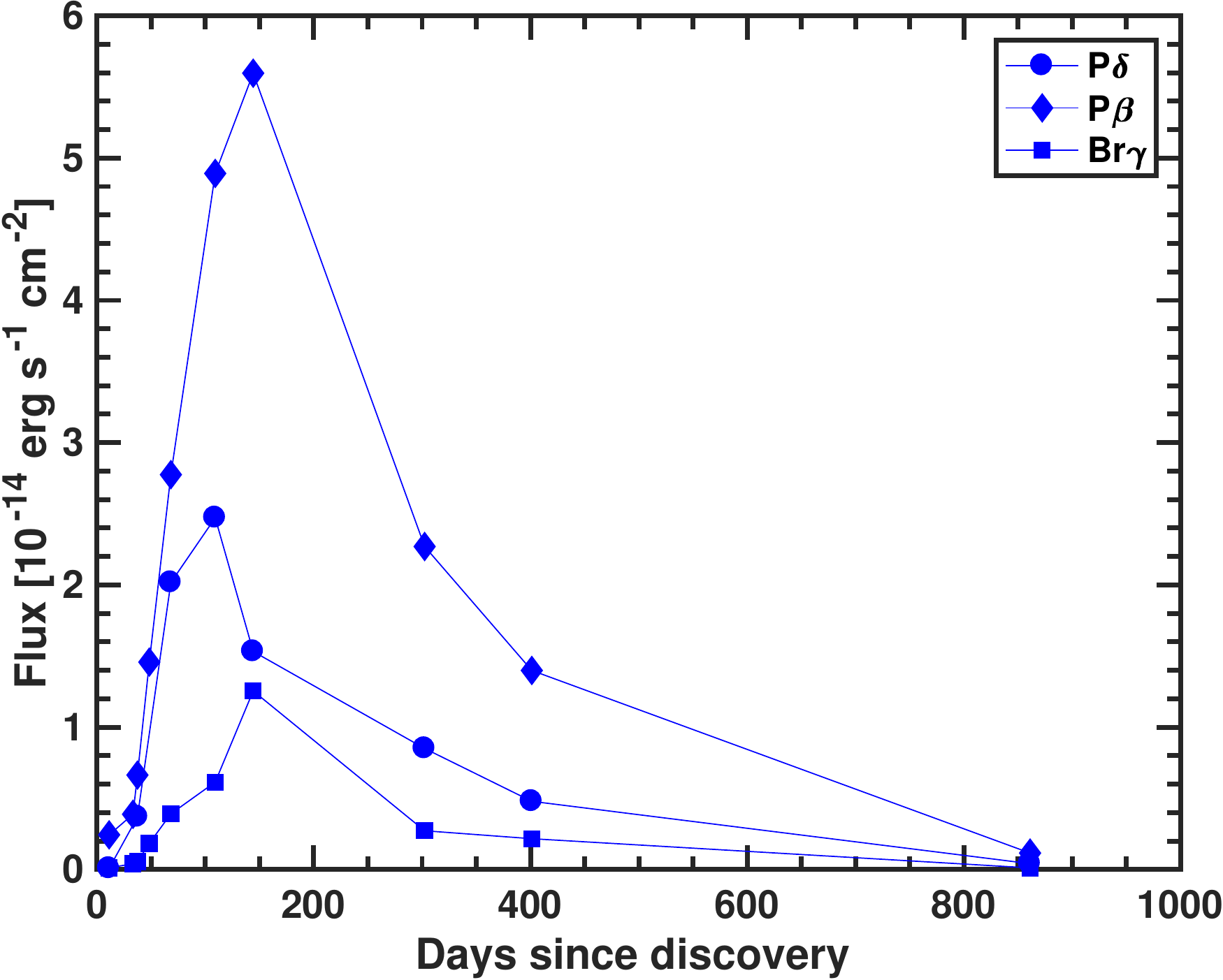}
\caption{\label{fluxNIR}Flux as a function of time for three bright hydrogen lines in the NIR spectra. Their evolution is rather similar.}
\end{figure}

\begin{figure}
\includegraphics[width=9cm]{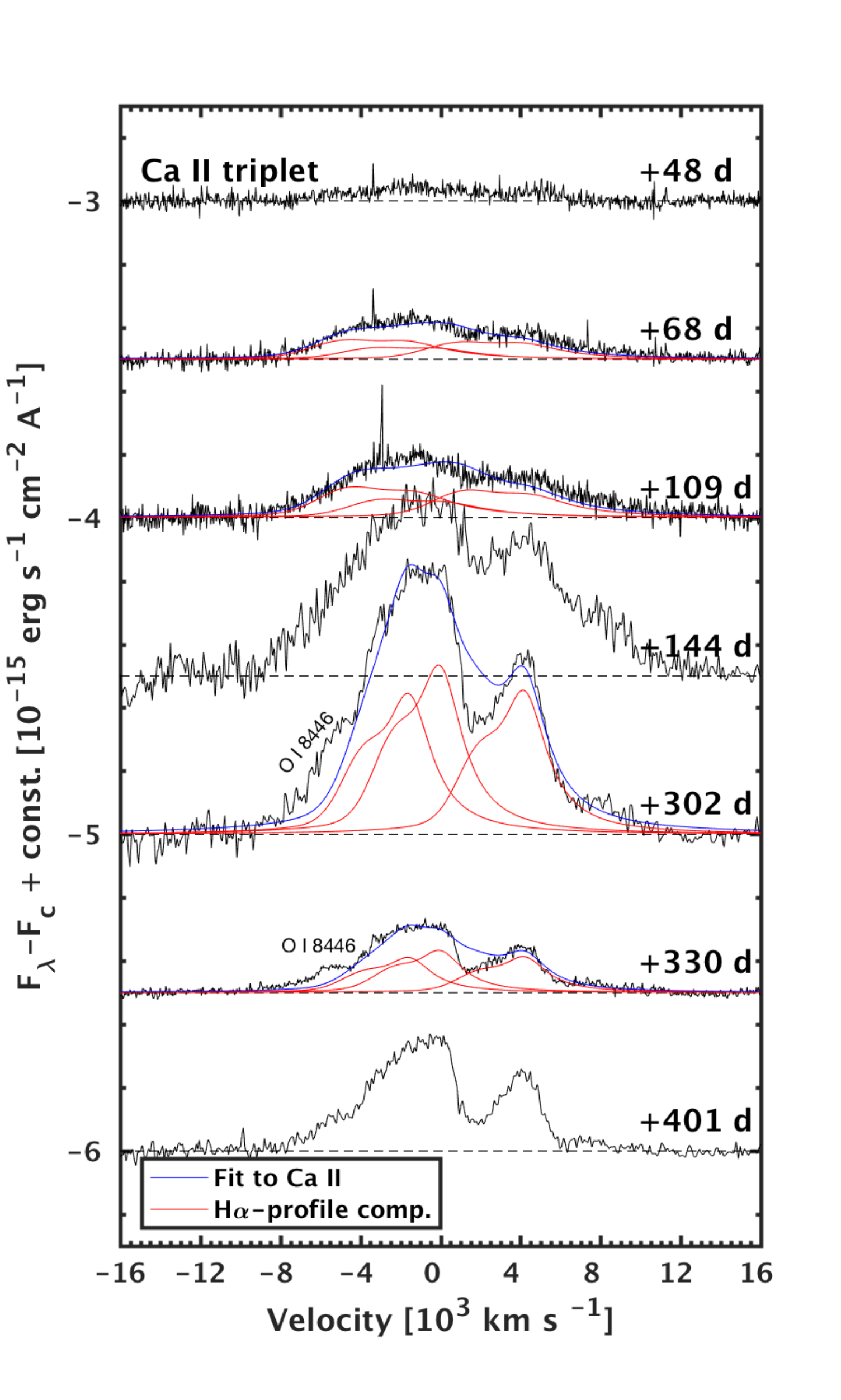}
\caption{\label{CaII}Continuum-subtracted and extinction corrected \ion{Ca}{ii} triplet profiles (black) fitted by a sum (blue) of three H$\alpha$ best-fit profiles from Fig.~\ref{fitHa} to reproduce the complex line shape. We followed the approach by A17 and confirmed that this is also a good fit for the \ion{Ca}{ii} triplet at +302~d and +330~d. The spectral phases are shown on the right next to each spectrum. An excess blueward of the best line fit is compatible with the emerging of \ion{O}{i}~$\lambda$8446, but there is also \ion{Fe}{ii} that can contribute to the flux in that range.}
\end{figure}

\begin{figure}
\includegraphics[width=18cm]{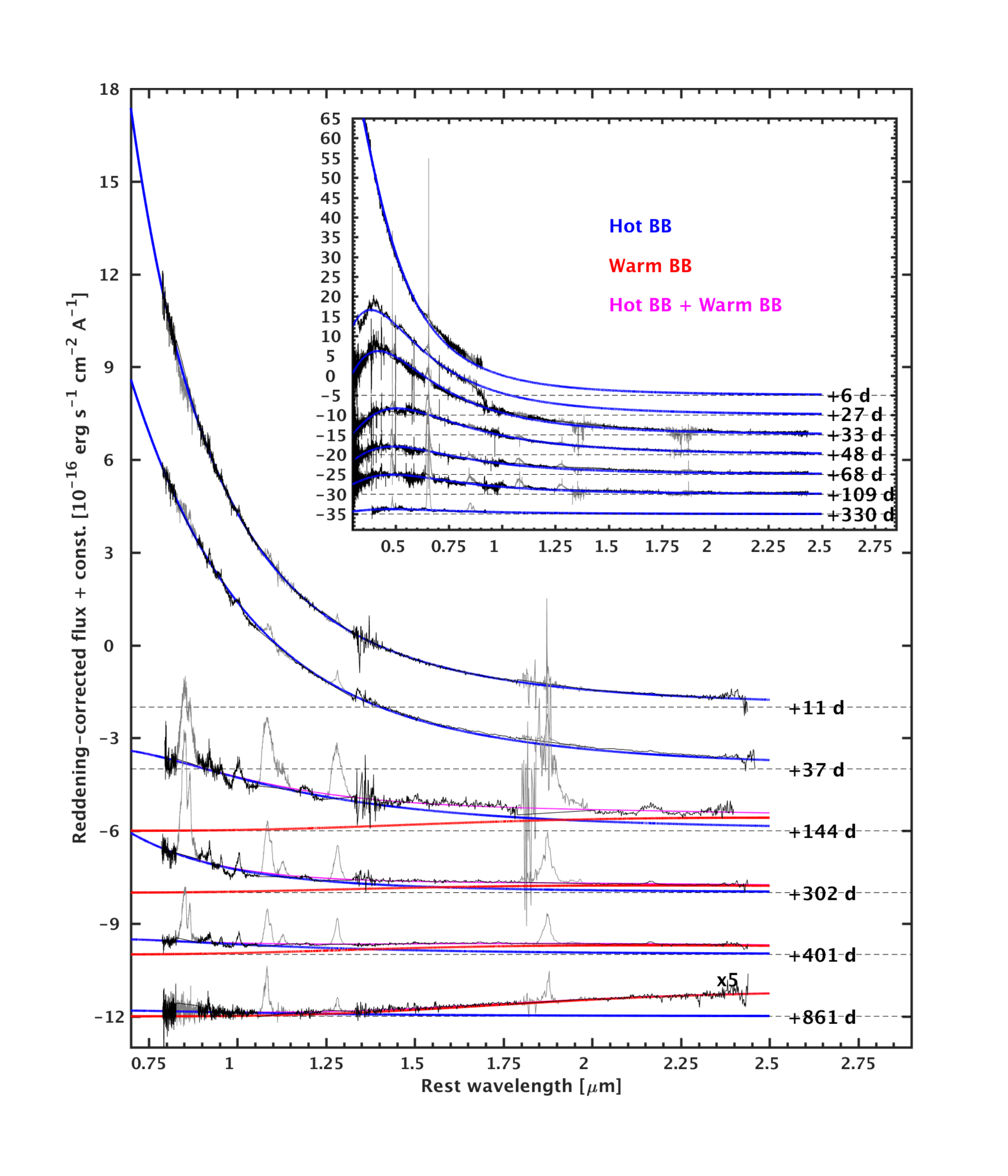}
\caption{\label{specfit}(Inset) Optical spectra after absolute flux calibration and extinction correction, fitted by a single BB function. The parts of the spectrum in light gray were excluded from the fit because of the presence of strong emission lines. (Main panel) NIR spectra after absolute flux calibration (using $J$ band) and extinction correction, fitted either by a single BB (at early epochs), or by a sum (magenta) of two BB components, one warm (red) and one hot (blue). The warm (red) component dominates at late epochs. The flux of the last spectrum was multiplied by five for a better visualization. }
\end{figure}

\begin{figure}
\includegraphics[width=18cm]{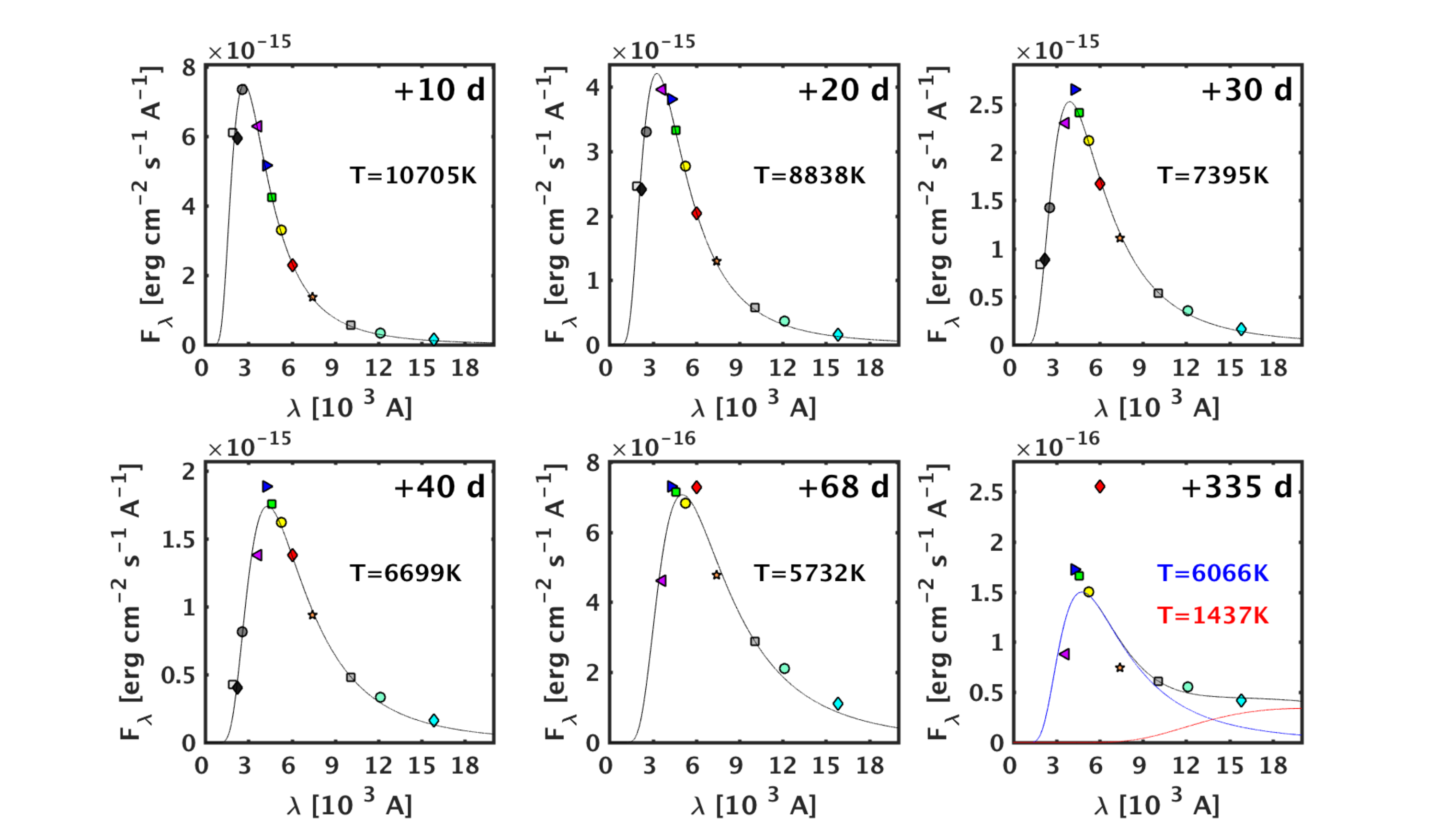}
\caption{\label{SEDfit}Example of SEDs of SN~2013L with the specific epochs corresponding to times  when  NIR photometry is available. Each flux point was obtained by converting the magnitudes into fluxes after having applied the appropriate extinction correction. The color and symbol code for each band is the same as in Fig.~\ref{LC}. A single BB function was fit to each SED until +68~d (black line), while a sum of two BB functions was used to fit  the   +335~d SED in order to reproduce the NIR excess (red line) and the optical BB shape (blue line). The $r$ band is not included in the fit due to the prevalent H$\alpha$ emission. The BB temperature is reported in each subpanel as well as the phase of the SED.}
\end{figure}

\begin{figure}
\includegraphics[width=18cm]{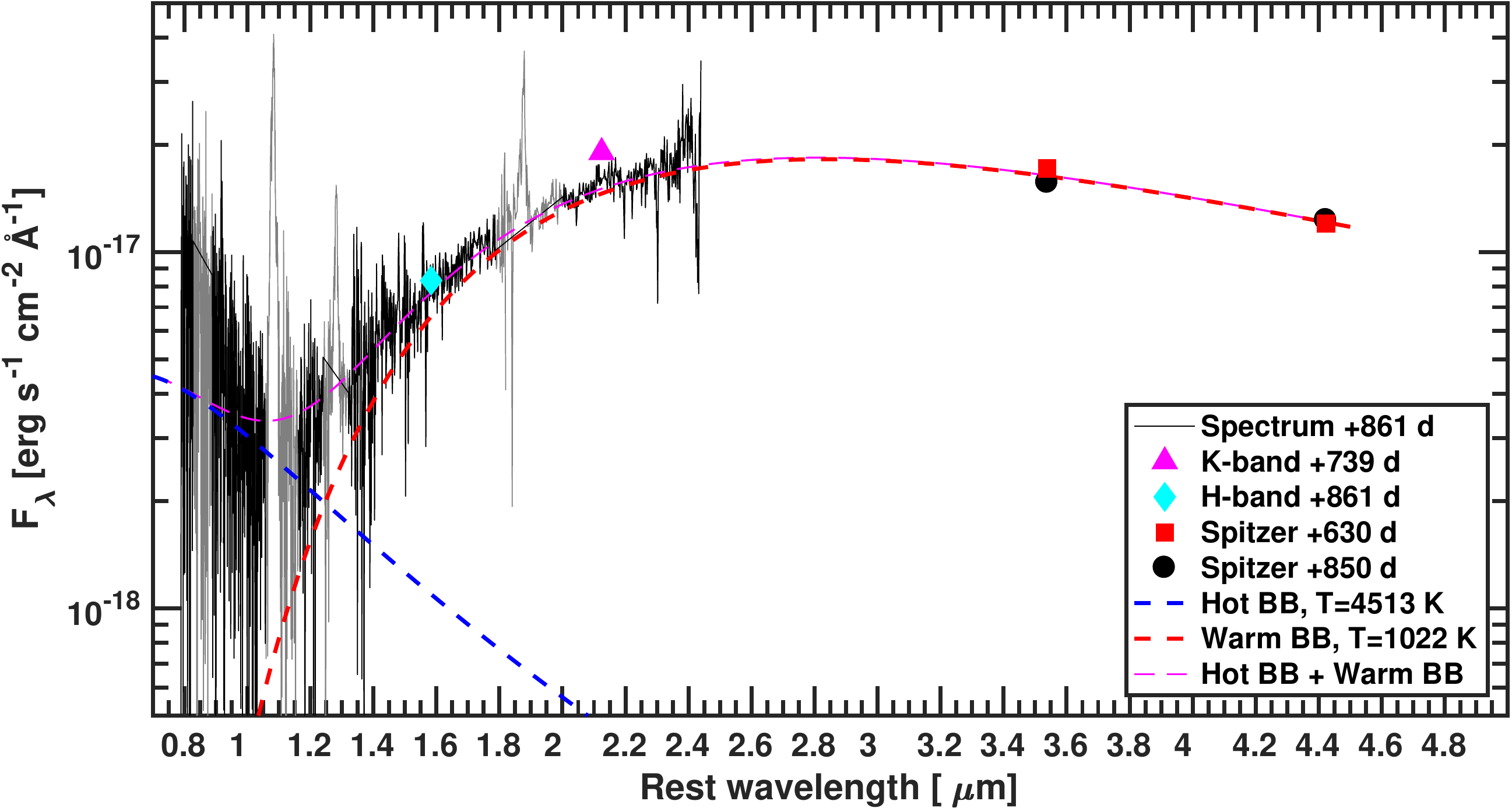}
\caption{\label{spitz}Late phase  $K_s$-band and  \textit{Spitzer}  photometry converted into monochromatic fluxes plotted together with our last epoch NIR spectrum. The SED exhibits a strong NIR and MIR excess that dominates  the SN emission two years after discovery. The NIR spectrum was scaled to match the $K_s$-band flux. We fit a sum of two BB functions (magenta dashed line) to reproduce the NIR spectrum and the MIR fluxes from the first Spitzer epoch. The warm BB component (red dashed line) dominating the NIR and MIR has a temperature of  1022~K and the hot component (blue dashed line) has a temperature of 4513~K.}
\end{figure}

\begin{figure}
\includegraphics[width=14cm]{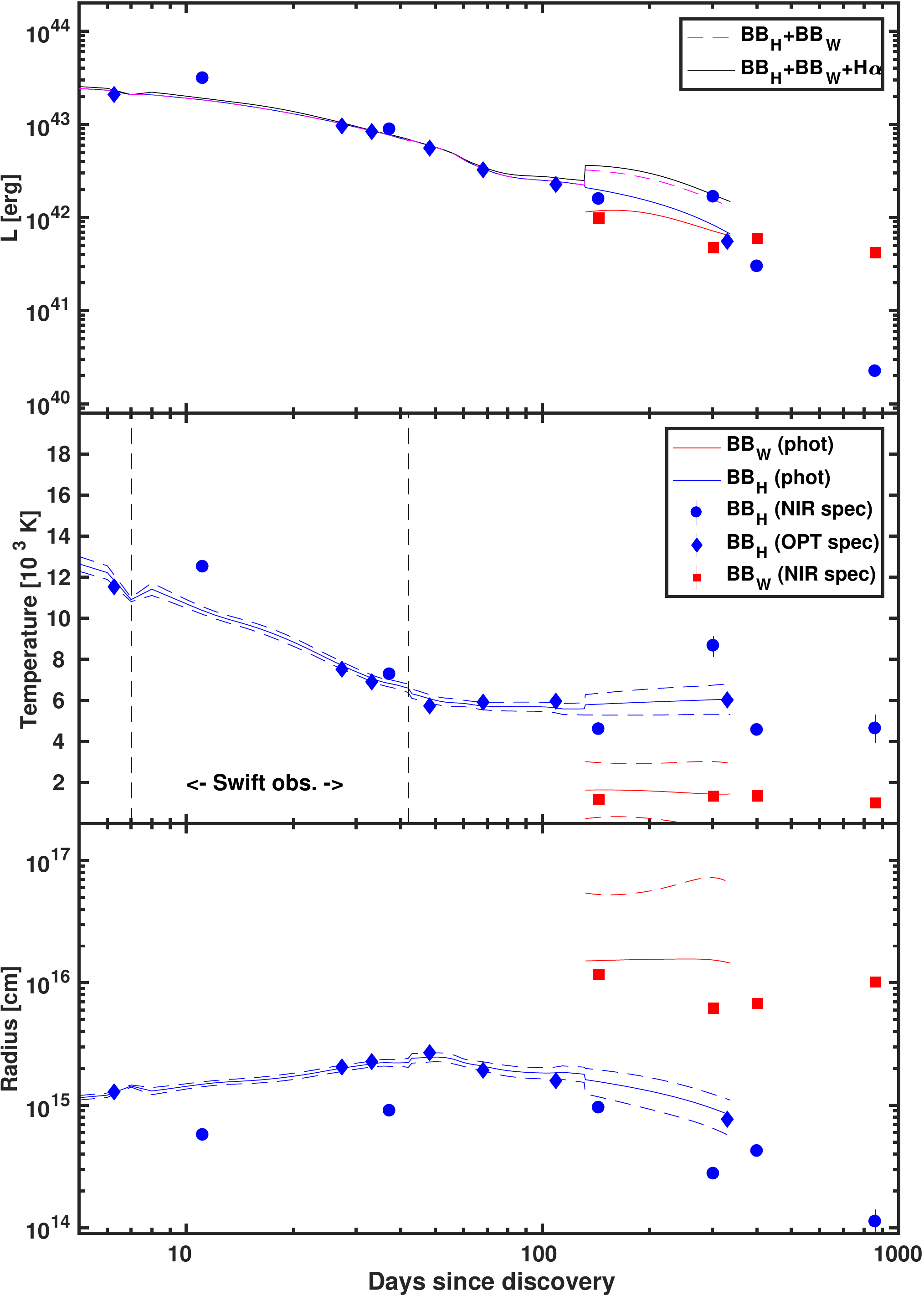}
\caption{\label{LTR}(Top panel) Bolometric luminosity of SN~2013L. From the BB fit to the photometry (solid lines), we derived the luminosity of the hot BB component (solid blue line, dominating at early epochs) and of the warm BB component (red solid line, dominating at late epochs and emerging at +132~d). Their sum (dashed magenta line) and their sum plus the H$\alpha$ emission neglected from the BB fit (black solid line) are also shown, the last one represents the total luminosity. We also obtained the corresponding luminosities from the BB fit to the optical (blue diamonds) and NIR spectra (blue dots and red squares). (Central panel) Temperature evolution of SN~2013L for the two different BB components fitting the SEDs of SN~2013L (symbols as in the top panel). In addition, the dashed lines mark the uncertainties on the temperature from the BB fit to the photometry. (Bottom panel) Radius evolution for the two different BB components fitting the SEDs of SN~2013L, symbols are the same as those in the top and central panels.}
\end{figure}

\begin{figure}
\includegraphics[width=18cm]{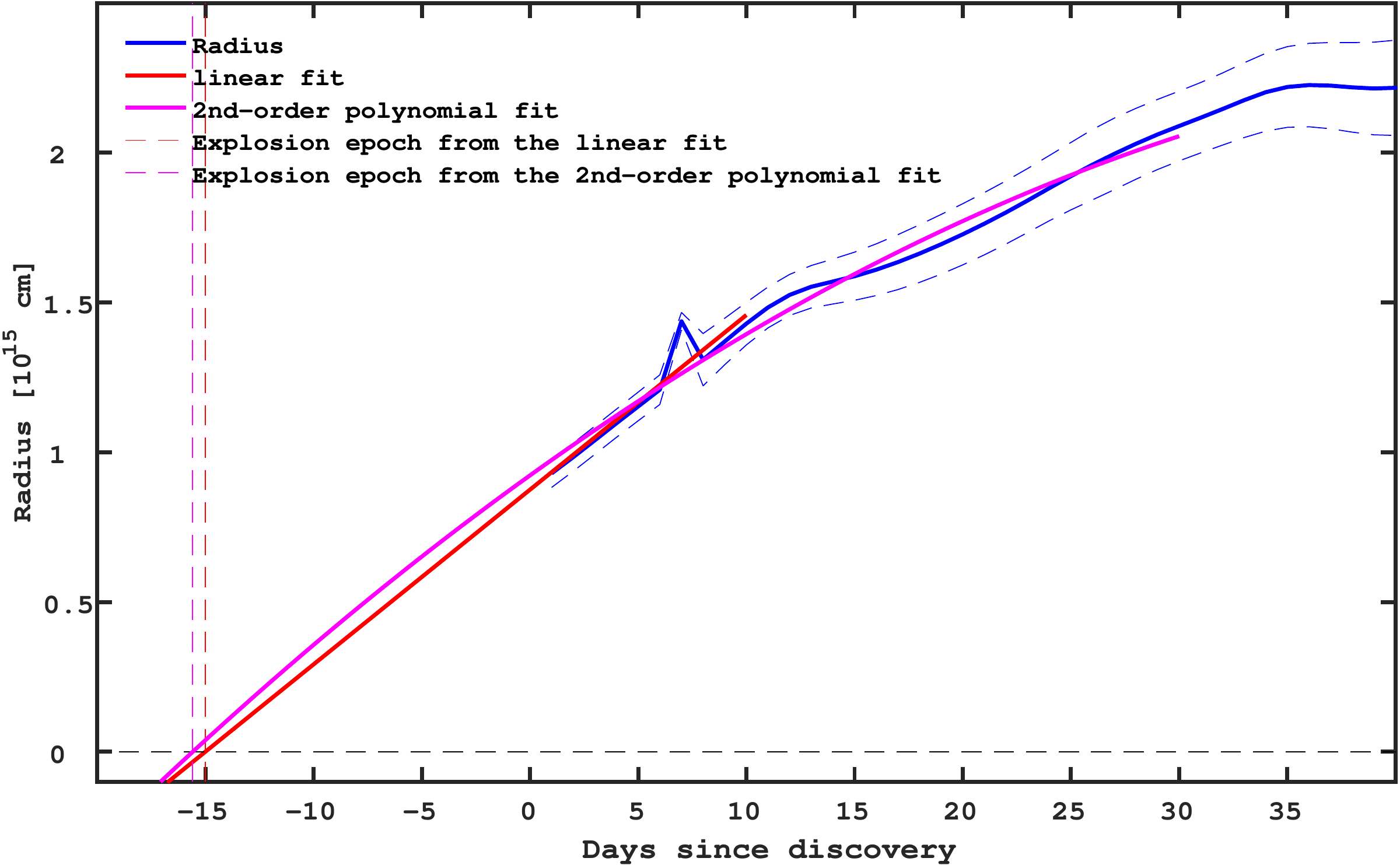}
\caption{\label{fitradius}Fit of the radius evolution used to derive the explosion epoch. A linear fit from the first 10 days (red) gives an explosion epoch of $-$15~d before explosion. A similar explosion epoch was obtained by fitting the first +30~d with a second-order polynomial (magenta).}
\end{figure}

The main characteristics of the spectra are the dominance of the Balmer lines, the wide line profiles, and the blue-shift of the lines. To illustrate the evolution of the line profiles, we show the H$\alpha$
line from +48~d to +1133~d after explosion in Fig.~\ref{lineHa_scale}.  
To clearly see the evolution of the shoulder, we normalized the flux to that of the flat part of the shoulder.

Figure~\ref{lineHa_scale} clearly demonstrates the gradual decrease of the velocity of the shoulder in the blue wing of the line, from $\sim 4000$ km~s$^{-1}$ at early epochs to a near disappearance of the shoulder at the last epochs. The figure also shows the separation of the line profile via two distinct components, a flat part with decreasing blue-shift and a central nearly symmetric line, with a fairly constant FWHM.  

To model the dynamic line profile, we note that a flat-topped line profile is indicative of a hollow emitting shell, moving with a high expansion velocity.  The observed profiles are, however, asymmetric and smooth with very extended wings, which are characteristic of Thomson electron scattering. To model the line profile, we used a Monte Carlo code, calculating the emitted spectrum after scattering for a given input spectrum. The code is a modified version of the code used for modeling the line profiles of SN~2010jl \citep{fransson14}; it calculates the emission including the scattering by the thermal electrons as well as a macroscopic bulk velocity. We emphasize that we do not aim to develop a complete self-consistent model here. This requires a full radiation-hydrodynamical code, which is beyond the scope of this paper. Here we mainly aim to use the line profiles and light curve as diagnostics of the main properties of the SN and, in particular, the shock velocity and mass-loss rate.

We assume that the H$\alpha$ photons are emitted from a spherical shell with an inner radius of $R_{\rm em~ in}$ and an outer radius of $R_{\rm em ~out}$, and that the velocity inside $R_{\rm em~ out}$ is homologous, that is, $V(r) = r/ t$. For example, a constant velocity shell would give a similar result. The natural interpretation is that the emitting shell is the cool, dense shell (CDS) resulting from 
the interaction region between the expanding ejecta and the dense CSM, as discussed in \citet{chevalier94} or \citet{dessart15} for SN 2010jl. The photons are then followed as they are scattered by the slow, ionized, but unshocked CSM outside of the shock front. In the standard scenario, we assume this CSM to have a density $\propto r^{-2 }$, which is characteristic of a constant velocity wind before the explosion, but we also discuss cases with a different dependence, which may apply for a time limited eruption with a varying mass-loss rate. Because the shell is narrow, the latter assumption is not very important. We discuss the density and temperature structure further below. 

The ionized CSM is mainly characterized by an optical depth $\tau_{\rm e}$ and a temperature $T_{\rm e}$. Because the dispersion in frequency by the incoherent electron scattering is $\Delta V \approx N_{\rm scatt}^{1/2} V_{\rm thermal} \propto \tau_{\rm e} T_{\rm e}^{1/2}$ (since $N_{\rm scatt} \approx \tau_{\rm e}^2$ ), the optical depth and temperature of the CSM are not independent. A different average CSM temperature would therefore give a similar result for the line profile for an optical depth scaling as $ \tau_{\rm e} \propto (T_{\rm e}/10^4 {\rm ~K})^{-1/2}$. 
 
In realistic models, the temperature depends on the distance from the shock. The calculations in  \cite{dessart15} show a temperature decreasing from $\sim 2\times 10^4$ K to $\sim 1\times 10^4$ K for the SN 2010jl model at 30 days.  Here, we model this by a temperature decreasing as $T_{\rm e} \propto  r^{-0.4}$ from the shock, but the line profile is not very sensitive to this assumption. Also for other parameters, such as the bound-free opacity, we use the  model by  \cite{dessart15} as a guide. It is important to note, however, that this model was used for a relatively low ejecta mass of $9.8 \Msun$, while a more massive ejecta may give different parameters.

The importance of scattering in the post-shock region and ejecta depends on the degree of thermalization there. Because of the high density and low temperature in the cooling gas in the CDS, the bound-free and free-free continuum absorption increases compared to electron scattering, as can be seen in the simulations of \cite{dessart15}. The exact level depends on the shock compression, which is sensitive to multidimensional effects, as well as magnetic fields and the specific ejecta model. The details of the shock properties and the radiation from this are, however,  not yet fully understood  \citep[see e.g., discussion in][]{waxman17}. 

In assuming a viscous shock and ignoring preacceleration of the CSM, the temperature of the shock is 
 \begin{equation}
T_{\rm s} = 3.4 \times 10^8   \left(  { V\over 5000 \kms}\right)^{2}      \ \rm K ,
\label{eq_shocktemp}
\end{equation} 
for solar abundances. \cite{dessart15}, however, find a considerably lower shock temperature of $\la 10^5 $ K, including the effects of radiation, although they note that the shock structure is not fully resolved in their standard calculation, while higher temperatures are obtained in high resolution simulations. The presence of hot gas with $T_{\rm e} \ga 10^8$ K from a viscous shock is consistent with the X-ray observations of SN 2010jl \citep{ofek14a,Chandra2012,Chandra2015}.  

For a CSM  with a constant mass-loss rate, $\dot M$, and with a wind velocity, $u_{\rm wind}$,  the density corresponds to  $ \rho(r)= \dot M /  (4 \pi u_{\rm wind} r^2)$. 
 For a more general CSM  with a power law density $\propto r^{-s}$, this corresponds to
  \begin{equation}
  \rho(r)={ \dot M \over  4 \pi u_{\rm wind} R_0^2 }  \left(  {R_0 \over r}\right)^{s}    \ ,
  \label{eq_csmdens}
\end{equation} 
 where $r_0$ is a reference radius. The radius of the shock is then
  \begin{equation}
  R_{\rm s} = R_0 (t/t_0)^{(n-3)/(n-s)} \ ,
  \label{eq_shradius}
\end{equation} 
  assuming a similarity solution can be applied \citep{chevalier82}. The shock velocity evolves as 
\begin{equation}
V(t) = V_{\rm 0} \left(  {t \over t_{\rm 0} }\right)^{(s-3)/(n-s)}      \ .
\label{eq_vel}
\end{equation} 
Here $V_0$ is the velocity at a reference time $t_0$.  For a CSM shell with
 an outer radius $R_{\rm out}$, we can take $R_0=R_{\rm out}$, or in terms of the time of the breakout of the shell, $t_{\rm b}$, $R_{\rm out} =   V(t_{\rm b} ) \ t_{\rm b}$.

 The cooling time of the shock is 
 \begin{equation}
\begin{split}
 t_{\rm cool}= 5.7  \left(  { \dot M \over 0.1 \ \ml}\right)^{-1}     \left(  {u_{\rm wind} \over 100 \ \kms}\right) \\
 \left(  {V \over 5000 \ \kms}\right)^{3}    \left(  {t \over 100 \  \rm days}\right)^{2}   \  \rm days ,\label{eq_csmdens}
  \end{split}
  \end{equation}
  
assuming free-free cooling dominates. Compton cooling may decrease this further. 
The gas therefore rapidly cools until the cooling is balanced by the radiative heating. Because the post-shock gas is in near pressure balance, the compression is $\sim T_{\rm s}/T_{\rm ps} \ga 10^4$, where $T_{\rm ps}$ is the post-shock temperature, set by the radiation field. In reality, instabilities and magnetic and radiation pressure probably limit the compression \citep{steinberg18}. In any case, the region is cool and dense, and thermalization by bound-free absorption is very important.

 The
total electron scattering depth of the CSM (if completely ionized, see below) is then
\begin{equation}
\tau_{\rm e \ CSM}=  \frac{ \dot M \kappa} { 4 \pi (s-1) u_{\rm wind}  {V(t_{\rm b} ) t_{\rm b} } }[ \left(\frac{{t_{\rm b}} }{t}\right)^{(s-1)(n-3)/(n-s)} - 1 ].
\label{eq_taucsm}
\end{equation}

Here, we ignore the contribution from the lower density CSM outside $R_{\rm out}$, which provides a small extra contribution. 
 
The swept up shocked CSM mass is 
\begin{equation}
 M_{\rm CDS}= {\dot M  \over  (3-s)  u_{\rm w} } \left( { R_{\rm b} \over R_{\rm s}}\right)^{s-2}  R_{\rm s} [1 - \left({R_{\rm in} \over R_{\rm s}}\right)^{3-s}]  \ ,\end{equation} 
where  $R_{\rm in}$ is the inner radius of the CSM. Also, in assuming this to be fully ionized, the electron scattering depth is 
\begin{equation}
\begin{split}
 \tau_{\rm e \ CDS}= {\dot M \kappa \over  4 \pi (3-s)  u_{\rm w} R_{\rm s}}   \left( { R_{\rm b} \over R_{\rm s}}\right)^{s-2} 
 [1 - \left( { R_{\rm in} \over R_{\rm s}}\right)^{3-s} ] \approx \\
 {\dot M \kappa \over  4 \pi (3-s)  u_{\rm w} t_{\rm b} V(t_{\rm b} )}   \left(\frac{t_{\rm b} }{t}\right)^{(s-1)(n-3)/(n-s)} ,\label{eq_taucds}
 \end{split}
 \end{equation}
 if $R_{\rm in} \ll R_{\rm s}$. 
 
 We treat the structure of the shocked gas as a region of hot gas with $T_{\rm e}=T_{\rm s}$ and thickness $t_{\rm cool} V_{\rm s}/4,$ and inside of this are a dense shell with optical depth $ \tau_{\rm e \ CDS}$ and a temperature $\sim 10^4 $ K. The optical depth of the hot gas immediately behind the shock is $\la 0.1$ for $V_{\rm s} \la 5000 \kms$ and it is not important for the scattering of the optical photons. Because of the need for thermalization in the CDS, the structure of the ejecta interior to the CDS is not important, and we assume the density to be constant and similar to that in \cite{dessart15}.
 
 The main free CSM parameters are therefore $\dot M, V(t_{\rm b} ), T_{\rm e}$, and $\epsilon$.  Finally, the line profile depends on the H$\alpha$ emissivity from the shocked gas as well as the CSM. If recombination dominates,  we expect the emissivity, $j$, to scale as $j \propto \rho^2 T_{\rm e}^{-0.942-0.031 \log(T_{\rm e}/10^4 \ {\rm K})}$ \citep{Draine11}. At high H$\alpha$ optical depth and electron density this is likely to be a rough approximation at best, especially in the CDS. We use this expression for the CSM, where the radial dependence of the emissivity is of some importance. However, the small radial extent of the CDS and its dominance for the luminosity means that the radial dependence is unimportant and the main parameter is the total luminosity relative to that of the CSM luminosity.  We illustrate the effect of this below. 
 
 The shape of the line profile depends on $\dot M, V(t_{\rm b} ), T_{\rm e}$, and $\epsilon$ in different ways. The most important aspect for the asymmetry of the line profile and also the main result of this modeling is the velocity of the shock, which determines the velocity shift of the shoulder in the blue wing of the profile. 

 The optical depth of the CSM, $\tau_{\rm e \  CSM}$, is the most important parameter for the slope of the blue wing shortward of the shoulder. Together with $V_{\rm s}$, this determines the mass-loss rate needed. 
There is, however, a degeneracy of the mass-loss rate and the outer radius, or equivalently $t_{\rm b}$, for a given $V_{\rm s}$ (Eqn. \ref{eq_taucsm}).

In our line profile calculations, we treated the ratio of absorption to electron scattering as a parameter, $\epsilon$. In the CSM   \cite{dessart15} find $\epsilon \sim 10^{-3}$; however, in the dense post-shock gas,  $\epsilon$ depends on the density and temperature in the CDS and it may be $\ga$1, resulting in strong thermalization of the H$\alpha$ photons. 
Thermalization in the CDS and ejecta are important, while for $\epsilon \sim 10^{-3}$  they are less important in the CSM. An argument for high values of $\epsilon$ in the CDS and ejecta come from the fact that multiple scattering between regions with high relative velocities, from either the CSM to the ejecta or within the ejecta, produces a strong red wing extending to high velocities \citep[e.g.,][]{fransson89}. This is not seen in SN 2013L. To suppress this wing, an efficient thermalization in the ejecta is required. In our models, we have assumed $\epsilon \ga 1$ in the CDS. 

\begin{figure}
$\begin{array}{cc}
\includegraphics[width=8cm]{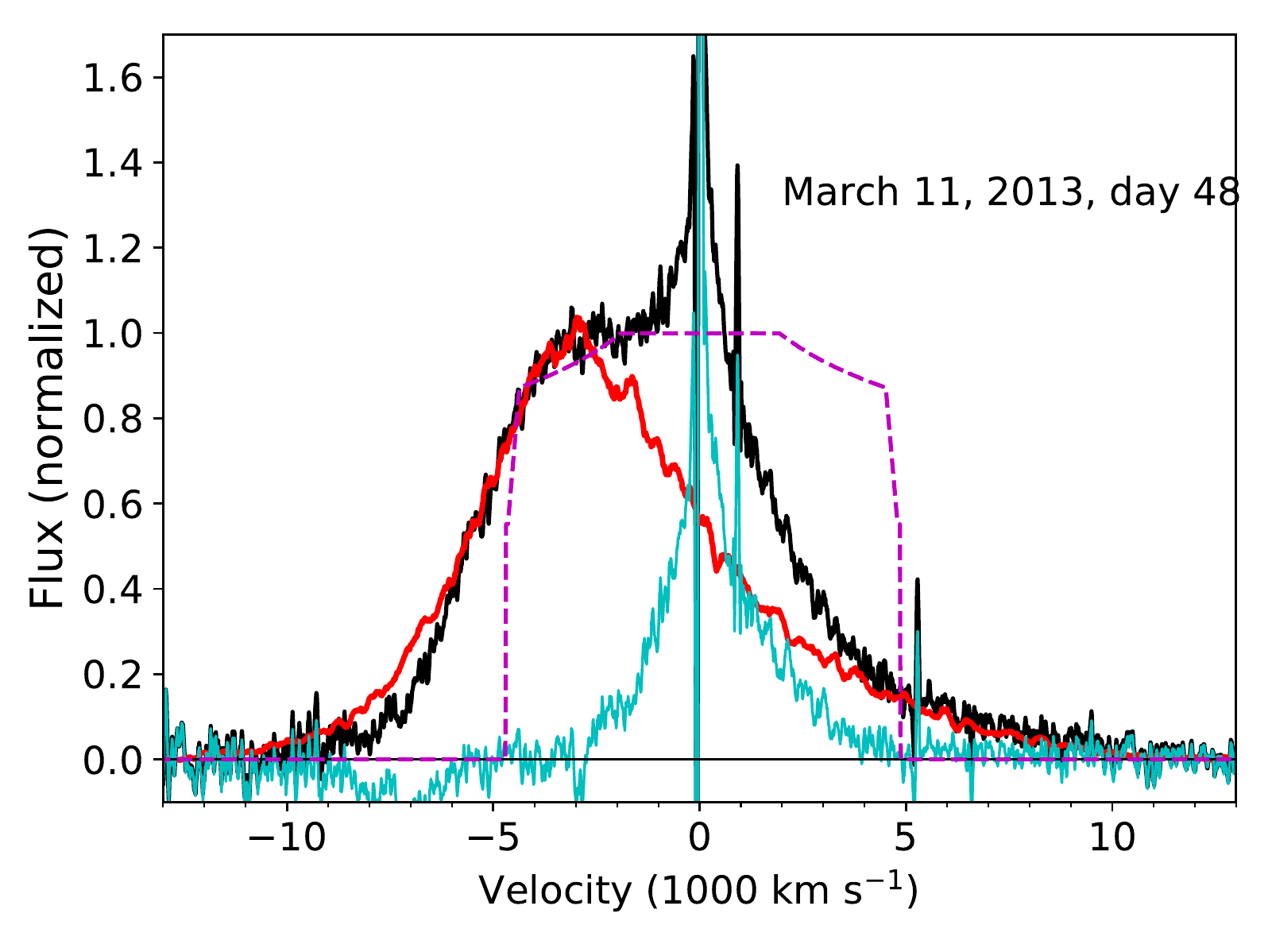} &
\includegraphics[width=8cm]{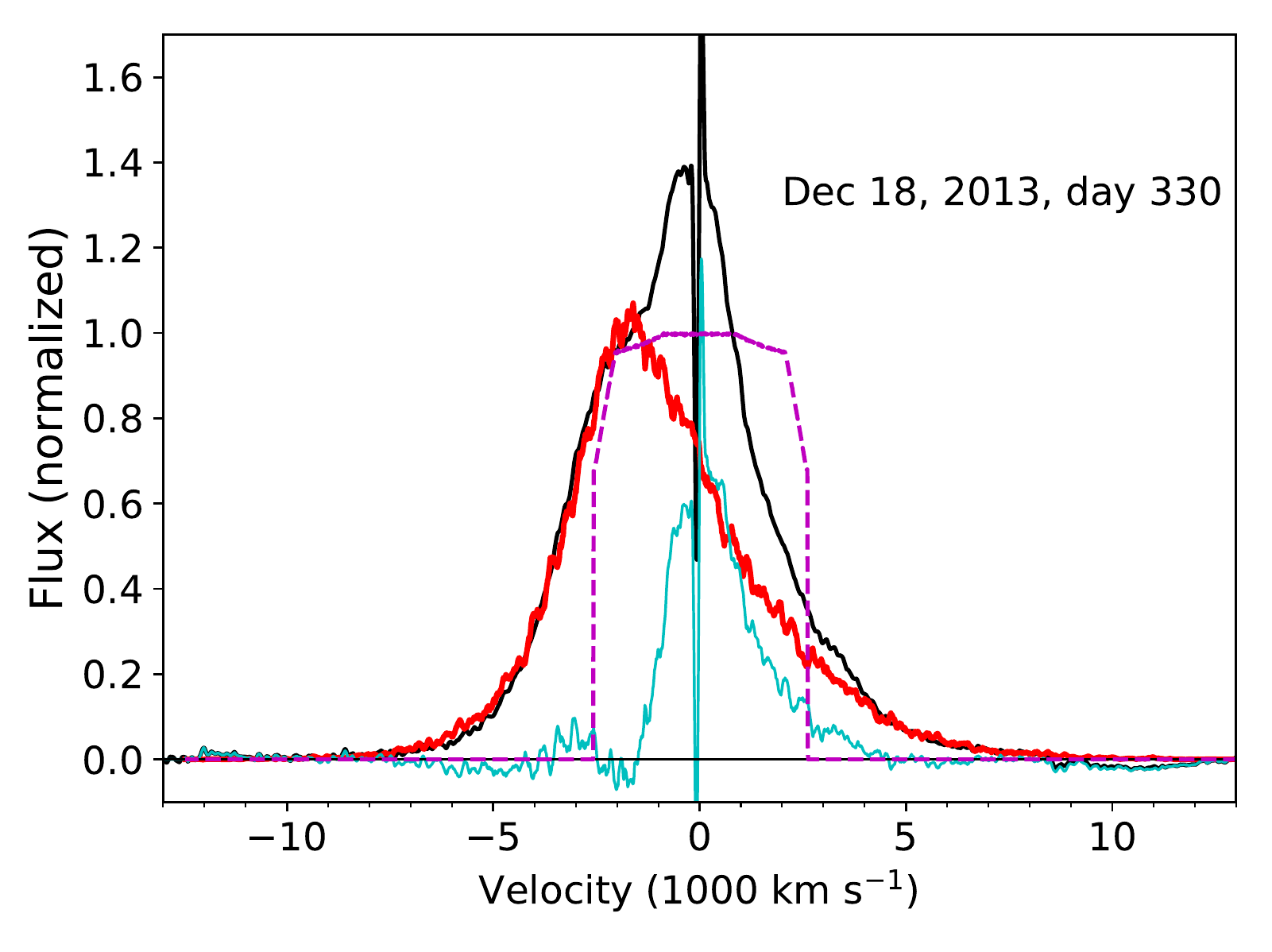}
\end{array}$
\caption{\label{Halpha_130311} The continuum subtracted H$\alpha$ line (shown in black) from days 48 (left) and 330 (right) together with the electron scattering profiles from the model with H$\alpha$ emission only from the ejecta and shocked region (red). The dashed magenta  lines show the box-like input emission profiles from a radially thin shell with velocity $4800  \kms$ and $2700  \kms$, respectively. The cyan lines show the residual between the model and the observed profile. Electron scattering and occultation by the photosphere result in the smooth, blue-shifted profile. We note the central excess from the models, indicating a different component.}
\end{figure}

In the left panel of Fig.~\ref{Halpha_130311}, we show the result of a simulation of the observed H$\alpha$ line at +48~d. The parameters for this simulation are $V_{\rm s}= 4800 \kms$, $\tau_{\rm e, \ CSM}=12.3$, and $\tau_{\rm e, \ CDS}=8.6$. 

 In the right panel, we also show a model for the line at +330~d. The corresponding parameters at this epoch are $V_{\rm s}= 2700 \kms$ and $\tau_{\rm e, \ CSM}=7.2$. In both simulations, we used $T_{\rm e} = 1.5 \times 10^4$ K at the shock, decreasing to 9000 K at the outer boundary of the CSM shell.  For the observed line profile, we subtracted the continuum by fitting a straight line between $\pm 15\,000 \kms$ from the line center. We did not attempt to model the narrow P~Cygni profile of the outer CSM, with velocity $\sim 120 \kms$.

From these figures, it is seen that this simple model with a few parameters can provide a very good fit to the observed line profile outside of $\pm 3000 \kms$, as shown by the cyan colored lines, which show the difference between the model and the observations. As illustrated by the magenta, dashed lines, showing the input emissivity, the line profile at these dates is dominated by the Doppler broadening from the macroscopic velocity of the CDS.  However, in the  center of the line, below velocities $\pm 3000 \kms$, there is a significant difference. This residual is interesting since it has a similar width as the central component, which becomes increasingly dominant at the later phases, and is very prominent by +330~d and later. We return to this component below.

The  agreement between the model and the observed line profiles shows that an expanding H$\alpha$ emitting shell, with electron scattering in a dense CSM outside the shell, provides a reasonable fit to the spectrum. While several of the parameters, in particular the temperature, emissivity, and thermalization parameter, $\epsilon$, are model dependent and require a more sophisticated NLTE calculation, as in \cite{dessart15}, the optical depth of the CSM and, most important for this paper, the shock velocity should be robust. The shoulder in the blue wing of the line therefore gives a measure of the velocity and its evolution with time. 

\begin{figure}
\includegraphics[width=10cm]{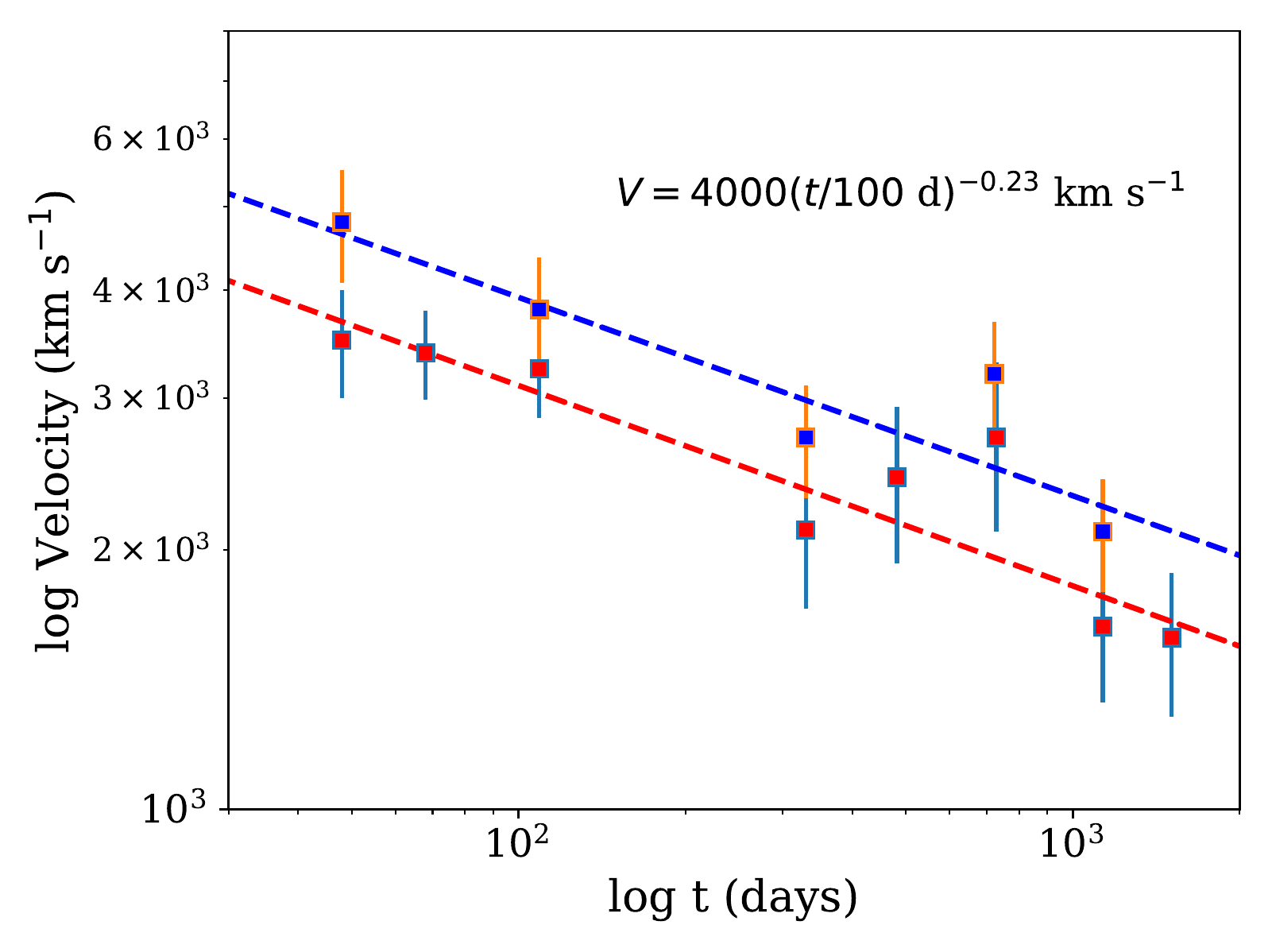}
\caption{\label{Halpha_vel} Red squares: Velocity evolution of the H$\alpha$-line blue shoulder as function of time. Blue squares: Velocity of the shell from the electron scattering model for selected dates. The dashed lines show a least square fit to these.}
\end{figure}

 In Fig.~\ref{Halpha_vel}, we show the observed velocity of the shoulder of the line profile for the different dates. For a number of selected dates, including the ones shown in Fig.~\ref{Halpha_130311}, we  also plotted the velocity of the H$\alpha$ emitting shell as  blue squares. For both sets of data, we fit a power law to the velocity as a function of time. For the velocity of the modeled shell before electron scattering, we find that 
\begin{equation}
\label{eq:Vel}
V_{\rm shell} = 4000 \left({t \over 100~ \rm{days}}\right)^{-0.23} \ \kms.
\end{equation}
The slope of the velocity of  the blue shoulder has the same time dependence, but a lower normalization, $V_{\rm break} = 3100~ (t / 100~ \rm{days})^{-0.23} \ \kms$. This difference is also illustrated Fig. \ref{Halpha_130311}, where it is seen that the electron scattering smooths the intrinsic line profile, causing the blue shoulder in the line profile to move to a lower velocity. 

\subsection{The location of spectral line formation \label{sec:whereline}}

From the above modeling, it is clear that most of the H$\alpha$ emission must originate from gas behind the shock. For a dense CSM, this is from the radiative forward shock \citep[e.g.,][]{rac_cf17}. This explains the high velocity needed for the majority of the photons, which are scattered by gas in this region and by gas that is external to the shock. The X-ray and UV emission from the shock is also the source of the ionization of the dense unshocked CSM that is needed for most of the electron scattering. This explains the major part of the line profile shown in Fig.~\ref{Halpha_130311}. 

The presence of a dense slow moving CSM is apparent from the narrow P~Cygni lines seen on top of the Balmer lines. 
From the spectra, H$\alpha$ was measured to have a BVZI of 120--240~km~s$^{-1}$.
This line component, as in other SNe~IIn, is formed from unshocked CSM that is excited by the  emission coming from the shock region. As a result, the BVZI value of this narrow component provides a measure of the CSM velocity. This component is, however, likely to be only the outer region of the CSM, or a separate extended  component with lower density, since the P~Cygni line would otherwise be smeared out by electron scattering (see below).

The origin of the low-velocity component below $\pm 3000 \kms$ (the central component of H$\alpha$), which is apparent as the residual in the simulations of the line profiles in Fig. \ref{Halpha_130311}, is especially interesting. The fact that the luminosity of this component approximately follows that of the broad component (see Fig.~\ref{fitHa3}) suggests that a similar excitation mechanism is occurring for both, that is, excitation by the radiation from the shock region. Excitation by radioactivity in the core is very unlikely, both because of the high luminosity, the time evolution, and the fact that the ejecta are opaque to electron scattering and bound-free absorption.  

We believe that the most likely explanation is emission coming not only from the high velocity shock, but also from H$\alpha$ emitting gas ahead of the shock, with low velocity that is characteristic of the unshocked CSM. The motivation for this comes from the symmetric residual in the modeling shown in Fig.~\ref{Halpha_130311}, which is characteristic in shape of an electron scattering line from a stationary medium. Referring to shock models, this is also a natural consequence of preionization of the unshocked gas by the UV and X-ray emission from the shock, which heats the gas to 10\,000 -- 20\,000 K \citep[e.g.,][]{dessart15}. This is partly the same region that is giving rise to the scattering of the H$\alpha$ emission from the shocked gas, discussed above. Because the emission is $\propto n_{\rm e}^2$, while the electron scattering depth is $\propto n_{\rm e}$, it is  the inner denser regions close to the shock that mainly contribute to the emission, while the scattering is done over a larger radius. The main difference between the two emitting components is the low velocity of the unshocked emitting gas. 
To model this quantitatively, we added an emitting component coinciding with the inner region of the ionized CSM, immediately outside the shock,  with emissivity as discussed in Sect. \ref{sec:linemodel}. 
The main requirement is that the H$\alpha$ emitting region should coincide with the electron scattering region and have a low velocity. These requirements are met by the preionization region in front of the shock. 

In Fig. \ref{Halpha_cds_csm}, we show the result of a model where we have adjusted the ratio of the H$\alpha$ luminosity of the unshocked and shocked gas to give the observed ratio of the narrow and broad components of the line. The velocity of the unshocked gas has been assumed to be similar to that inferred for the narrow P~Cygni absorption, $\sim 100 \kms$. As is apparent from the figure, this model greatly improves the agreement between the observations and model and now includes the full line profile, with the exception of the narrow P~Cygni profile below $\sim 100 \kms$. The latter must come from gas, which is optically thin to electron scattering, and most likely from gas located outside of the high-density CSM.

Besides this general agreement with the observations, the model provides several interesting results. One is that the optical depth to electron scattering required to produce the necessary smoothing of the input emission only  slowly decreases with time. In Sect.~\ref{sec:shockdisc}, we provide possible explanations to this result. 

The \ion{O}{i}~$\lambda$11287 line profile is similar to H$\alpha$, although the blue extension is somewhat smaller (see Fig.~\ref{OIvsHa}). This line may arise as a result of a recombination from \ion{O}{ii}, collisional excitation, or fluorescence. If recombination dominates, one expects the strength of other high excitation lines to be similar, in particular the \ion{O}{i}~$\lambda$7774 line. This line is, however, very weak if it is detected at all (see Fig.~\ref{lineID}), and we therefore exclude this mechanism. Collisional excitation is also unlikely since this would also result in comparable line flux for the \ion{O}{i}~$\lambda$7774 line.

A likely explanation for the excitation is fluorescence by the Lyman-$\beta$ line. This was discussed by \cite{grandi75} for AGNs and has also been invoked for SNe and, in particular, in the case of 
SN~1987A. This process results in the line  \ion{O}{i}~$\lambda$11287 and then  \ion{O}{i}~$\lambda$8446, with similar flux values. The latter line is, however, blended with the \ion{Ca}{ii} triplet as well as a possible strong \ion{Fe}{ii} line (Fig.~\ref{lineID}), but it is consistent with the observations. Therefore these lines are likely to come from the hydrogen-rich ejecta, and not from processed gas. 
We also note that the calcium lines are likely produced in the same regions as H$\alpha$ since the \ion{Ca}{ii} lines can be reproduced using a combination of H$\alpha$ profiles. 

Finally, we comment on the origin of the H$\alpha$ emission during the first month of SN~2013L. As shown in Fig.~\ref{fitHa}, the H$\alpha$ is symmetric until +27~d. The emission from the shock at these early phases is thermalized due to the large optical depth of the CSM. Only at +27~d do the photons from the shock escape without thermalization and the line profile shows asymmetry. Therefore, at very early phases, the H$\alpha$ emission comes from the unshocked gas, where photons with $\tau_e \la 10$ can escape.

\subsection{A CSM-interaction model}
\label{sec:csmmodel}

 After $\sim 20-30$ days, we found from the line profile fitting that $\tau_{\rm e} \la 10$. The diffusion time scale is therefore $t_{\rm diff} \approx \tau_e R/c \approx 7 (v/c) t \approx t/10 $, so $t_{\rm diff} \ll t$. At times later than a month, it should therefore be a good approximation to ignore diffusion for the light curve. This is also supported by the simulations of \citet[][see their Fig. 3]{dessart15}, where the energy input from the shock is compared to the bolometric luminosity.  
 
 If radiative and with 100\% efficiency (see below), the luminosity from the forward shock is given by
\begin{equation}
L = 2 ~\pi ~\rho_{\rm CSM}  r_{\rm sh}^2  V_{\rm sh}^3 = {1\over 2} {\dot M\over u_{\rm wind}}V_{\rm sh}^3 \ ,
\label{eq_lum}
\end{equation}
assuming a steady mass-loss rate for the CSM. In 
Fig. \ref{fig:lum_13L_10jl_massloss}, we plotted this luminosity as a function of time, assuming a shock velocity from Eq. (\ref{eq:Vel}). For comparison purposes, we also plotted the bolometric luminosity of SN 2010jl from \citet{fransson14}. In this fit for SN 2010jl, we used the same velocity dependence as for SN 2013L, since a shock velocity could not be derived from the line profiles in that case. The assumed velocity is, however, compatible with that inferred from X-ray observations (\citealp{ofek14a}). For the velocity, we used $u_{\rm wind}=120 \kms$ for SN 2013L (Sect.~\ref{sec_Balmer}) and $u_{\rm wind}=105 \kms$ for SN~2010jl \citep{fransson14}. As seen in Fig. \ref{fig:lum_13L_10jl_massloss}, this simple model gives a good fit to the observations for both SNe up to $\sim 300$ days, after which the observed light curve becomes steeper. We note that the points that depart most from the model are the luminosities derived from using only NIR spectra, which are also the ones affected most by the uncertain extrapolation into the optical range. 

With these assumptions, we derived the mass-loss rates of $\dot M= 1.7 \times 10^{-2}\mll$ for SN~2013L compared to $\dot M = 9 \times 10^{-2}\mll$ for SN~2010jl. Because the CSM velocity of these two SNe were very similar, the mass-loss ratio also corresponds to a similar ratio for the densities. 

Extrapolating back to the explosion and assuming a shock velocity of $V_{\rm s}(t) = V_0 (t_0/t)^\alpha$ (as in Eq. \ref{eq:Vel}), the total CSM mass swept up by the shock is

\begin{equation}
\label{eq:Mass2}
M_{\rm shell} = {\dot M V_0 t_2 \over (1-\alpha) u_{\rm wind}} \left({t_0 \over t_2}\right)^\alpha   .\\ 
\end{equation}
 Here, $t_2$ is the time the interaction effectively stops, which we consider to be 350 days when the break in the bolometric light curve is seen. With $
V_0 = 4000 \kms$,  $t_0 = 100$ days,  $\alpha =0.23,$ and the above values for $\dot M$ and $u_{\rm w}$, we find $M_{\rm shell} =0.64 \Msun$. The duration of this high mass-loss phase is $\sim M_{\rm shell}/\dot M \approx 42$ years. A similar duration can be obtained from the drop in luminosity at $\sim 350$ days and the shock velocity at this epoch, $t_{\rm eruption} \approx t_2 V(t_2) / u_{\rm w} \approx 350 \times2700 / 105 = 25$ years.
 
The mass-loss rate found here is larger than the rate found by A17, but it is similar to those found for other SN~IIn progenitors. At +33~d, A17 found a mass-loss rate of 0.34--1.75$\times$10$^{-3}$~M$_{\odot}$~yr$^{-1}$, using $\dot{M}=2~u_{\rm wind}~L(H\alpha)~\epsilon({\rm H}\alpha)^{-1}~V_{\rm sh}^{-3}$, where $\epsilon({\rm H}\alpha)$ is the fraction of the bolometric luminosity emerging in the H$\alpha$ line. Their assumed value for this, $\epsilon({\rm H}\alpha) = 0.1-0.5,$ is very high compared to what is found here, $\epsilon({\rm H}\alpha) = 6\times 10^{-3}$. Also, their velocity, $V_{\rm sh}=$2696~km~s$^{-1}$ (from the electron scattering wings of H$\alpha$) is low by a factor of $\sim 2$ compared to what we find. Together, this explains the discrepancy.
 
  A problem with the above mass-loss rate is that the optical depth to electron scattering of the CSM is lower than estimated from the line profiles in Sect. \ref{sec:linemodel}. Even at +48 days, the optical depth corresponds to only $\sim 1.7$. This is even more accentuated by the fact that the lines are still optically thick at +330 days and even the day +728 observation shows a clear electron scattering wing (Fig. \ref{fitHa3}). This is close to or after the drop in luminosity, which may be interpreted as the break out of the CSM by the shock and consequently a drop in the CSM column density.
 
A  caveat to the above mass-loss rate and total mass estimates is that multidimensional effects likely  affect the radiative efficiency for radiative shocks through thin-shell instabilities, creating a corrugated
shock interface \citep{kee14,steinberg18}. This is found to decrease the ratio between the X-ray luminosity and the total luminosity from the shocks, but it is not clear if this changes the total radiative efficiency. A larger fraction of UV and soft X-rays actually increases the absorbed fraction of the radiation because of the larger photoabsorption cross sections.  If the radiative efficiency is lowered by a factor of $\eta,$ this increases the mass-loss rate and total mass by a factor of $\eta^{-1}$. An anisotropic mass-loss rate has a similar effect. For the simple case of a conical outflow with solid angle, $\Omega$,  the mass-loss rate inferred from the luminosity increases by a factor of $4 \pi / \Omega$ for a fixed luminosity. Both from observations of LBVs, such as Eta Car \citep{smith06}, and from observations of SN~2010jl \citep{fransson14, patat11}, for example, there is  evidence for asymmetries in the CSM.

A further caveat is  that we may have underestimated the bolometric luminosity due to bolometric corrections in the UV and IR, or due to reddening. We have already discussed this in Sect. \ref{sec:bolo} with regard to the UV, which may be severely underestimated. In addition, some of the X-rays from the shock may escape without being thermalized, depending on the shock velocity. 
 
The cross section for photoelectric absorption can be approximated by $\sigma_{\rm abs}  \approx \sigma_{\rm T} (E/10 \ \rm keV)^{-3}$. The effective optical depth for thermalization is $\tau_{\rm therm} \approx \sqrt{\tau_{\rm abs}  \tau_{\rm e} } \approx  \tau_{\rm e}  (E/10 \ \rm keV)^{-3/2} $.  For $ \tau_{\rm e}=5,$ one finds $\tau_{\rm therm} \approx 1$ at $\sim 30$ keV, which is similar to the shock temperature for a shock velocity of 5000 $\kms$  (Eq. \ref{eq_shocktemp}). The thermalization of the X-rays may, therefore, only be fulfilled for the low energy part of these. As discussed earlier,  the shock properties and the radiation are, however,  not yet fully understood.

We  include these effects in a parameter $\eta$, similar to that invoked by \cite{ofek14a}. A more general form of Eq. (\ref{eq_lum}), for an arbitrary  density slope, is then
\begin{equation}
L(t) = {1\over 2} \eta {\dot M\over  u_{\rm wind}}  V_{\rm 0}^{3}  \left(  {t\over t_{\rm 0}}\right)^{[n(2-s )+ 6s -15]/(n-s))}      \ .
\label{eq_lum_v2}
\end{equation} 

We can estimate the velocity, optical depth, and luminosity from the observations. 
At $t_0 = 48$ days, we estimated $V_0 = 4800 \kms$, $\tau_{\rm e}(t_0) = 8$ and $L(t_0) = 6 \times 10^{42} \ergs$. 
 
We assumed that $n=5$ and varied the power law index of the density law $s$, which for this $n$ is only important for the evolution of $\tau_{\rm e}$. The time of the breakout is not well determined. For a spherical geometry, one may consider this to be the time of the drop in luminosity, 
 $\sim 350$ days. Deviations from this are discussed below. A calculation with these parameters is shown in the left panel of Fig. \ref{fig:lum_tau_vel} for $s=0, 1$ and $2$.
 The most interesting result is that to reconcile both the luminosity, optical depth, and velocity, a mass-loss rate corresponding to $\sim   0.15 \mll$ is necessary while the efficiency parameter is low, $\eta \sim 0.14$ for $s=2$. 
 
 As shown in the second panel, the sharp drop in density at 350 days, however, results in a very low value of  $\tau_{\rm e}$, which is in contradiction to the optical depth at 330 days (Sect. \ref{sec:linemodel}). What we believe to be the most likely way of reconciling both a value of  $\tau_{\rm e} \ga 1$ and a drop in luminosity would be if the CSM is asymmetric, as discussed above. 
 
 An asymmetric CSM would result in different times for the shock break-out and a steepening of the light curve, as shown in  \cite{vanmarle10} and \cite{vlasis16}. The optical depth of the CSM in the directions of large column density may, however, still be large corresponding to a larger value of $R_{\rm out}$, that is, $t_{\rm b}$ and $\tau_{\rm e}$ in this direction (Eqns.~\ref{eq_vel} -\ref{eq_lum_v2}). In reality, an asymmetric CSM would also affect the light curve and velocity. To model this, a multidimensional hydrodynamic calculation would be necessary, as discussed in \cite{vanmarle10} and \cite{vlasis16}. To illustrate the effect on $\tau_{\rm e,}$ we show one calculation using $t_{\rm b} = 700$ days. The longer  $t_{\rm b}$ results in a less steep decrease  of $\tau_{\rm e}$, as expected, and a lower mass-loss rate (because $R_{\rm out}$ increases). For $s=0,$  the value of $\tau_{\rm e}$ is nearly constant. To get the same luminosity, the efficiency has to increase (see Fig.~\ref{fig:lum_tau_vel}).
 
 As seen from Eqns. (\ref{eq_vel} -\ref{eq_lum_v2}), the value of $R_{\rm out}$, that is, $t_{\rm b}$, only affects $\tau_{\rm e}$.  In reality, an asymmetric CSM would also affect the light curve and velocity. To model this, a multidimensional hydrodynamic calculation would be necessary, as discussed in \cite{vanmarle10} and \cite{vlasis16}. To illustrate the effect on $\tau_{\rm e}$, we show one calculation using $t_{\rm b} = 700$ days. The longer  $t_{\rm b}$ results in a less steep decrease  of $\tau_{\rm e}$, as expected, and a lower mass-loss rate (because $R_{\rm out}$ increases). For $s=0,$  the value of $\tau_{\rm e}$ is nearly constant. To get the same luminosity, the efficiency has to increase (see Fig.~\ref{fig:lum_tau_vel}). 

\section{Discussion}
\label{sec:discussion}

\subsection{The progenitor system}
\label{sec:progenitor}

A first indication of the nature of the progenitor system of SN~2013L can be obtained from the observed narrow H$\alpha$ component, which provides a direct measure of the unshocked CSM wind velocity. The origin of this CSM is likely to be the stellar winds of the progenitor that enriched the environment over its evolutionary lifetime. 
A wind velocity around 120--240~km~s$^{-1}$ is consistent with a luminous blue variable wind \citep[see e.g.,][]{kiewe12,taddia13IIn,smith14}, but also with a wind from a yellow hypergiant (YHG; \citealp{smith14,smith15iqb}), as is also noted by A17. 
Typical red supergiant winds have lower velocities ($\sim$10 km~s$^{-1}$; \citealp{smith09_RSG}), whereas typical Wolf-Rayet wind velocities are higher ($\gtrsim$1000 km~s$^{-1}$; \citealp{pastorello07}). 

The mass-loss rate is at least 
about $1.7 \times$ 10$^{-2}$~M$_{\odot}$~yr$^{-1}$ and from our discussion in the previous section, it is more likely to be $\sim 0.15 \ml$. For these respective mass-loss rates, this is consistent with an LBV progenitor that is capable of expelling a large amount of mass, from $0.4 - 0.7\Msun$ up to 3.8 - 6.3 $\Msun$ in 25 -- 42 yrs in this case. The most well-studied case of this is $\eta$ Carinae, which expelled about 10 M$_{\odot}$ in 30 yrs during the great eruption that occurred in the 19$^{th}$ century \citep{smith13eta}. \cite{woosley17} discusses this case in the context of pulsational pair instability SNe. We note that in the case of SN~2013L, there is  no direct indication of advanced nuclear burning, which is as expected for this type of event.
 A YHG progenitor can be considered if the mass-loss rate is closer to 10$^{-3}$~M$_{\odot}$~yr$^{-1}$ rather than 10$^{-2}$~M$_{\odot}$~yr$^{-1}$.

Our spectral model does not require strong asymmetry of the CSM to reproduce the H$\alpha$ profile, but rather a combination of occultation and electron scattering. However, we cannot exclude the possibility of asymmetry in the CSM, as in the case of $\eta$ Carinae and its complex CSM characterized by polar lobes \citep[see, e.g.,][]{smith13eta}. In fact, as we discuss in Sect. \ref{sec:csmmodel}, there are indications
in favor of this from the light curve and line profiles.

\begin{figure}
$\begin{array}{cc}
\includegraphics[width=8cm]{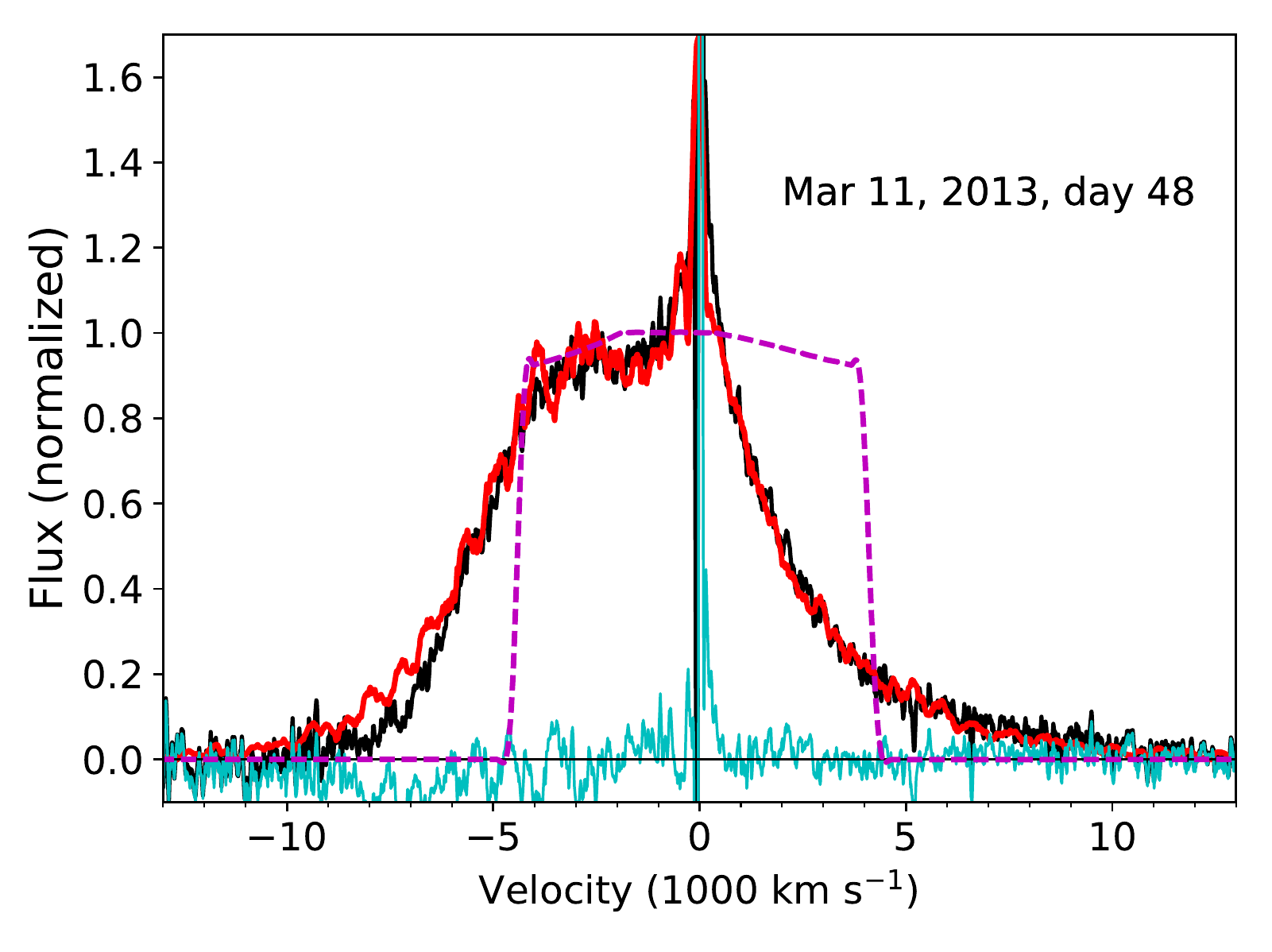}&
\includegraphics[width=8cm]{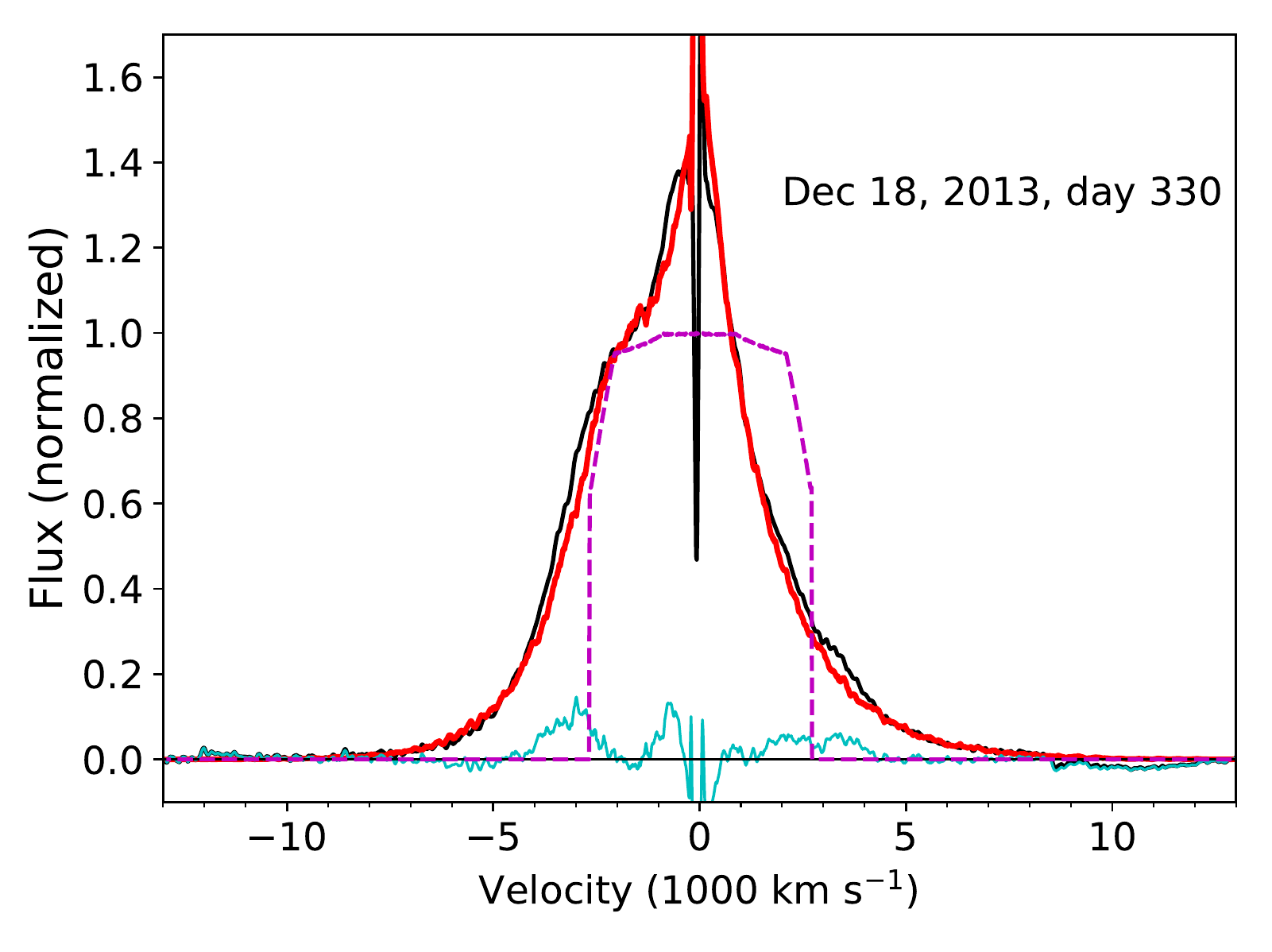}
\end{array}$
\caption{\label{Halpha_cds_csm} Same as Fig. \ref{Halpha_130311}, but now also including H$\alpha$ emission from the unshocked medium preionized by the shock, which can be seen as the narrow magenta line on top of the broad from the shocked gas (normalized to one at maximum).  The magenta, dashed profile shows the line profile \textit{without} electron scattering or continuum thermalization.  We note that the residual close to the line center is now much lower at both epochs. The steep profile on the red side is a result of the comparatively narrow electron scattering profile from the CSM and the damping of the red wing from the scattering from the CDS and the ejecta. }
\end{figure}

\begin{figure}
\centering
\includegraphics[width=12cm]{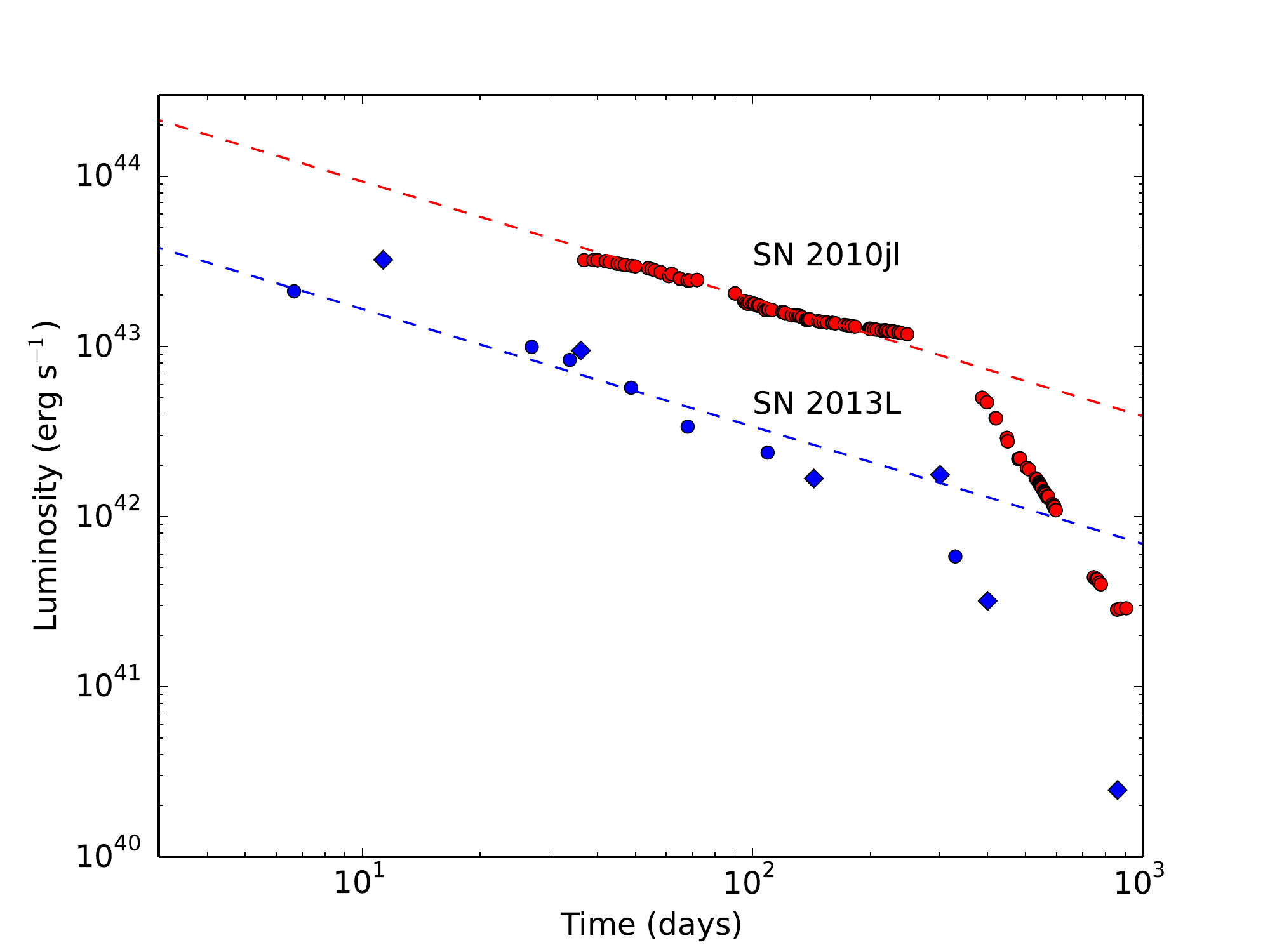}
\caption{ Bolometric luminosity as a function of time for SN 2013L together with SN 2010jl. The dashed lines represent the  luminosity predicted for a model with the velocity dependence from Eq. (\ref{eq:Vel}) and mass-loss rates $\dot M= 1.7 \times 10^{-2}\mll$ and $\dot M= 9 \times 10^{-2}\mll$, respectively. For SN 2013L the blue dots refer to luminosities derived from optical spectral fits, while the diamonds refer to NIR fits.}
\label{fig:lum_13L_10jl_massloss}
\end{figure}

The large CSM mass indicated by our analysis is also consistent with the strong deceleration of the H$\alpha$ emitting CDS, with $V_{\rm shell} \propto t^{-0.23}$ (Fig. \ref{Halpha_cds_csm}, Eq. \ref{eq:Vel}), showing that a large fraction of the kinetic energy of the ejected material is radiated. Such, or even more extreme scenarios have been invoked by \cite{dessart16} and \cite{chugai16} for SN 1994W and SN 2011ht. In these simulations, the ejecta mass is much lower than the CSM mass, resulting in, in some cases, complete deceleration of the fast ejecta and an initially luminous SN, dropping fast at $\sim 100$ days. In the case of SN 2013L, the velocity is still $\ga 1700 \kms$ at $\sim 1500$ days (Fig. \ref{Halpha_cds_csm}) and the drop in luminosity is slower than in SN 1994W. However, the need for a very massive CSM is probably required, although the ejecta mass may be higher than for SN 1994W.

\subsection{Dust in SN~2013L}
\label{sec:dustdisc}

In SN~2013L, there are clear signs, at least at late epochs, of luminous NIR and MIR emission, which is commonly associated with dust emission. The dust could already be present in the unshocked CSM and from there it could easily drive a light echo where the SN optical and UV emission is redistributed to longer wavelengths, but it could also be associated with newly formed  dust in the SN ejecta \citep[e.g.,][]{stritzinger12,gall14, chugai18}. In the case of SN~2010jl, an origin in the inner ejecta has been argued against \citep{fransson14,sarangi18}. Here we attempt to determine where the dust is located in SN~2013L and characterize the nature of the NIR emission. 

The NIR and MIR emission was modeled by fitting a single BB function to the NIR and MIR part of the SED that corresponds to the warm component discussed in Sect.~\ref{sec:bolo}. We find that between +132~d and +861~d,  the luminosity ($\sim$10$^{42}$~erg~s$^{-1}$), the temperature (1300~K), and the radius (10$^{16}$~cm) of this warm component are rather constant. 

\begin{figure}
\includegraphics[width=18cm]{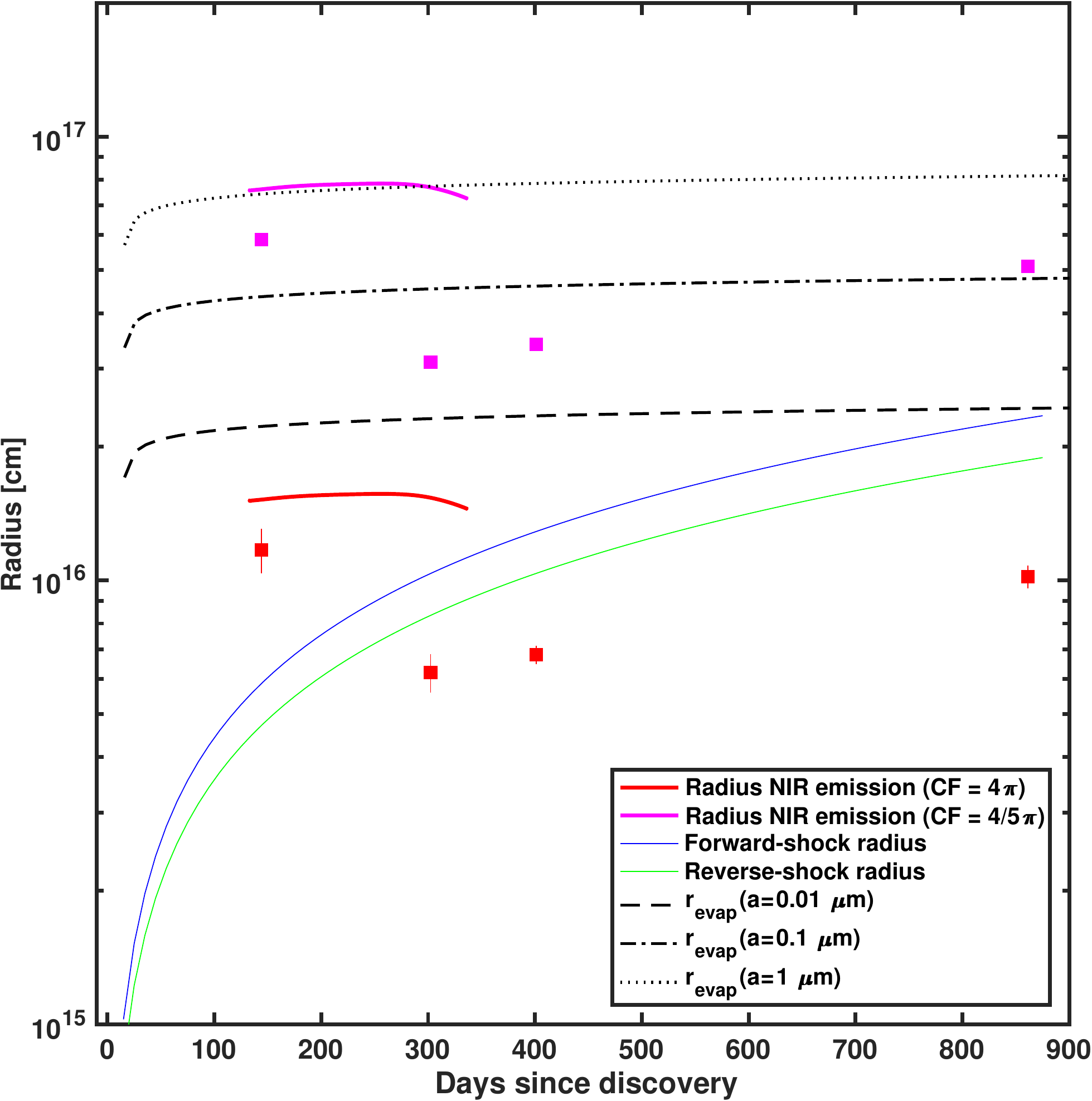}
\caption{\label{radius_dust}Radius of the warm BB component (NIR emission) assuming it originated from dust distributed spherically with full covering factor (red) and with a smaller covering factor (4$\pi$/5, magenta). The lines are from the SEDs fit and the symbols are from the spectral fit with BBs. The radius of the forward shock as obtained from the spectral modeling of H$\alpha$ (see Fig.~\ref{Halpha_vel}) is shown in blue. 
The velocity of the reverse shock from which we derived its radius (shown in green) is taken to be 1000 km~s$^{-1}$ slower than that of the forward shock. The evaporation radius for the dust, which depends on the dust grain size (see legend), is shown in black. If the NIR emission is due to pre-existing dust, this must be distributed with a covering factor of $\lesssim$ 4$\pi$/5 otherwise the ejecta would have already reached and destroyed it around 300 days. Furthermore, the grain size of the dust must be $<$0.1 $\mu$m otherwise the large luminosity of the SN would have vaporized it (this is assuming a covering factor of 4$\pi$/5). }
\end{figure}

In Fig.~\ref{radius_dust}, we plotted the radius of the warm BB component already shown in Fig.~\ref{LTR}.
If produced by spherically-distributed dust with a covering factor ($\Omega_{\rm CF}$) of 4$\pi$,
then the BB radius is approximately the dust radius. If the dust is distributed with a smaller covering factor (see the cases of SNe~2005ip and 2006jd, \citealp{stritzinger12}), the dust radius is larger than the BB radius. 
If the dust is located at a certain radius from the SN, it is necessary check if the forward shock could reach and destroy the dust.
We adopted the velocity of the shell from Fig.~\ref{Halpha_vel} as a measure of the velocity of the forward shock, which evolves as a power law of index $-$0.23.
The radius reached by the forward shock (blue lines) intercepts that of the dust already at $\sim$ +300~d if we assume $\Omega_{\rm CF} =4\pi$ (red lines and squares). 
For a smaller covering factor, $\Omega_{\rm CF} =4\pi/5$, the dust is not reached by the forward shock, and it can survive and lead to echo emission. We also checked whether or not the high SN luminosity could vaporize the dust using the technique adopted in \citet{stritzinger12}.  For a covering factor $\Omega_{\rm CF} =4\pi/5$, the dust should consist of grains larger than $\sim$0.01$\mu$m in order to survive. For a $\Omega_{\rm CF}=4\pi,$ the dust would be  vaporized by the high SN luminosity, regardless of its grain size.  
We find that it is entirely plausible that the dust in SN~2013L is pre-existing and it covers only a fraction of the $4\pi$ surface. 

Given the uncertainty on the dust radius, there might also be the possibility that the dust is formed, with $\Omega_{\rm CF}=~4\pi$, within the cold dense shell between the forward shock and the reverse shock, which we have assumed to move 1000 km~s$^{-1}$ slower than the forward shock, see green lines in Fig.~\ref{radius_dust}. However, we do not see an increasing depression on the red side of the emission lines, such as H$\alpha$, which would be a sign of newly formed dust. 
In \citet{stritzinger12}, SN~2005ip and SN~2006jd show this progressive depression on the red side of H$\alpha$ at relatively early epochs. There is also the possibility that the dust did not form close to the line forming regions, and therefore we did not observe the depression. 
Furthermore, the high dust temperature at late epochs might point to dust formation in the ejecta at these late phases \citep[see][for the case of SN~2010jl]{gall14, chugai18}. 

\subsection{The shock structure}
\label{sec:shockdisc}

In Fig. \ref{fig:lum_13L_10jl_massloss}, we show the bolometric light curves of SN~2010jl and SN~2013L. Although the luminosity of SN 2010jl was a factor of $\sim 5$ higher, the general shape is similar with a break  $\sim 300 - 400$ days and a similar slope of the light curve, both before and after the break. While the paper by \cite{dessart15} aimed to model SN~2010jl to a large extent, many of the results are therefore also relevant to SN 2013L.

\begin{figure}
\centering
\includegraphics[width=12cm]{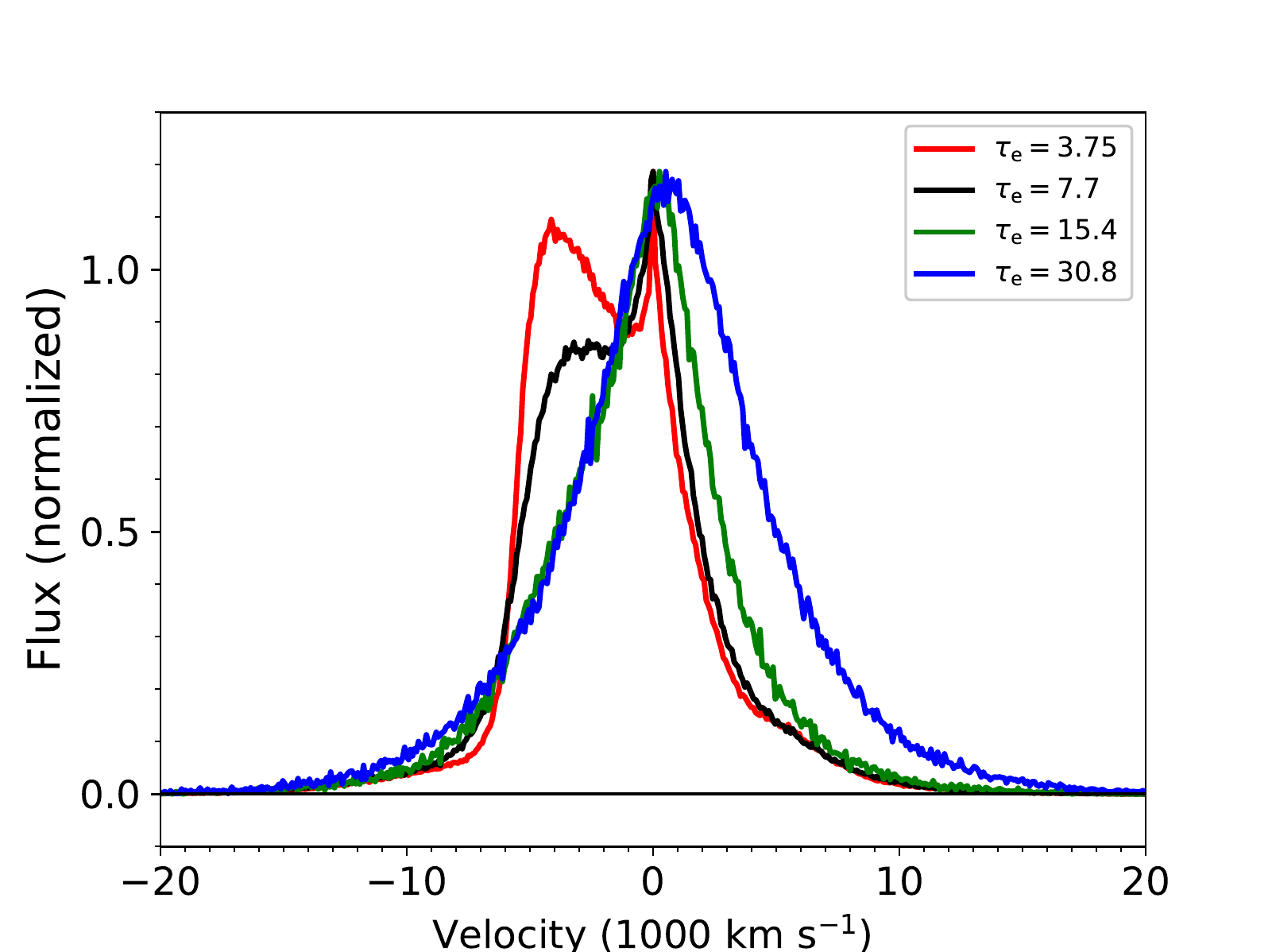}
\caption{Line profiles for different optical depth of the CSM. The black with $\tau_{\rm e}=7.9$ corresponds to the best fit model at +48 days. The velocity of the emitting shell behind the shock is in all cases $V_{\rm em ~in}=5390 \kms$ and $V_{\rm em ~out}=6160 \kms$. One notes the gradual transition from highly skewed profiles to symmetric as the optical depth increases. In all cases, an electron temperature of 15\,000 K is assumed.}
\label{fig:line_profiles}
\end{figure}

\begin{figure}
\centering
\includegraphics[width=6.5cm,trim={0 0 2cm 0}, clip]{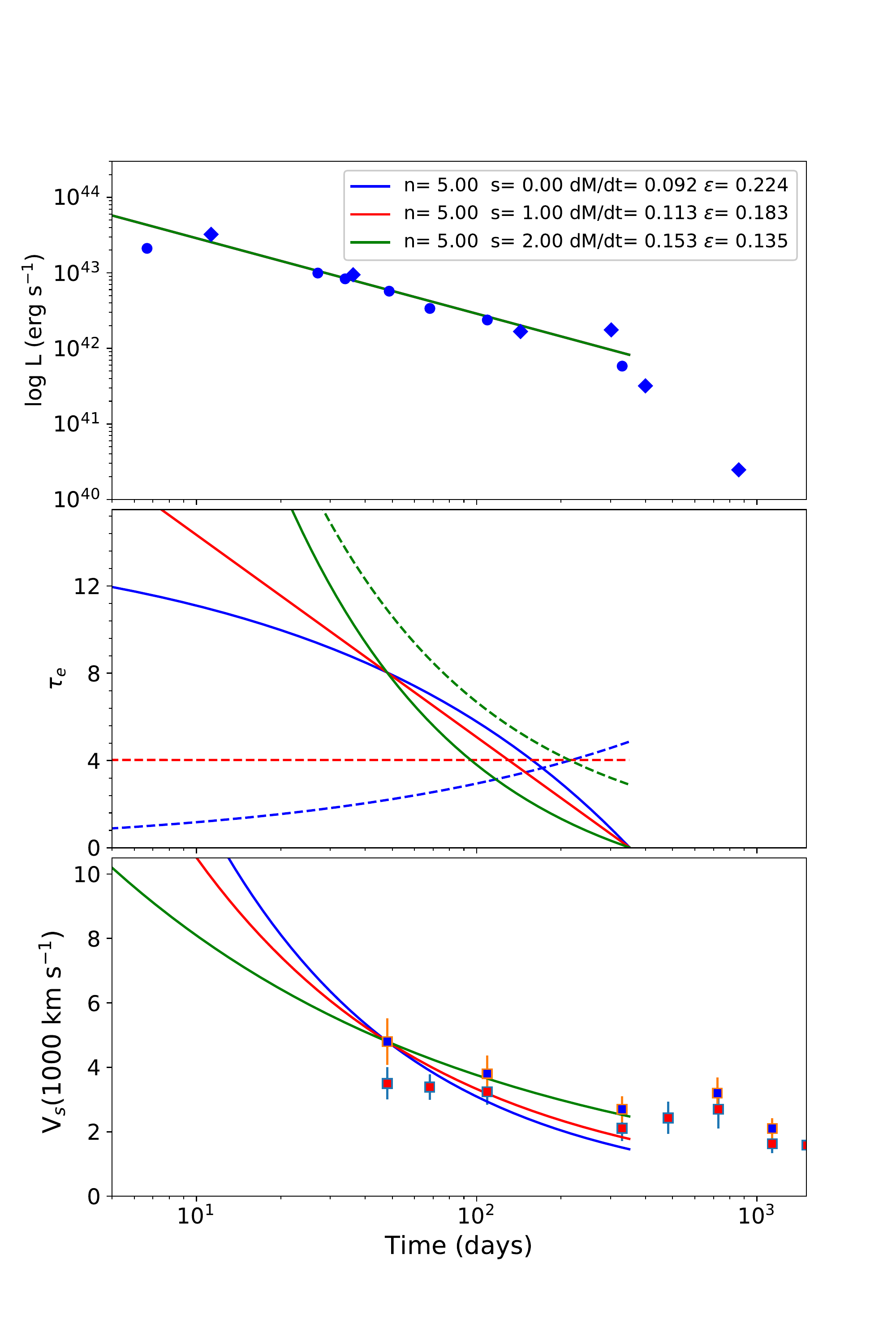}
\includegraphics[width=6.5cm,trim={0 0 2cm 0}, clip]{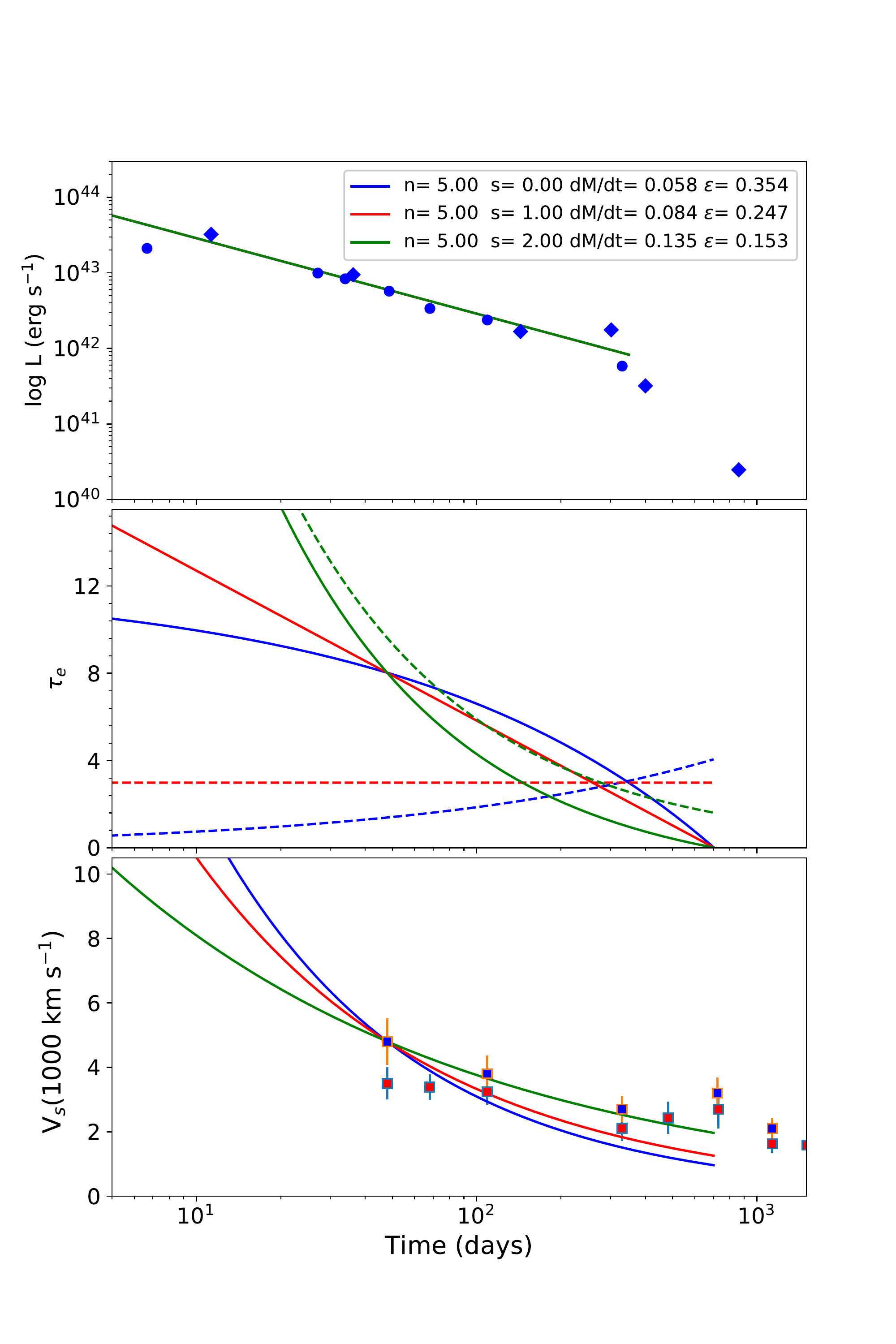}
\caption{Models for the luminosity, electron scattering optical depth of the CSM (solid lines) and CDS (dashed lines), and velocity based on the similarity solution for different values of the density power law index $s$. The left panels assume a break out time $t_{\rm b} = 350$ days while the right panels have $t_{\rm b} = 700$ days. The mass-loss rate and efficiency parameters are given in the legends for the different $s$. }
\label{fig:lum_tau_vel}
\end{figure}

From the modeling of the line profiles, we find several interesting results, which give information about the structure of the shock and dense CSM. 
A unique feature of SN~2013L is that we have direct information about the shock velocity from the skewed line profile. One may then ask what the difference is between SN 2013L and other Type IIn SNe, which show more symmetric line profiles. As Fig. \ref{fig:line_profiles} shows, this is mainly a result of a lower optical depth to electron scattering of the dense unshocked CSM from the shock. In this figure, we show the line profile for several optical depths, including the one used for the fit at day 48 in Fig. \ref{Halpha_cds_csm} $\tau_{\rm e}=12.3$. As in these calculations, we used a maximum electron temperature $T_{\rm e}=1.5 \times 10^4$ K.  One can obtain the same line profiles for other temperature profiles, as seen from scaling $\tau_{\rm e} \propto T_{\rm e}^{-1/2}$ for a constant temperature medium.

As we see from Fig. \ref{fig:line_profiles}, once $\tau_{\rm e}\gtrsim 10,$ the shoulder in the line profile disappears and the line becomes increasingly symmetric as the optical depth increases further. 
The most likely explanation for the usually symmetric lines is, therefore, that for most other bright Type IIn SNe, the shock has been at such a large optical depth to electron scattering that the velocity information has been lost. 

Another special feature of SN~2013L is that we can identify two different components, one central, symmetric and one broad, skewed profile. We interpret these to come from the preshock and postshock regions, respectively. As is seen in Fig.~\ref{fig:line_profiles}, for low optical depths (e.g., the $\tau_{\rm e}=3.75$ profile), these components may give rise to a double peak structure, one narrow component from the preshock gas, and a broader component from the postshock. At the other extreme, we note the almost perfectly symmetric line for $\tau_{\rm e}=30.8$. This is in spite of the high shock velocity and also the occultation by the photosphere, which is assumed to coincide with the inner boundary of the emitting region, here at $V_{\rm ph} = V_{\rm em~ in}= 5390 \kms$. At lower optical depths, the occultation explains the depression on the red side of the line. 

A somewhat surprising result that was found is that the optical depth to electron scattering required to produce the necessary smoothing of the input emission is only weakly dependent on time,  $\tau_{\rm e}=12.3$ at +48~d, compared to $\tau_{\rm e}=7.2$ at +330~d. The difference is hardly significant and depends on the assumed temperature of the gas, for example. Naively, one could expect this to decrease by a large factor as the column density of the CSM in front of the shock decreases. As discussed in \citet[][see their Eq.~17]{fransson14}, if ionization and not density bounded, the optical depth of the region ionized by the shock radiation is for a an $\rho \propto r^{-2}$ CSM mainly sensitive to the shock velocity and not to the density of the CSM. The lower density results in a lower shock luminosity for a radiative shock ($\propto n_{\rm CSM} V_{\rm s}^3 $), but also in a lower total recombination rate ($\propto n_{\rm CSM}^2 \Delta r$). The optical depth, $\tau_{\rm e} \propto n_{\rm CSM} \Delta r$, of the  ionized region is therefore sensitive to the shock velocity, $\tau_{\rm e} \propto V_{\rm s}^3$, but independent of density. Because $V_{\rm s}$ is only slowly decreasing with time, the electron scattering depth is expected to decrease slowly. 

 For SN~2010jl, \citet{dessart15}  find  $\tau_{\rm e} \approx 5-10$ between 30 - 100 days, decreasing to unity at $\sim 350$ days. The fact that this model shows a decrease in the optical depth probably means that the ionized region is density bounded.

The $\tau_{\rm e} \propto V_s^3$ scaling may also explain why Type IIn SNe in many cases have an electron scattering depth so large that we do not see any velocity shift.  As discussed in Sect. \ref{sec:csmmodel}, an anisotropic, dense CSM may resolve the problem with the drop in the light curve, but $\tau_{\rm e}$ that is still high.

\section{Conclusion}
\label{sec:conclusion}

\begin{figure}
\centering
\includegraphics[width=18cm]{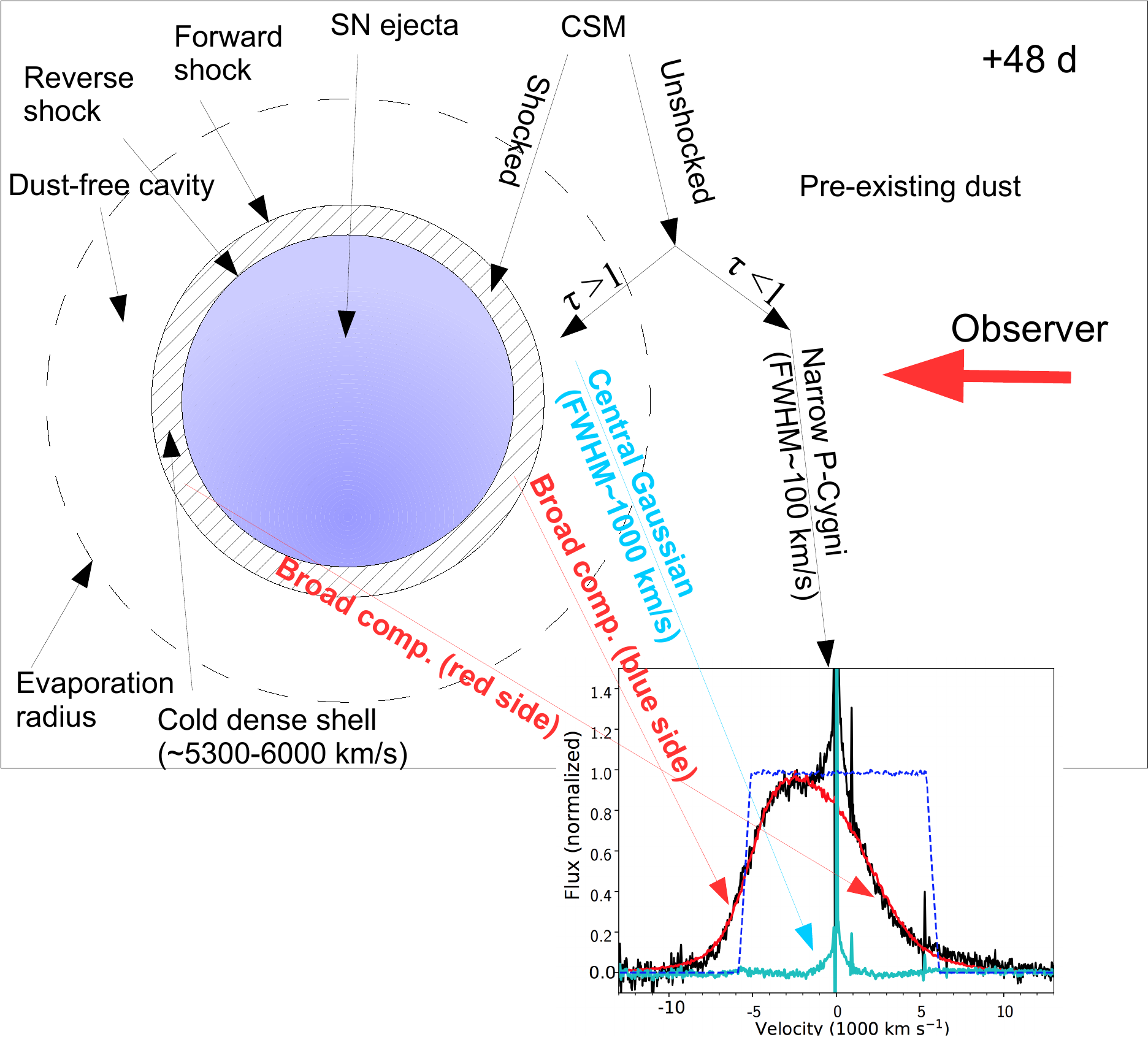}
\caption{\label{CSMscheme_4}Schematic view of SN~2013L interacting with its CSM, as inferred from our analysis.}
\end{figure}

To conclude our analysis, we schematize the most likely CSM interaction scenario for SN~2013L in Fig.~\ref{CSMscheme_4}. Here the ejecta of a LBV (or YHG) star interacts with its CSM and thereby produces both a forward and a reverse shock. The spectral model provides a good fit of the H$\alpha$ profile without invoking asymmetry of the CSM, which we represent as spherically symmetric, which is likewise for the ejecta. The high luminosity of the SN destroys all the dust within the evaporation radius. Outside the evaporation radius, pre-existing dust reprocesses the UV and optical light of the SN into the NIR and MIR excess that we observe. The dust is distributed with a small covering factor, which might hint toward some asymmetry in the outer CSM. The H$\alpha$ broad component (red line in the H$\alpha$ inset) is produced in the cold-dense shell between forward and reverse shock (dashed area) and the red side of this emission is occulted by the ejecta. The broad, boxy emission of the shell (blue line in the H$\alpha$ inset) is broadened by Thomson scattering due to CSM located outside the shock. The central component (cyan in the H$\alpha$ inset) is produced by narrow emission in the CSM directly outside the forward shock which is
again broadened by Thomson scattering in the CSM. 
The excited, unshocked, outer CSM produces the narrow H$\alpha$ component, which reveals a slow wind (120--240 km~s$^{-1}$) that is compatible with a LBV (or a YHG) star. Our mass-loss-rate estimates favor a LBV progenitor star. We find strong arguments from the light curve and electron scattering wings for an anisotropic CSM.

\begin{acknowledgements}
We thank the referee for their  comments which helped to improve the paper, especially concerting the modeling. 
We thank Jennifer Andrews for having shared with us the late-time spectra of SN~2013L.
F.T. and M.D.S. acknowledge support for a project grant provided by the Independent Research Fund Denmark (IRFD). F.T. and M.D.S. also acknowledge support from the VILLUM FONDEN under experiment grant 
28021.
F.T. and J.S. gratefully acknowledge the support from the Knut and Alice Wallenberg Foundation. C.F. acknowledges support for the Swedish Research Council. 
M.D.S. acknowledges support from  a research grant (13261) from VILLUM FONDEN.
The Oskar Klein Centre is funded by the Swedish Research Council.
The CSP-II has been supported by the National Science Foundation under grants AST1008343, AST1613426, AST1613455, and AST1613472, and also in part by a grant from the Danish Agency for Science and Technology and Innovation through a Sapere Aude Level 2 grant (PI M.D.S.). 
The Swift Optical/Ultraviolet Supernova Archive (SOUSA) is supported by NASA's Astrophysics Data Analysis Program through grant NNX13AF35G. 
This research has made use of the NASA/IPAC Extragalactic Database (NED) which is operated by the Jet Propulsion Laboratory, California Institute of Technology, under contract with the National Aeronautics and Space Administration.

\end{acknowledgements}

\bibliographystyle{aa} 

\def\colhead#1{#1}

\clearpage
\onecolumn

\begin{table*}
\centering
\caption{Optical photometry of the local sequences for SN~2013L in the standard system.\tablenotemark{a}\label{SN13L_opt_locseq}}
\begin{tabular}{ccccccccc}
\hline\hline
\colhead{ID} &  
\colhead{$\alpha$ (2000)} &  
 \colhead{$\delta$ (2000)} &  
   \colhead{$B$} &  
    \colhead{$V$} &  
    \colhead{$u^\prime$} &  
    \colhead{$g^\prime$} &  
   \colhead{$r^\prime$} &  
\colhead{$i^\prime$}\\
\hline
  1 & 176.372940 & $-$50.566639 & 14.085(009)& 13.564(012)& 14.909(017)& 13.779(008)& 13.432(007)& 13.320(010)\\
  2 & 176.420593 & $-$50.632069 & 15.631(022)& 14.548(010)& 17.259(081)& 15.046(015)& 14.197(009)& 13.857(009)\\
  3 & 176.448257 & $-$50.577740 & 16.060(022)& 15.407(018)& 16.993(032)& 15.694(016)& 15.224(013)& 15.057(013)\\
  4 & 176.415695 & $-$50.644630 & 16.866(036)& 16.172(023)& 17.808(068)& 16.480(029)& 15.967(018)& 15.766(021)\\
  5 & 176.474594 & $-$50.619598 & 17.240(055)& 16.431(030)& 18.459(097)& 16.781(026)& 16.189(019)& 15.972(015)\\
  6 & 176.288162 & $-$50.547031 & 17.699(083)& 16.877(043)& 19.023(131)& 17.256(045)& 16.607(035)& 16.372(035)\\
  7 & 176.435425 & $-$50.628880 & 19.067(071)& 17.724(059)& 18.882(078)& 18.358(066)& 17.143(033)& 16.646(038)\\
  8 & 176.278427 & $-$50.581089 & 18.180(025)& 17.479(064)& 19.078(107)& 17.794(049)& 17.252(033)& 17.028(048)\\
  9 & 176.351624 & $-$50.602581 & 18.319(148)& 17.706(079)& 19.175(095)& 18.030(056)& 17.533(054)& 17.314(043)\\
 10 & 176.417496 & $-$50.562401 & 18.988(050)& 18.080(098)& \ldots     & 18.515(055)& 17.758(038)& 17.467(071)\\
 11 & 176.274002 & $-$50.595940 & 17.639(103)& 17.391(042)& 18.770(060)& 17.448(037)& 17.415(065)& 17.481(073)\\
 12 & 176.376709 & $-$50.653801 & 19.092(042)& 18.114(105)& 20.204(115)& 18.582(096)& 17.832(068)& 17.550(083)\\
 13 & 176.394547 & $-$50.654530 & 19.970(038)& 18.624(094)& \ldots     & 19.203(106)& 18.076(085)& 17.601(058)\\
 14 & 176.468704 & $-$50.540169 & 19.102(075)& 18.173(092)& 20.451(194)& 18.608(100)& 17.842(060)& 17.585(066)\\
 15 & 176.309906 & $-$50.574009 & \ldots     & 19.697(077)& \ldots     & 20.394(049)& 19.003(092)& 17.559(076)\\
 17 & 176.342255 & $-$50.543720 & 16.666(035)& 15.964(019)& 17.668(062)& 16.267(027)& 15.756(016)& 15.563(015)\\
 18 & 176.293945 & $-5$0.667351 & 15.999(022)& 15.157(014)& 17.388(019)& 15.532(019)& 14.892(013)& 14.690(012)\\
 19 & 176.322266 & $-$50.618359 & 17.310(040)& 16.521(036)& 18.468(103)& 16.868(026)& 16.272(019)& 16.062(023)\\
 20 & 176.484634 & $-$50.526569 & \ldots     & \ldots     & \ldots     & 20.639(148)& 19.856(096)& 18.523(074)\\
 21 & 176.406937 & $-$50.592899 & 18.120(089)& 17.457(076)& 19.107(135)& 17.749(043)& 17.238(044)& 17.061(041)
\end{tabular}
\tablenotetext{a}{Values in parenthesis are 1-$\sigma$ uncertainties and correspond to a root mean square (rms) of the instrumental errors of the photometry obtained over a minimum of three photometric nights.}
\end{table*}

\clearpage
\onecolumn
\begin{deluxetable}{cccccc}
\tablecolumns{6}
\tablewidth{0pt}
\tablecaption{NIR photometry of the local sequences for SN~2013L in the standard system.\tablenotemark{a}\label{SN13L_nir_locseq}}
\tablehead{
\colhead{ID} &
\colhead{$\alpha (2000)$} &
\colhead{$\delta (2000)$} &
\colhead{$Y$} &
\colhead{$J$} &
\colhead{$H$} }
\startdata
101 & 11:45:38.03 & $-$50:34:20.8 & 11.583(016) & 11.414(017) & 11.078(018)  \\ 
102 & 11:45:27.47 & $-$50:34:29.0 & 13.217(011) & 12.829(003) & 12.332(024)  \\ 
103 & 11:45:36.91 & $-$50:36:11.3 & 14.119(011) & 13.753(009) & 13.317(012)  \\ 
104 & 11:45:37.61 & $-$50:34:49.5 & 14.205(018) & 13.970(005) & 13.701(032)  \\ 
105 & 11:45:19.92 & $-$50:36:31.8 & 14.224(010) & 13.835(003) & 13.318(026)  \\ 
106 & 11:45:30.63 & $-$50:35:38.9 & 14.386(020) & 14.105(006) & 13.760(038)  \\ 
107 & 11:45:26.68 & $-$50:34:55.4 & 14.549(008) & 14.225(009) & 13.860(030)  \\ 
108 & 11:45:34.50 & $-$50:36:50.5 & 14.645(013) & 14.371(004) & 14.078(026)  \\ 
109 & 11:45:31.15 & $-$50:35:24.9 & 14.784(011) & 14.455(011) & 14.050(044)  \\ 
110 & 11:45:34.94 & $-$50:35:52.6 & 14.972(015) & 14.716(011) & 14.459(051)  \\ 
111 & 11:45:24.04 & $-$50:37:18.4 & 15.036(024) & 14.582(030) & 13.899(019)  \\ 
112 & 11:45:35.32 & $-$50:36:02.4 & 15.486(040) & 15.168(022) & 14.834(075)  \\ 
113 & 11:45:29.55 & $-$50:35:13.7 & 15.457(021) & 15.311(031) & 15.135(063)  \\ 
114 & 11:45:35.20 & $-$50:34:46.9 & 15.619(036) & 15.326(016) & 14.967(042)  \\ 
115 & 11:45:36.77 & $-$50:34:58.7 & 15.628(032) & 15.251(029) & 14.793(043)  \\ 
116 & 11:45:27.75 & $-$50:37:16.2 & 15.684(026) & 15.379(034) & 14.874(054)  \\ 
117 & 11:45:28.67 & $-$50:36:16.8 & 15.701(026) & 15.373(020) & 15.017(024)  \\ 
118 & 11:45:24.91 & $-$50:34:30.9 & 15.770(032) & 15.459(003) & 15.074(034)  \\ 
119 & 11:45:39.29 & $-$50:36:46.4 & 15.843(011) & 15.405(035) & 14.918(053)  \\ 
120 & 11:45:21.80 & $-$50:35:09.1 & 15.867(034) & 15.505(004) & 15.142(036)  \\ 
121 & 11:45:30.05 & $-$50:36:28.7 & 15.827(016) & 15.313(019) & 14.725(040)  \\ 
122 & 11:45:35.90 & $-$50:35:10.9 & 15.899(032) & 15.448(017) & 14.885(026)  \\ 
123 & 11:45:24.27 & $-$50:35:24.5 & 15.868(018) & 15.349(015) & 14.781(051)  \\ 
124 & 11:45:24.36 & $-$50:36:09.0 & 16.670(090) & 16.270(017) & 15.924(127)  \\ 
\enddata
\tablenotetext{a}{Note. -- Values in parenthesis are 1-$\sigma$ uncertainties that correspond to the rms of the instrumental errors of the photometry obtained over a minimum of three nights observed relative to standard star fields.}
\end{deluxetable}

\clearpage
\onecolumn
\begin{deluxetable}{lccccccc}
\tabletypesize{\scriptsize}
\tablewidth{0pt}
\tablecaption{Optical photometry of SN~2013L in the Swope "natural" photometric system.\label{tab:phot}}
\tablehead{
\colhead{JD}&
\colhead{Phase$^{*}$}&
\colhead{$u$}&
\colhead{$g$}&
\colhead{$r$}&
\colhead{$i$}&
\colhead{$B$}&
\colhead{$V$}\\
\colhead{(days)}&
\colhead{(days)}&
\colhead{(mag)}&
\colhead{(mag)}&
\colhead{(mag)}&
\colhead{(mag)}&
\colhead{(mag)}&
\colhead{(mag)}}
\startdata
2456316.71 &      2.21   &  15.678(012) & 15.595(009) & 15.638(009) & 15.746(009) & 15.692(008) & 15.596(008) \\ 
2456317.81 &      3.31   &  15.683(012) & 15.575(010) & 15.600(010) & 15.703(008) & 15.679(010) & 15.570(008) \\ 
2456318.78 &      4.28   &  15.698(012) & 15.563(010) & 15.586(010) & 15.677(015) & 15.678(010) & 15.552(010) \\ 
2456319.75 &      5.25   &  15.710(013) & 15.562(011) & 15.561(010) & 15.648(013) & 15.683(009) & 15.534(010) \\ 
2456320.71 &      6.21   &  15.733(011) & 15.561(009) & 15.547(009) & 15.617(008) & 15.686(008) & 15.528(008) \\ 
2456321.72 &      7.22   &  15.760(012) & 15.561(009) & 15.539(010) & 15.600(011) & 15.691(008) & 15.539(008) \\ 
2456322.80 &      8.30   &  15.781(014) & 15.584(010) & 15.524(010) & 15.584(011) & 15.701(010) & 15.527(009) \\ 
2456323.77 &      9.27   &  15.878(014) & 15.577(010) & 15.536(011) & 15.581(013) & 15.731(009) & 15.531(010) \\ 
2456324.80 &      10.30   &  15.857(012) & 15.601(008) & 15.541(009) & 15.567(009) & 15.752(008) & 15.547(008) \\ 
2456325.76 &      11.26   &  15.894(013) & 15.623(010) & 15.550(010) & 15.565(012) & 15.769(010) & 15.559(008) \\ 
2456326.80 &      12.30   &  15.949(012) & 15.643(009) & 15.545(011) & 15.576(015) & 15.796(009) & 15.572(008) \\ 
2456327.76 &      13.26   &  15.992(011) & 15.670(009) & 15.564(009) & 15.584(009) & 15.831(008) & 15.589(008) \\ 
2456328.72 &      14.22   &  16.050(013) & 15.723(007) & 15.573(009) & 15.578(006) & 15.876(008) & 15.599(008) \\ 
2456330.70 &      16.20   &  16.154(012) & 15.746(010) & 15.598(011) & 15.597(012) & 15.940(010) & 15.638(010) \\ 
2456336.82 &      22.32   &  16.495(014) & 15.941(009) & 15.710(011) & 15.665(011) & 16.150(013) & 15.802(012) \\ 
2456338.88 &      24.38   &  16.601(015) & 16.005(009) & 15.759(012) & 15.698(017) & 16.247(012) & 15.850(011) \\ 
2456339.77 &      25.27   &  16.648(014) & 16.040(009) & 15.774(011) & 15.718(015) & 16.265(013) & 15.887(010) \\ 
2456340.90 &      26.40   &  16.753(019) & 16.079(010) & 15.791(012) & 15.725(012) & 16.256(024) & 15.940(016) \\ 
2456341.90 &      27.40   &  16.792(024) & 16.114(009) & 15.809(011) & 15.761(011) & 16.366(013) & 15.938(010) \\ 
2456344.77 &      30.27   &  16.965(014) & 16.218(008) & 15.886(011) & 15.807(011) & 16.479(012) & 16.032(011) \\ 
2456352.74 &      38.24   &  17.411(019) & 16.488(012) & 16.060(014) & 15.946(018) & 16.772(014) & 16.260(013) \\ 
2456355.70 &      41.20   &  17.570(021) & 16.602(015) & 16.108(016) & 16.009(014) & 16.884(015) & 16.352(013) \\ 
2456358.85 &      44.35   &  17.718(018) & 16.700(010) & 16.202(013) & 16.075(012) & 17.008(016) & 16.431(012) \\ 
2456366.69 &      52.19   &  \ldots      & \ldots      & \ldots      & \ldots      & 17.310(015) & \ldots      \\ 
2456369.75 &      55.25   &  18.204(024) & 17.080(013) & 16.447(014) & 16.307(022) & \ldots      & \ldots      \\ 
2456370.91 &      56.41   &  18.227(047) & 17.105(009) & 16.453(011) & 16.324(011) & 17.419(074) & 16.797(022) \\ 
2456373.76 &      59.26   &  \ldots      & 17.241(012) & 16.577(013) & 16.453(012) & 17.580(016) & 16.934(012) \\ 
2456380.84 &      66.34   &  18.638(044) & 17.487(016) & 16.747(015) & 16.674(021) & 17.826(018) & 17.204(019) \\ 
2456388.82 &      74.32   &  \ldots      & 17.664(013) & 16.885(012) & 16.856(014) & \ldots      & \ldots      \\ 
2456407.78 &      93.28   &  \ldots      & 17.807(024) & 16.953(014) & 17.056(014) & \ldots      & \ldots      \\ 
2456424.68 &      110.18   &  19.124(034) & 17.891(011) & 16.968(012) & 17.156(017) & 18.252(015) & 17.622(009) \\ 
2456435.58 &      121.08   &  19.286(059) & 17.915(013) & 16.974(010) & 17.220(012) & 18.282(018) & 17.667(014) \\ 
2456438.58 &      124.08   &  19.238(069) & 17.903(012) & 16.965(011) & 17.208(014) & 18.268(021) & 17.672(012) \\ 
2456445.58 &      131.08   &  19.255(032) & 17.936(009) & 16.970(011) & 17.264(016) & 18.320(013) & 17.693(009) \\ 
2456448.63 &      134.13   &  19.236(034) & 17.939(010) & 16.965(009) & 17.277(014) & 18.326(011) & 17.712(009) \\ 
2456611.86 &      297.36   &  \ldots      & \ldots      & \ldots      & \ldots      & 19.214(019) & 18.657(038) \\ 
2456632.86 &      318.36   &  \ldots      & 18.987(021) & \ldots      & \ldots      & \ldots      & \ldots      \\ 
2456636.84 &      322.34   &  20.369(067) & 19.020(017) & 17.836(023) & 18.639(027) & \ldots      & \ldots      \\ 
2456649.78 &      335.28   &  20.530(097) & 19.111(023) & 17.926(020) & 18.719(036) & 19.443(027) & 18.917(019) \\ 
2456651.83 &      337.33   &  20.545(081) & 19.136(023) & 17.938(030) & 18.743(055) & 19.439(025) & 18.897(018) \\ 
2456678.87 &      364.37   &  \ldots      & 19.362(030) & 18.104(023) & \ldots      & \ldots      & \ldots      \\ 
2456679.87 &      365.37   &  20.668(111) & \ldots      & \ldots      & 18.987(034) & \ldots      & \ldots      \\ 
2456707.74 &      393.24   &  \ldots      & 19.543(033) & 18.305(020) & \ldots      & \ldots      & \ldots      \\ 
2456743.61 &      429.11   &  \ldots      & 19.821(040) & 18.556(023) & 19.477(072) & \ldots      & \ldots      \\ 
2456759.64 &      445.14   &  \ldots      & 20.031(056) & 18.649(024) & 19.693(072) & \ldots      & \ldots      \\ 
2456787.52 &      473.02   &  \ldots      & 20.270(108) & 18.755(043) & \ldots      & \ldots      & \ldots      \\ 
2456791.50 &      477.00   &  \ldots      & 20.416(062) & 18.902(037) & \ldots      & \ldots      & \ldots      \\ 
2456979.85 &      665.35   &  \ldots      & \ldots      & 20.473(065) & \ldots      & \ldots      & \ldots      \\ 
2457010.82 &      696.32   &  \ldots      & \ldots      & 20.654(129) & \ldots      & \ldots      & \ldots      \\ 
  \enddata
    \tablenotetext{*}{Days since discovery.}
\end{deluxetable}

\begin{deluxetable}{lccccc}
\tabletypesize{\scriptsize}
\tablewidth{0pt}
\tablecaption{Near-infrared photometry of SN~2013L in the du Pont natural photometric system.\label{tab:photnir}}
\tablehead{
\colhead{JD}&
\colhead{Phase$^{**}$}&
\colhead{$Y$}&
\colhead{$J$}&
\colhead{$H$}&
\colhead{$K_s$}\\
\colhead{(days)}&
\colhead{(days)}&
\colhead{(mag)}&
\colhead{(mag)}&
\colhead{(mag)}&
\colhead{(mag)}}
\startdata 
2456318.75 & 4.25    &  15.331(006) &  15.216(005) &   15.136(009)  & \ldots        \\
2456319.81 & 5.31    &  15.292(006) &  15.184(008) &   15.101(016)  & \ldots        \\
2456320.77 & 6.27    &  15.266(006) &  15.084(009) &   15.047(015)  & \ldots        \\
2456325.88*& 11.38   &  15.130(009)*&  15.028(011)*&   14.999(010)* & \ldots        \\
2456326.78 & 12.28   &  15.168(006) &  15.011(008) &   14.888(016)  & \ldots        \\
2456342.87 & 28.37   &  15.224(006) &  14.997(005) &   14.812(015)  & \ldots        \\
2456343.77 & 29.27   &  15.224(006) &  15.003(009) &   14.816(017)  & \ldots        \\
2456346.73 & 32.23   &  15.228(007) &  15.020(010) &   14.827(016)  & \ldots        \\
2456348.68 & 34.18   &  15.269(006) &  15.057(005) &   14.736(009)  & \ldots        \\
2456351.74 & 37.24   &  15.320(008) &  15.060(006) &   14.795(015)  & \ldots        \\
2456354.71 & 40.21   &  15.328(006) &  15.062(006) &   14.823(012)  & \ldots        \\
2456382.77 & 68.27   &  15.898(006) &  15.573(009) &   15.232(016)  & \ldots        \\
2456648.83 & 334.33  &  17.576(012) &  \ldots      &   16.276(017)  & \ldots        \\
2456649.77 & 335.27  &  \ldots      &  17.017(014) &   \ldots       & \ldots        \\
2456725.63 & 411.13  &  18.018(016) &  17.249(017) &   16.430(025)  & \ldots        \\
2457030.79 & 716.29  &  19.761(049) &  \ldots      &   \ldots       & \ldots        \\
2457032.80 & 718.30  &  19.732(067) &  \ldots      &   \ldots       & \ldots        \\
2457053.86*& 739.36  &  19.746(014)*&  19.258(022)*&   17.625(014)* & 15.925(162)*  \\
2457092.65 & 778.15  &  20.056(148) &  19.154(094) &   17.604(063)  &  \ldots       \\
2457112.69 & 798.19  &  \ldots      &  19.466(107) &   17.878(050)  &  \ldots       \\
2457201.47 & 886.97  &  \ldots      &  \ldots      &   18.106(069)  &  \ldots       \\
  \enddata
    \tablenotetext{**}{Days since discovery.}
  \tablenotetext{*}{Data from the Magellan Baade ($+$FourStar) telescope.}
\end{deluxetable}

\begin{deluxetable}{lccccccc}
\tabletypesize{\scriptsize}
\tablewidth{0pt}
\tablecaption{UVOT photometry of SN~2013L.\label{tab:photUVOT}}
\tablehead{
\colhead{MJD}&
\colhead{Phase$^{*}$}&
\colhead{$uw2$}&
\colhead{$um2$}&
\colhead{$uw1$}&
\colhead{$U$}&
\colhead{$B$}&
\colhead{$V$}\\
\colhead{(days)}&
\colhead{(days)}&
\colhead{(mag)}&
\colhead{(mag)}&
\colhead{(mag)}&
\colhead{(mag)}&
\colhead{(mag)}&
\colhead{(mag)}}
\startdata
56320.98   &   6.98 & 15.068(062)  & 15.067(087)  & 14.789(054) &  14.691(047) & 15.780(057) & 15.481(135)  \\  
56322.57   &   8.57 & 15.273(067)  & $\cdots$     & 14.870(054) &  $\cdots$   & $\cdots$    & $\cdots$     \\  
56322.85   &   8.85 & 15.357(072)  & 15.133(063)  & 14.965(063) &  14.755(051) & 15.748(059) & 15.521(067)  \\  
56324.49   &  10.49 & 15.615(073)  & 15.344(064)  & 15.100(064) &  14.849(052) & 15.733(058) & 15.606(068)  \\  
56326.40   &   12.4 & 15.924(074)  & 15.600(064)  & 15.319(064) &  14.960(054) & 15.928(061) & 15.676(066)  \\  
56328.30   &   14.3 & 15.944(074)  & 15.693(064)  & 15.399(065) &  15.044(055) & 15.903(061) & 15.656(067)  \\  
56330.27   &  16.27 & 16.098(080)  & 15.857(069)  & 15.619(071) &  15.163(062) & 15.989(067) & 15.614(076)  \\  
56332.91   &  18.91 & 16.386(078)  & 16.186(068)  & 15.820(068) &  15.299(061) & 16.063(063) & 15.701(068)  \\  
56336.83   &  22.83 & 16.812(094)  & 16.542(081)  & 16.149(082) &  15.581(076) & 16.243(073) & 15.789(085)  \\  
56338.13   &  24.13 & 17.053(090)  & 16.801(080)  & 16.326(077) &  15.762(071) & 16.334(068) & 15.860(074)  \\  
56340.73   &  26.73 & 17.186(094)  & 16.969(081)  & 16.550(082) &  15.819(071) & 16.391(068) & 15.861(074)  \\  
56342.59   &  28.59 & 17.563(102)  & 17.225(084)  & 16.781(085) &  16.054(073) & 16.514(069) & 15.973(074)  \\  
56344.39   &  30.39 & 17.742(105)  & 17.476(089)  & 16.811(083) &  16.115(072) & 16.534(067) & 16.029(073)  \\  
56346.43   &  32.43 & 17.909(112)  & 17.523(090)  & 16.947(086) &  16.160(073) & 16.574(068) & 16.126(075)  \\  
56348.53   &  34.53 & 18.106(122)  & 17.890(104)  & 17.362(101) &  16.356(076) & 16.766(071) & 16.167(076)  \\  
56350.64   &  36.64 & 18.331(163)  & 17.979(130)  & 17.342(120) &  16.508(093) & 16.843(084) & 16.165(092)  \\  
56352.77   &  38.77 & 18.197(168)  & 18.155(155)  & 17.229(126) &  16.447(100) & 16.880(093) & 16.184(102)  \\  
56354.54   &  40.54 & 18.532(153)  & 18.228(124)  & 17.515(109) &  16.654(083) & 16.934(075) & 16.198(078)  \\  
56356.88   &  42.88 & 18.565(170)  & 18.428(147)  & 17.646(127) &  16.937(100) & 16.964(081) & 16.261(087)  \\ 
\enddata
\tablenotetext{*}{Days since discovery.}
\end{deluxetable}

\begin{deluxetable}{ccccc}
\tabletypesize{\scriptsize}
\tablewidth{0pt}
\tablecaption{Journal of Spectroscopic Observations of SN~2013L.\label{tab:spectra}}
\tablehead{
\colhead{Date (UT)}&
\colhead{JD}&
\colhead{Phase$^{*}$}&
\colhead{Telescope+Instrument}&
\colhead{Range}\\
\colhead{}&
\colhead{(days)}&
\colhead{(days)}&
\colhead{}&
\colhead{(\AA)}}
\startdata
        28.27 Jan 2013 &  2456320.77   & 6.27    &  NTT+EFOSC$^{**}$ &  3649--9245   \\ 
         01.24 Feb 2013 &  2456324.74   & 10.24   &  Clay+MIKE    &  3300--9400       \\
         02.14 Feb 2013 &  2456325.64   & 11.14   &  Baade+FIRE   &  8022--24801  \\ 
        18.35 Feb 2013 &  2456341.85   & 27.35   &  Dupont+BC    &  3287--9491   \\
        24.17 Feb 2013 &  2456347.67   & 33.17   &  VLT+Xshooter$^{**}$ &  3050--24790  \\ 
        28.28 Feb 2013 &  2456351.78   & 37.28   &  Baade+FIRE   &  8021--25016  \\
        11.22 Mar 2013 &  2456362.72   & 48.22   &  VLT+Xshooter$^{**}$ &  3050--24790  \\ 
        31.13 Mar 2013 &  2456382.63   & 68.13   &  VLT+Xshooter$^{**}$ &  3050--24790  \\ 
        11.06 May 2013 &  2456423.56   & 109.06  &  VLT+Xshooter$^{**}$ &  3050--24790  \\
        15.10 Jun 2013 &  2456458.60   & 144.10  &  Baade+FIRE    &  8092--24812  \\
        20.35 Nov 2013 &  2456616.85   & 302.35  &  Baade+FIRE   &  7997--24810  \\ 
        18.33 Dec 2013 &  2456644.83   & 330.33  &  Clay+MagE   &  3973--9438   \\  
        27.24 Feb 2014 &  2456715.74   & 401.24  &  Baade+FIRE   &  8022--24807  \\
19 May 2014 & 2456796.5 & 482  & Magellan+IMACS$^{**}$ &6488--6866$^{***}$  \\
20 Jan 2015 & 2457042.5 &728  & Magellan+IMACS$^{**}$ &6488--6866$^{***}$  \\
02.13 Jun 2015 &  2457175.63   & 861.13  &  Baade+FIRE   &  8031--24817  \\   
05 Mar 2016 & 2457452.5 &1133 & Magellan+IMACS$^{**}$ &6488--6866$^{***}$ \\
11 Mar 2017 & 2457823.5 &1509 & Magellan+IMACS$^{**}$ &6488--6866$^{***}$  \\
     
\enddata
\tablenotetext{*}{Days since discovery. }
\tablenotetext{**}{Public available spectra published by A17.}
\tablenotetext{***}{Spectra from  A17.}
\end{deluxetable}

\begin{deluxetable}{cccccc}
\tabletypesize{\scriptsize}
\tablewidth{0pt}
\tablecaption{Peak epochs and magnitudes of SN~2013L.\label{tab:peak}}
\tablehead{
\colhead{Filter}&
\colhead{Peak epoch$^{*}$}&
\colhead{Peak magnitude}&
\colhead{Peak absolute magnitude}&
\colhead{$\Delta m_{80}$}&
\colhead{$\Delta m_{80-300}$}\\
\colhead{}&
\colhead{(days)}&
\colhead{(mag)}&
\colhead{(mag)}&
\colhead{(mag)}&
\colhead{(mag)}}
\startdata
$uw2$ & $<$+7   & $<$15.07    & $<-$19.95 & $\cdots$  &  $\cdots$ \\
$um2$ & $<$+7   & $<$15.07    & $<-$19.91 & $\cdots$  &  $\cdots$ \\
$uw1$ & $<$+7   & $<$14.79    & $<-$20.14 & $\cdots$  &  $\cdots$ \\
$u$   & $<$+1   & $<$15.68    & $<-$19.13 & $>$3.27 & 1.32  \\
$B$   &  +4     &    15.68$\pm$0.01    &    $-$19.07$^{+0.24}_{-0.16}$ & 2.40 & 1.15  \\
$g$   &  +6     &    15.56$\pm$0.01    &    $-$19.15$^{+0.24}_{-0.16}$ & 2.17 & 1.12  \\
$V$   &  +8     &    15.53$\pm$0.01    &    $-$19.11$^{+0.24}_{-0.16}$ & 1.96 & 1.18  \\
$r$   &  +8     &    15.53$\pm$0.01    &    $-$19.06$^{+0.24}_{-0.16}$ & 1.40 & 0.76  \\
$i$   & +11     &    15.57$\pm$0.01    &    $-$18.94$^{+0.24}_{-0.16}$ & 1.38 & 1.51  \\
$Y$   & +17     &    15.16$\pm$0.02    &    $-$19.26$^{+0.24}_{-0.16}$ & 0.62 & 1.31  \\
$J$   & +19     &    14.99$\pm$0.02    &    $-$19.39$^{+0.24}_{-0.16}$ & 0.51 & 1.17  \\
$H$   & +28     &    14.86$\pm$0.03    &    $-$19.49$^{+0.24}_{-0.16}$ & 0.36 & 0.87  \\
\enddata
\tablenotetext{*}{Days since discovery. The errors on the peak absolute magnitudes include and are dominated by the error on the adopted distance to the host galaxy.}
\end{deluxetable}

\clearpage

\clearpage

\clearpage

\end{document}